\documentclass[12pt]{article}

\usepackage[margin=1in]{geometry}

\usepackage[parfill]{parskip}
\usepackage[utf8]{inputenc}
\usepackage{natbib}
\usepackage{amsmath,amssymb,amsfonts,amsthm,mathabx}
\usepackage{subfiles}
\usepackage{cancel}

\usepackage{bbold}

\newcommand{\Beta}{\mbox{\boldmath $\beta$}}
\newcommand{\Eta}{\mbox{\boldmath $\eta$}}

\newcommand{\diag}[1]{ \text{diag}\{#1\}}

\newcommand{\bel}[1]{\begin{equation}\label{#1}\begin{aligned}}

\newcommand{\eel}{\end{aligned}\end{equation}}

\newcommand{\be}{\begin{equation*}\begin{aligned}}

\newcommand{\ee}{\end{aligned}\end{equation*}}
\newtheorem{theorem}{Theorem}
\newtheorem{remark}{Remark}

\usepackage{graphicx}

\usepackage{subcaption}
\usepackage{float}

\usepackage{environ}

\newif\ifshowproof
\showprooffalse

\NewEnviron{Proof}{%
    \ifshowproof%
        \begin{proof}%
            \BODY
        \end{proof}%
    \fi%
}%

\renewcommand{\t}[1]{\text{#1}}

\newenvironment{psmallmatrix}{\left(\begin{smallmatrix}}{\end{smallmatrix}\right)}

\showprooftrue

\begin{document}

\title{Transport Monte Carlo:\\High-Accuracy Posterior Approximation\\ via Random Transport}
\def\spacingset#1{\renewcommand{\baselinestretch}%
{#1}\small\normalsize} \spacingset{1}
\author{ Leo L. Duan\thanks{Department of Statistics, University of Florida, Gainesville, FL, email: li.duan@ufl.edu} }
\date{}
\maketitle

\medskip

{\bf Abstract:}
In Bayesian applications, there is a huge interest in rapid and accurate estimation of the posterior distribution, particularly for high dimensional or hierarchical models. In this article, we propose to use optimization to solve for a joint distribution (random transport plan) between two random variables,  $\theta$  from the posterior distribution and $\beta$ from the simple multivariate uniform. Specifically, we obtain an approximate estimate of the conditional distribution $\Pi(\beta\mid \theta)$ as an infinite mixture of simple location-scale changes; applying the Bayes' theorem, $\Pi(\theta\mid\beta)$ can be sampled as one of the reversed transforms from the uniform, with the weight proportional to the posterior density/mass function. This produces
independent random samples with high approximation accuracy, as well as nice theoretic guarantees. Our method shows compelling advantages in performance and accuracy, compared to the state-of-the-art Markov chain Monte Carlo and approximations such as variational Bayes and normalizing flow. We illustrate this approach via several challenging applications, such as sampling from multi-modal distribution, estimating sparse signals  in high dimension, and soft-thresholding of a graph with a prior on the degrees.


\newpage

\section{Introduction}
The Bayesian framework is routinely used to impose model regularization and obtain uncertainty quantification. As the posterior distribution often does not have a closed-form, it is common to rely on the Monte Carlo estimation. The Markov chain Monte Carlo (MCMC) has been the most popular method due to the ability to alternatively update one part of the parameter each time; often, each conditional update is easy to carry out, such as having tractable full conditional form. As a side effect, this creates a Markov chain dependency among the collected samples; to reduce this effect, one can filter down the collected Markov chain (often known as ``thinning''), by keeping the samples that are a few iterations apart and discarding the ones in between.

A primary challenge is that modern Bayesian applications often face complications  such as high dimensionality or hierarchical structure, the above
computing strategy can become very inefficient: since each update corresponds to a small local change, the Markov chain will still be highly auto-correlated, even after a sizeable amount of thinning. This is known as the low effective sample size problem, or slow mixing of Markov chains. This issue has been well known for a long time in the community, yet it was formally studied only until recently. See \cite{rajaratnam2015mcmc} on the failing of convergence rate guarantee in high dimension, \cite{johndrow2019mcmc} on the case of imbalanced categorical data, \cite{duan2018scaling} on the need to calibrate the step size for data augmentation, etc. For a recent survey on this issue, see \cite{robert2018accelerating}. This issue has motivated a large literature of new Markov chain methods, using different proposing algorithms such as those originating from physics, to make the new state less correlated to the current one. Examples include Metropolis-adjusted Langevin algorithm \citep{roberts1996exponential}, Hamiltonian Monte Carlo \citep{neal2011mcmc}, piecewise deterministic \citep{bierkens2019zig}, or continuous-time MCMC \citep{fearnhead2018piecewise}.

At the same time, there is a sizeable literature focusing on sampling approaches, that bypass the use of Markov chains; they are capable of generating independent random samples. For example, the approximate Bayesian computation \citep{beaumont2009adaptive} rejection algorithm samples a parameter from the prior, simulates a set of data and compare with actually observed ones, and accept the sampled parameter if the data divergence is small; the variational Bayes \citep{blei2017variational} approximates the posterior with another simple distribution, such as one assuming independence for a multivariate parameter. Despite their popularity,
a primary concern is that there is a non-negligible gap (that is, a positive statistical distance even under idealized condition) between the target posterior and the approximation ---  this gap could impact  the accuracy of the uncertainty quantification
such as the covariance estimation. For the discussion and some remedy on those issues, see \cite{giordano2018covariances}.

Among these approaches, a particularly distinctive approach involves searching for a ``transport map'', an invertible mapping between the posterior and a ``reference'' distribution, a relatively easy-to-simulate distribution such as a multivariate normal. The pioneering work was proposed by \cite{el2012Bayesian}. By parameterizing the transport map as monotonic, the transformed distribution of the reference can be obtained in a closed-form via the change-of-variable. Then one could estimate the working parameters in the mapping via minimizing a divergence between the target posterior and the transformed reference. Compared to the other approaches, a major advantage is that if an invertible solution does exist, then in theory, there is no approximation error, and the algorithm would have very high accuracy.

The crux of the problem is how to parameterize the invertible transform with sufficient flexibility? This has generated a large class of interesting work. \cite{parno2018transport} proposed to parameterize the invertible transform via a lower-triangular mapping (where the $k$th output variable depends on the first $k$ input variables), which approximates the flexible Knothe-Rosenblatt rearrangement transform between two probability measures.   \cite{spantini2018inference} further imposed sparse or decomposable structure on these transport maps, assuming that there is a low-dimensional coupling between two high-dimensional variables. \cite{doucetgibbs2021} proposed to use an ordinary differential equation to parameterize the transport map.

At the same time, there is a machine learning literature, commonly known as normalizing flow, that attempts to automate this parameterization procedure (\cite{pmlr-v37-rezende15,dinh2016density,papamakarios2017masked}; among others). Specifically, the customized mapping is replaced by an approximating neural network that has a guarantee in its invertibility. Despite some empirical success, the large number of working parameters in the neural network pose a  challenge to scale up for high dimensional posterior estimation. A recent theoretic study  \citep{pmlr-v108-kong20a} formally demonstrated a curse-of-dimensionality result, that the depth of the neural network needs to grow at a polynomial rate of the dimension of the target distribution; hence it is a demanding computational problem, with the potential solution that could be provided by an infinite-depth neural network \citep{chen2018neural}. 

We are largely inspired by the rapid development in this field; nevertheless, here we explore a quite different alternative: instead of relying on one sophisticated transport map, we consider an infinite mixture of maps, where each map can be as simple as a location-scale change. This can be viewed as ``a wide but shallow model'' for the transport problem instead of ``one deep model''.

This mixture of maps framework has two major benefits. (i) It forms a non-deterministic joint distribution (a transport {\em plan}) between the reference and the posterior distributions, hence can be used to connect the two, even if there is a measure dimension discrepancy. For example, if the target is a degenerate normal distribution in a subspace of $\mathbb{R^p}$   (such as in the variable selection model, most elements of the model parameter will be fixed to zero), while the reference is a non-degenerate $p$-variate normal. Notably, we can also transport a continuous reference to a discrete target distribution (that is, of zero-dimensional measure). In these cases, the transport map methods are not suitable, due to the lack of solutions in invertible maps. (ii) The flexibility of the infinite mixture allows us to use a very simple parameterization for each component map, leading to both computational ease and tractability for the theoretic analysis. In our method, the transport plan can be estimated using optimization; then, each new approximate posterior sample is generated as a random draw from several candidates, where each candidate is calculated via the fast transform from a reference sample. We call this algorithm the ``Transport Monte Carlo''.

The idea of using transport plans (or, ``couplings'') is commonly seen in the optimal transport literature [see Chapter 6 of \cite{ambrosio2008gradient}], in which one searches for the best plan that minimizes a given transportation cost function. There has been a rich class of methods, such as \cite{solomon2015convolutional,NEURIPS2019_f0935e4c}, as well as efficient algorithms, such as \cite{cuturi2013sinkhorn}. Nevertheless, since we are considering the posterior sampling problem, there is no cost function; hence we only need to solve for {\em one} plan (among many solutions) that connects the reference and the target distributions. Therefore, our focus is different. We will discuss potentially interesting connections at the end of the article.

The rest of the article is organized as follows: in Section 2, we introduce the transport plan and discuss its parameterization; in Section 3, we describe the algorithmic details; in Section 4, we establish the theoretical properties, including asymptotic guarantee, approximation error due to finite samples; in Section 5, we compare our approach with the state-of-art Hamiltonian Monte Carlo algorithms; in Section 6, we demonstrate the performance through several challenging posterior estimation tasks.

\section{Transport Monte Carlo}

\subsection{Two Types of Transport: Deterministic versus Random}

In order to properly introduce the Transport Monte Carlo approach, we first define some notations and give a brief review of the relevant transport concepts. Let $\theta\in {\Theta}$ be a parameter of interest, $\Pi_0(\theta)$ the prior density/mass function, $y$ the data and $L(y;\theta)$ the likelihood. Our interest is the random variable from the posterior, associated with the measure $\mu_\theta: \mathcal{B}(\Theta) \to \mathbb R_+$, with $\mathcal{B}$ the Borel $\sigma$-algebra:
 \[
         \theta\sim\Pi(\theta; y)  = \{z(y)\}^{-1}{L(y;\theta)\Pi_0(\theta)},
 \]
 where $z(y)= \int_{\Theta} L(y;\theta)\Pi_0( \theta)\textup{d} \theta$ or $z(y)= \sum_{\Theta} L(y;\theta)\Pi_0( \theta)$ is the normalizing constant.
  Let  $\beta \in {\Beta}$ be a random variable  from the reference distribution,
 \[
         \beta \sim \Pi_r(\beta),
 \]
 where $ \Pi_r(\beta)$ is the density/mass of another proper measure $\mu_{\beta}: \mathcal{B}(\Beta) \to \mathbb R_+$. For the ease of notation, we use $\Pi$ to denote both distribution and density/mass function, and we will use the name ``posterior distribution'' and ``target distribution'' interchangeably.

To introduce the transport idea, consider the earth mover's intuition: imagine a discrete distribution as a pile of earth, scattered at locations $\beta$'s and each containing mass $\Pi_r(\beta)$ (for continuous distribution, we can imagine each location as a small neighborhood around $\beta$). Our goal is to move the earth to locations $\theta$'s so that each contains mass $\Pi(\theta;y)$.

A simple strategy is known as the Monge transport \citep{monge1781memoire}: at location $\beta$, we move {\em all} the mass there to new location $T(\beta)$, with $T$ a deterministic transform $T:\Beta\to \Theta$, so that we have  $\mu_{\theta}(\mathcal A) =\mu_{\beta}[\{x:T(x)\in \mathcal A\}]$ for all $\mathcal A\in \mathcal{B}(\Theta)$.
However, there are two issues --- (i) for some combination of the reference and target distributions, the Monge transport may not exist; that is, there is not a feasible $T$ to make $\mu_{\theta}(\mathcal A) =\mu_{\beta}[\{x:T(x)\in \mathcal A\}]$. As a classic toy example, it is impossible to use a one-to-one transform for changing $\mu_{\beta}$ from a point mass at zero to a Bernoulli $\mu_{\theta}$   [supported at one with probability $p_1$ and zero with probability $(1-p_{1})$].
 (ii) the parameterization of $T$ is often a challenging task, especially when $\Pi(\theta;y)$ is complicated.

\begin{figure}[H]
     \centering
        \begin{subfigure}[b]{.48\textwidth}
         \centering
         \includegraphics[scale=1, width=\textwidth]{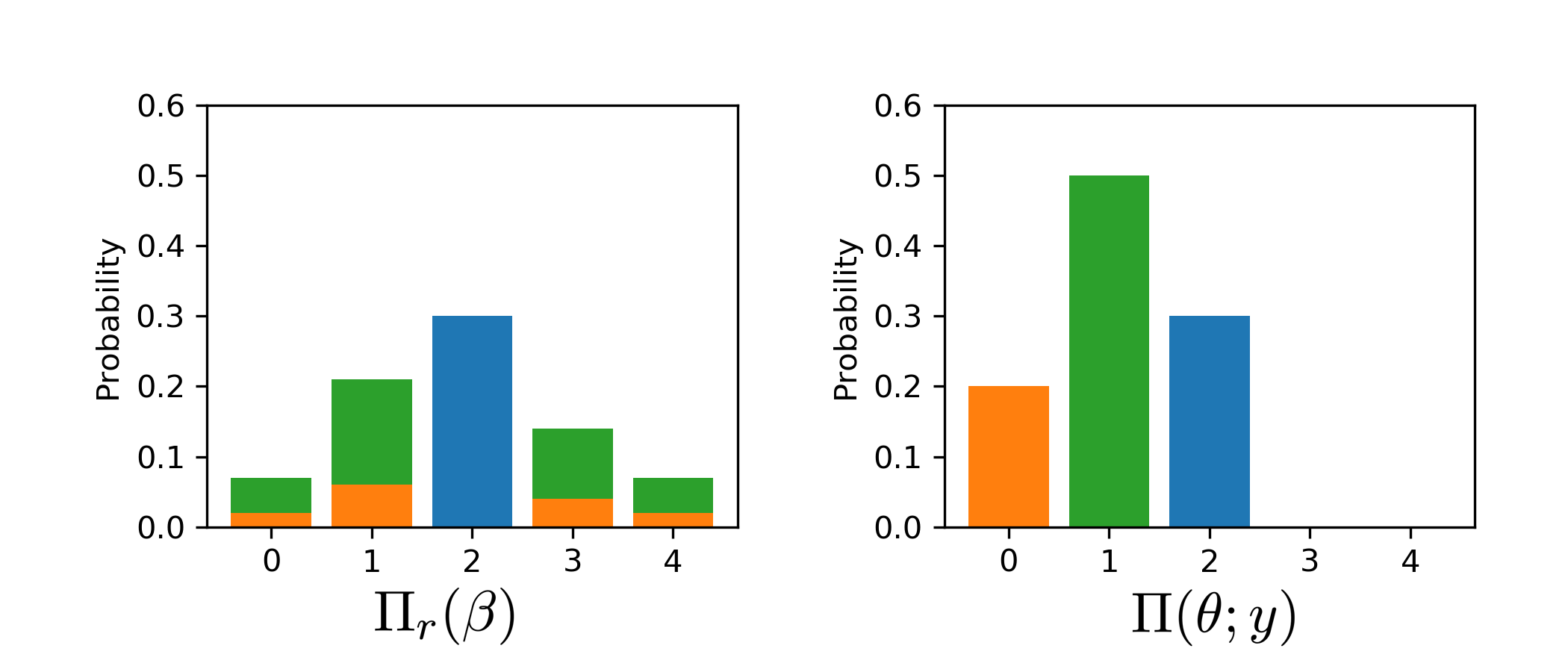}
         \caption{A transport plan to change a discrete distribution supported at 5 points to one at 3 points.}
     \end{subfigure}
     \quad
             \begin{subfigure}[b]{.48\textwidth}
         \centering
         \includegraphics[ width=\textwidth]{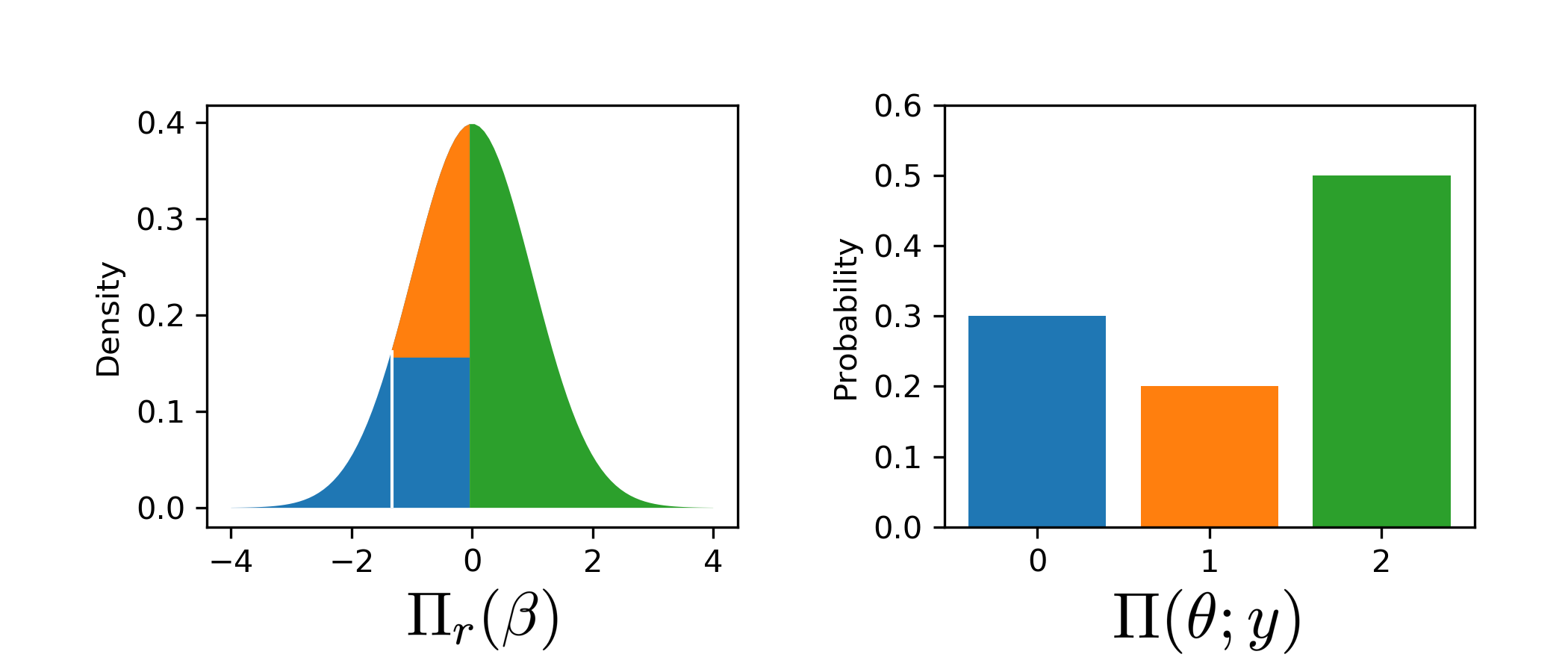}
         \caption{A transport plan to change a continuous distribution to a discrete supported at 3 points.}
     \end{subfigure}
     \caption{Two examples  showing the flexibility of  the random transport plan (the Kantorovich transport): in each panel, the colors in left figure represent the conditional distribution $\Pi(\theta\mid \beta)$, each is moved into the  block of the matching color in the right figure. In both examples, it is impossible to use an invertible transform $T$ to change $\Pi_r(\beta)$ to $\Pi(\theta;y)$.
  \label{fig:monge_vs_kantorovich}}
\end{figure}

These limitations of the Monge transport have motivated another one called the Kantorovich transport \citep{kantorovich1942translocation}:
at location $\beta$, instead of moving all the mass in the same way, we split the mass there into smaller units according to a conditional probability distribution $\Pi(\theta\mid \beta)$ for possible values $\theta_1,\theta_2,\ldots$, then moving each unit to $\theta_1,\theta_2,\ldots$ accordingly. For the above point-mass-to-Bernoulli transport, we can take $\text{pr}(\theta=1\mid \beta=0)=p_{1}$ and $\text{pr}(\theta=0\mid
\beta=0)=1-p_{1}$. Figure~\ref{fig:monge_vs_kantorovich} shows two more sophisticated examples.

The Kantorovich transport always exists --- after all, it is equivalent to finding a joint distribution between $\theta$ and $\beta$, also known as a  transport plan:
 \bel{eq:random_transport}
(\theta,\beta)  \sim \Pi(\theta, \beta), \text{ such that}\quad  \int_{\Beta} \Pi(\theta, \beta) \textup{d}\beta = \Pi(\theta; y), \int_{\Theta} \Pi(\theta, \beta) \textup{d}\theta = \Pi_r(\beta),
\eel
for continuous $(\beta,\theta)$; and  for discrete ones, the integrals are replaced with summations.


\subsection{Random Transport Plan as an Infinite Mixture}
Our goal is to use the transport plan for the posterior sampling. To be clear, we focus on the target posterior distribution with a density/mass function fully known except for some normalizing constant, and we will choose a reference $\Pi_r$ easy to sample from; hence, the only unknown part is the transport plan. For simplicity, we will focus on both $\beta$ and $\theta$ as continuous random variables from now on, with extension to the discrete $\theta$ deferred to a later section.
Without loss of generality, we assume both $\beta$ and $\theta$ are $p$-element vectors.

We want to find an approximate solution to the transport plan $\Pi(\beta, \theta)$ so that it is amenable to a tractable computation. On the surface, it may be tempting to start with $\beta\sim\Pi_r$ and find an approximation to 
$\Pi(\theta\mid \beta)$, so that the $\theta$-marginal density is close  to $\Pi(\theta;y)$; however, it is quite challenging to ensure the parameterization to  
$\Pi(\theta\mid \beta)$ is flexible enough.

Instead, we use the other factorization:  starting with the exact marginal $\theta\sim \Pi(\theta;y)$,  we use an approximate parameterization to  $\Pi(\beta\mid \theta)$ first; afterward, an application of the Bayes' theorem gives us $\Pi(\theta\mid \beta)$ that is proportional to the posterior density $\Pi(\theta;y)$ ---  intuitively, the reverse conditioning gives a calibration similar to importance weighting.

Specifically, we approximate the exact conditional kernel $\Pi(\beta\mid\theta)$ by an infinite mixture (for clarity, we will use $\tilde \Pi(.)$ to denote an approximation):
\bel{eq:rev_transport}
\tilde \Pi( \beta\mid \theta) = \sum_{k=1}^{\infty}w_k(\theta)\delta\{\beta
- T^{-1}_k(\theta)\},
\eel
where $\delta$ is the Dirac delta, representing a point mass distribution at $T^{-1}_k(\theta)$, $\int_{\Beta} \delta\{\beta
- T^{-1}_k(\theta)\} \textup d \beta=1$ and $\delta\{\beta
- T^{-1}_k(\theta)\}=0$ if $\beta\neq T^{-1}_k(\theta)$;  $w_k(\theta)\ge 0$ and $\sum_{k=1}^{\infty} w_k(\theta)=1$. This infinite mixture approximation was inspired by Bayesian non-parametric approximation of the conditional density \citep{dunson2007Bayesian}; nevertheless, the difference is that instead of treating $\beta$ as some predictor-based linear transform $x^{\rm T}\theta$, we set $T_k$ to be an invertible and differentiable transform.

We can view  $\beta\sim\tilde\Pi(\beta\mid \theta)$ as an augmented random variable drawn from $\{T^{-1}_k(\theta)\}_{k}$ with probability $w_k(\theta)$.
Although the conditional $\Pi(\beta\mid \theta)$ is a discrete distribution, when integrating over $\theta$, its marginal becomes a continuous  distribution:
\bel{eq:mixture_form}
\tilde \Pi(\beta) & = \int_{\Theta}  \tilde \Pi( \beta\mid \theta) \Pi(\theta; y) \textup{d} \theta \\
= &
\sum_{k=1}^{\infty}   w_k \{\ T_k(\beta)\} \Pi\{T_k(\beta); y\}
{|\textup{det} \nabla T_k(\beta)|} 1\{ T_k(\beta)\in \Theta \},
\eel
where the second line uses the change-of-variable in Dirac delta $ \delta\{\beta
- T^{-1}_k(\theta)\}=
{|\textup{det} \nabla T_k(\beta)|} \\ \delta\{\theta- T_k(\beta)\}$ and $\int_X f(x) \delta(x-y) \textup{d} x=  f(y) 1(y\in X)$, as well as the Fubini's theorem for exchanging summation and integration; $1(E)$ is an indicator function taking value $1$ if the event $E$ holds, or $0$ otherwise. In addition, regarding those $y\not\in X:f(y)=\infty$, we have $\int_X f(x) \delta(x-y) \textup{d} x=0$.

In the theory section, we will show that \eqref{eq:mixture_form} can well approximate some very simple continuous distributions, such as the multivariate uniform $\Pi_r(\beta)\sim \text{Uniform} \{(0, 1)^p\}$. Applying the Bayes' theorem, we obtain the reverse conditional  for sampling $\theta$ given $\beta$:
\bel{eq:rev_conditional}
\tilde\Pi\{\theta= T_k(\beta)\mid \beta\}\ &= \frac{ \Pi(\theta;y)\tilde\Pi( \beta\mid \theta){1}\{\theta= T_k(\beta)\}\ }{\tilde\Pi(\beta)}\\
& =\frac{ w_k \{ T_k(\beta)\} \Pi\{T_k(\beta); y\}
{|\textup{det} \nabla T_k(\beta)|} 1\{ T_k(\beta)\in \Theta \}}{
\sum_{k=1}^{\infty}   w_k \{ T_k(\beta)\} \Pi\{T_k(\beta); y\}
{|\textup{det} \nabla T_k(\beta)| 1\{ T_k(\beta)\in \Theta \}}
}\\
&:= v_k(\beta)
\eel
which is a discrete distribution drawn from the set $\{T_k(\beta)\}_k$. We  denote this conditional probability by $v_k(\beta)$ for convenience. To clarify, the union of the ranges of $T_k$'s does not need to cover the whole parameter space $\Theta$, but the high posterior density region; and  the \eqref{eq:rev_conditional} is conditioned on $y$ as well, and we omit $y$ for the ease of notation.
\begin{remark}[Difference from a mixture-based variational approximation]
It is important to distinguish $\tilde\Pi[\theta= T_k(\beta)\mid \beta]$ from a variational approximation using the  mixture $\sum_{k=1}^{\infty} v^*_k \delta\{\theta- T_k(\beta)\}$, with $v^*_k$ some constant weight that $\sum_k v^*_k=1$. In our case, the $v_k(\beta)$ in \eqref{eq:rev_conditional} is proportional to $\Pi(\theta;y)$, hence automatically favoring a transform $T_k$ that generates higher posterior density. This substantially reduces the burden to parameterize $T_k$.
\end{remark}
To develop an algorithm that we call Transport Monte Carlo (TMC), we optimize the working parameters in $\{(w_k, T_k)\}_{k}$ to match $\tilde \Pi(\beta)\approx\Pi_r(\beta)$, then sample $(\beta,\theta)$ via:
\bel{eq:generator}
& \beta \stackrel{iid}\sim \Pi_r, \\
& c\sim \text{Categorical}\{v_1(\beta),v_2(\beta),\ldots, \},\\
& \theta =T_c(\beta).
\eel
That is, $(\beta, \theta) \sim \Pi_r(\beta)\tilde \Pi(\theta\mid \beta)$. Note that the samples of $\theta$ are completely independent.

 \begin{remark}
If we could sample $\beta\sim \tilde\Pi(\beta)$ in the first step, then we would obtain the exact marginal
$\int \tilde\Pi(\beta)\tilde \Pi(\theta\mid \beta) \textup{d}\beta = \Pi(\theta; y)$. Because of the substitution, the samples from \eqref{eq:generator} are posterior approximation, with the error from the discrepancy between  $\tilde \Pi(\beta)$ and $\Pi_r(\beta)$.
\end{remark}

\subsection{Parameterizing the Mixture Weight and Transform}

Our next task is to parameterize $w_k(\theta)$ and $T_k(\beta)$.
Thanks to \eqref{eq:rev_conditional}, we can use some very simple form for $T_k$ --- this not only reduces the computing cost, but also allows more tractable theoretic analysis later on. In this article, we choose the element-wise location-scale change:
\bel{eq:affine_transform}
& T_k(\beta) = s_k \odot \beta + m_k,
\eel
where $s_k\in \mathbb{R}^p_+,m_k\in \mathbb{R}^p$, and $ \odot$ is the element-wise product. Accordingly, the Jacobian determinant is $\textup{det} \nabla T_k(\beta)=\prod_{j=1}^p s_{k,j} $, with $(s_{k,1}, \ldots, s_{k,p})=s_k$.

For the mixture weight, to satisfy $\sum_{k=1}^{\infty}w_k(\theta)=1$ while including a dependency on $\theta$, we use a multinomial logistic function,
\bel{eq:logistic_link}
& w_k(\theta)
= \frac{ b_k\exp(a_k^{\rm T}\theta) }{\sum_{k'=1}^{\infty}b_{k'}\exp(a_{k'}^{\rm T}\theta)}
,
\eel
where $a_k\in \mathbb{R}^p$, each $b_k\ge 0$. As \eqref{eq:logistic_link} is invariant a re-scaling of $b_k$'s, we further constrain $\sum_{k=1}^{\infty}b_k=1$, making $(b_1,b_2,\ldots)$ a probability vector. This allows us to efficiently deal with the infinite dimensionalty,
by treating $(b_1,b_2,\ldots)$ as the weights from a Dirichlet process, equivalent to the limit form of a finite $K$-element Dirichlet distribution
\be
(b_1,\ldots,b_K)\sim \text{Dir}(\alpha/K,\ldots,\alpha/K)
\ee
 as $K\to\infty$, where $\alpha>0$ is the concentration parameter. As a well-known property of Dirichlet process, $(b_1,b_2,\ldots)$ will have only a few elements away from zero, hence shrinking most of $w_k$'s close to zeros and effectively using only a few $T_k$'s.

To understand the geometric intuition behind \eqref{eq:affine_transform} and \eqref{eq:logistic_link}, we can focus on the most likely draw in \eqref{eq:generator}
\(\hat c(\beta)=\arg\max_k w_k \{ T_k(\beta)\} \Pi\{T_k(\beta); y\}
{|\textup{det} \nabla T_k(\beta)|},\)
which varies with the value of $\beta$. Therefore, we can treat $\hat c(\beta)$ as if a classifier with input $\beta$.

\begin{figure}[h]
     \centering
        \begin{subfigure}[b]{.31\textwidth}
         \centering
         \includegraphics[scale=1, width=\textwidth]{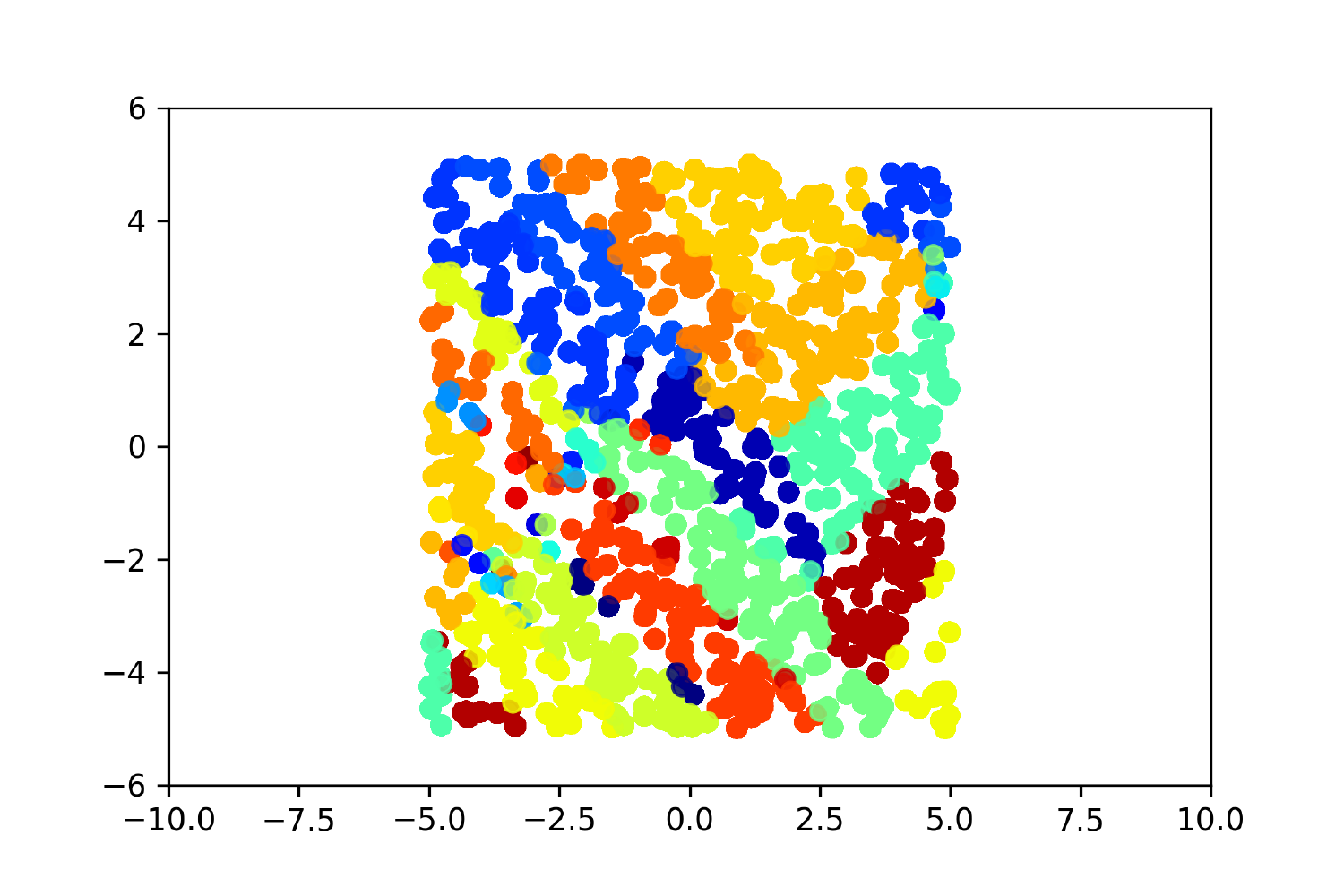}
         \caption{Reference samples  $\beta\sim \text{Uniform}( [0,1]^2)$,  colored by the most probable transform  $\hat c=\arg\max_k\text{Pr}\{\theta= T_k(\beta)\}$.}
     \end{subfigure}
     \quad
             \begin{subfigure}[b]{.31\textwidth}
         \centering
         \includegraphics[ scale=1, width=\textwidth]{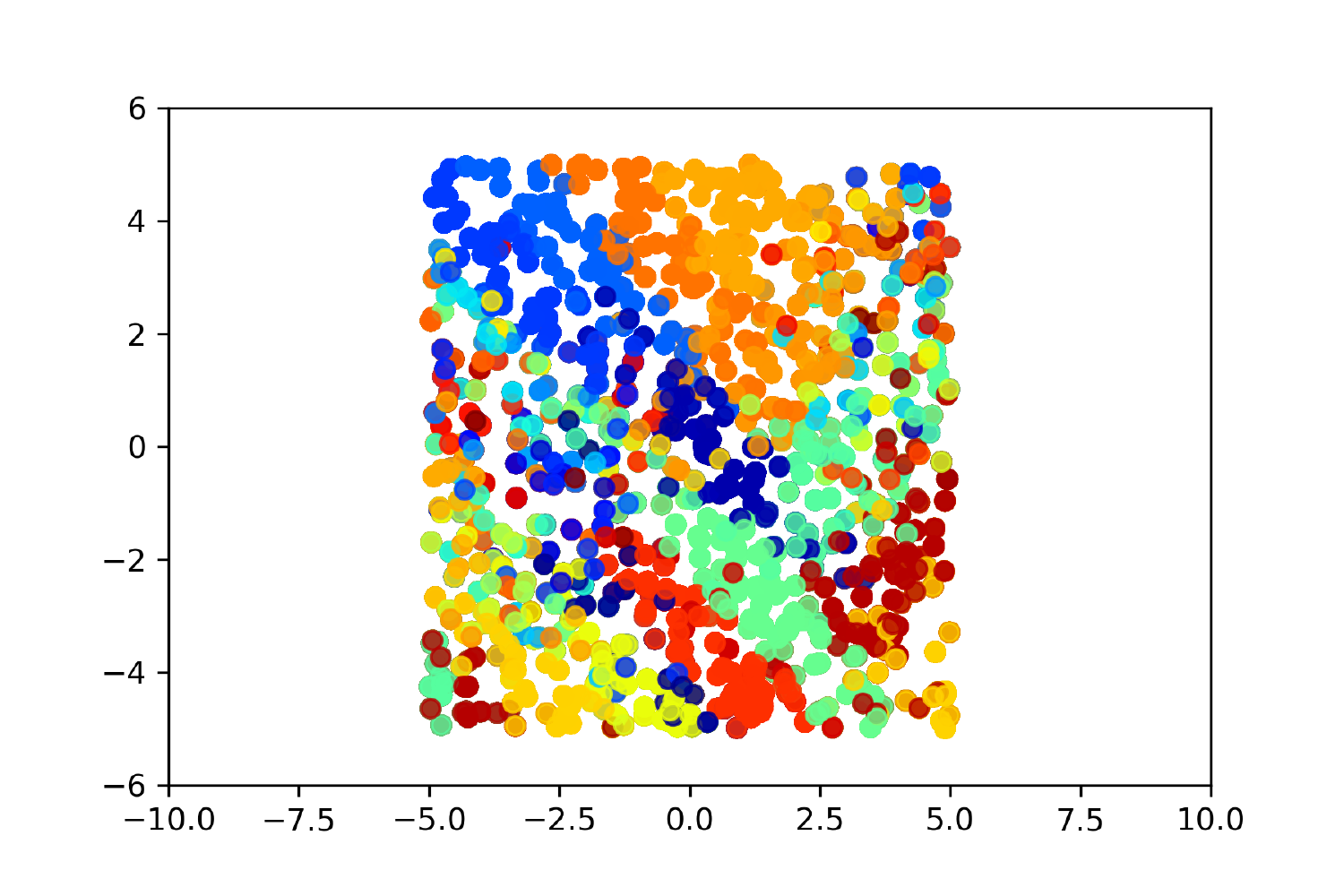}
         \caption{The reference samples colored by the drawn $c$, each $\beta$ will go through $T_c$, a location-scale change.}
     \end{subfigure}
     \quad
             \begin{subfigure}[b]{.31\textwidth}
         \centering
         \includegraphics[ width=\textwidth]{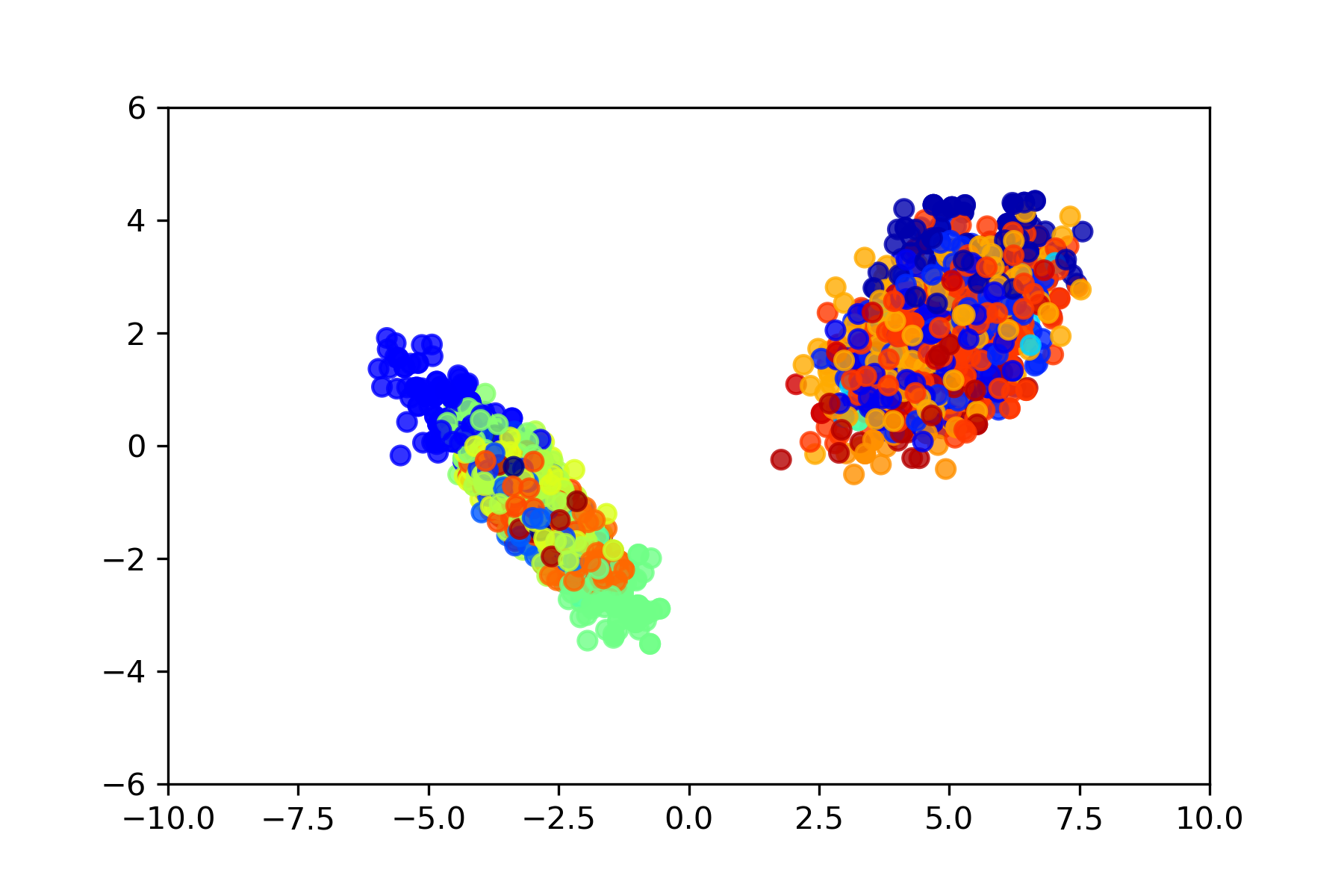}
         \caption{After drawing $c$ (shown in color) and the simple transform, the produced $\theta$ follows a bivariate normal mixture.}
     \end{subfigure}
         \caption{Simulation shows transporting a uniform $\beta$ into a two-component normal mixture $\theta$: each point of $\beta$ randomly draws a latent categorical variable
$c$ (panel b), then goes through a simple location-scale change $\theta= s_c \odot \beta + m_c$. The distribution of $c$ (panel a) and the working parameters $s_c$ and $m_c$ are estimated via optimization. \label{fig:mix_of_normals}} 
 \end{figure}

To illustrate this, we use an example of sampling the target form a two-component normal mixture in $\mathbb R^2$:
\(
\theta \sim 0.5  \; N( \begin{psmallmatrix}
  -3 \\-1
\end{psmallmatrix},
\begin{psmallmatrix}
    1 & -0.9 \\-0.9 & 1
  \end{psmallmatrix}
)+
0.5  \; N( \begin{psmallmatrix}
    5 \\ 2
  \end{psmallmatrix},
  \begin{psmallmatrix}
      1 & 0.5 \\0.5 & 1
    \end{psmallmatrix}
  ).
\)
with uniform reference $\Pi_r(\beta)\sim \text{Uniform} \{(0,1)^2\}$. After the optimization, we plot the randomly sampled $\beta$ and transported $\theta$ in Figure~\ref{fig:mix_of_normals}.
Panel(a) shows the partitioning of the space using $\hat c$.
The logistic function help divide the space of $\Beta$ into small local regions, where in each region, the points are most likely to go through $T_{\hat c}$.
Panels (b) and (c) show the randomly drawn $c$ for each point and the obtained sample $T_c(\beta)$.

 \subsection{Transport to Discrete Posterior}
Now we extend the transport plan to a discrete $\theta \sim \Pi(\theta;y)$. Since the continuous transform is much more convenient to deal with, we use an embedding strategy by considering another continuous latent variable $\eta\in \Eta$, such that its space $\Eta$ can be partitioned into disjoint subsets $\{\mathcal A_\theta \}_\theta$ and each $\mathcal A_\theta$ corresponds to a unique value of $\theta$. To give a concrete example, if our parameter of interest is  binary $\theta=\{0,1\}$, then we can use two intervals $\mathcal A_0=(-1,0)$ and $\mathcal A_1=(0,1)$ as the embedding sets. Since the subsets are disjoint, if we know $\eta$, we can recover the corresponding  $\theta$ via finding the enclosing set of $\eta$, we denote this reverse lookup by $\theta= R(\eta)$. In the above binary example, we can use $R(\theta)=\lceil \eta \rceil$, the ceiling funciton. For a categorical $m\in\mathbb{Z}_+$, one can similarly use $\mathcal A_m=(m-1,m)$ as the embedding. For more advanced examples, see \cite{nishimura2020discontinuous}, \cite{NIPS2013_a7d8ae45}.

Further, if we choose each $\mathcal A_\theta$ to have a unit volume, we can  assign a  uniform conditional density $\Pi(\eta\mid \theta)=  1(\eta\in \mathcal A_\theta)$. This leads to the marginal density,
\be
\Pi(\eta)= \sum_{\theta\in\Theta}\Pi(\theta;y)\Pi(\eta \mid \theta) =   \sum_{\theta\in\Theta}\Pi(\theta;y) 1(\eta\in \mathcal A_\theta) 
=   \Pi \{ R(\eta);y\} ,
\ee
with its support $\{\eta:R(\eta)\in \Theta\}$; the summation disappears because
for a given $\eta$, $1(\eta\in A_\theta)=1$ only when $\theta=R(\eta)$, and is $0$ for $\theta\neq R(\eta)$. We can now instead consider the random transport plan between $\eta$ and a continuous reference $\beta$, and transform $\eta$ to $\theta$ later:
\be
&(\eta, \beta)\sim \Pi(\eta, \beta)  \text{ such that}\quad  \int_{\Beta} \Pi(\eta,  \beta) \textup{d}\beta = \Pi(\eta), \int_{\Eta} \Pi(\eta, \beta) \textup{d}\eta = \Pi_r(\beta).
\ee
Similar to \eqref{eq:rev_transport}, we approximate the conditional $\Pi( \beta\mid \eta)$ by
\(
\tilde\Pi( \beta\mid \eta)= \sum_{k=1}^{\infty}w_k(\eta)\delta\{\beta
- T^{-1}_k(\eta)\},
\)
and integrating over $\eta$ gives the approximate marginal:
\be
&        \tilde\Pi(\beta) =
\sum_{k=1}^{\infty} w_k \{ T_k(\beta)\}\Pi [
 R\{T_k(\beta)\};y] {|\textup{det} \nabla T_k(\beta)|} 1\{ T_k(\beta)\in \Eta \}.
\ee
 After minimizing the difference between the $\tilde\Pi(\beta)$ and $\Pi_r(\beta)$. Using $\beta\sim \Pi_r$, the reverse conditional distribution for sampling $\theta$ is:
 \be
 \tilde\Pi[\theta= R\{T_k(\beta)\}\mid \beta ]=\frac{ w_k \{ T_k(\beta)\} \Pi [ R \{T_k(\beta)\}; y]
 {|\textup{det} \nabla T_k(\beta)|} 1\{ T_k(\beta)\in \Eta \} }{
 \sum_{k=1}^{\infty}   w_k \{ T_k(\beta)\} \Pi  [ R[\{T_k(\beta)\}; y]
 {|\textup{det} \nabla T_k(\beta)| 1\{ T_k(\beta)\in \Eta \}}.
 }
 \ee
In the data application, we will use this method to solve a challenging graph estimation problem. Due to the high similarity in methodology to the continuous cases, for conciseness, we will focus on continuous $\theta$ in the following discussion.

\section{Transport Monte Carlo Algorithm}

\subsection{Algorithm}

We design the Transport Monte Carlo (TMC) to be a two-stage algorithm: (i) optimization to estimate the working parameters in the mixture weights and location-scale transforms,
(ii) sampling independent $\beta$, and using the random transport to obtain $\theta$. We keep those two stages separate since the optimization is the time-consuming step; given an estimated transport plan, the sampling is easy to carry out rapidly.

{\noindent \bf Optimization:}  With $\{w_k, T_k\}_{k}$ fully parameterized, we can now minimize the Kullback-Leibler (KL) divergence $\mathbb{E}_{\beta \sim \Pi_r (\beta)}  \log \{{ \Pi_r(\beta) }/{\tilde \Pi(\beta)}\}$, so that $\Pi_r(\beta)\approx\Pi(\beta)$.

The total loss, including the Dirichlet process regularization on $b_k$ is
\bel{eq:loss}
Loss =& \mathbb{E}_{\beta \sim \Pi_r (\beta)}  \log \{{ \Pi_r(\beta) }/{\tilde \Pi(\beta)}\}- (\alpha/K-1) \sum_k \log b_k
       \\
      = &  - \mathbb{E}_{\beta \sim \Pi_r (\beta)}   \log
        \sum_{k=1}^{K}   w_k\{T_k(\beta)\}{L\{y; T_k(\beta)\}\Pi_0\{T_k(\beta) \}\ \prod_{j=1}^p s_{k,j} } 1\{ T_k(\beta)\in \Theta \}\\&-   \sum_{k=1}^K (\alpha/K-1)  \log b_k+\text{constant},
\eel
where the working parameters to optimize are $\{s_k, m_k, a_k, b_k\}_{k}$. To allow tractable computation, we use a truncation at $K$, as an approximation to the infinite dimension Dirichlet distribution \citep{ishwaran2002exact}. This leads to an effective number of working parameters $K(3p+1)$. In this article, we use $K=100$ in most of our examples.

To minimize the loss function, we use the stochastic gradient descent method. Since the expectation may be intractable, we draw a batch of $\beta_l \sim \Pi_r (\beta)$ for $l=1,\ldots,n_b$, then calculate the gradient based on this batch and carry out a gradient descent step on the parameters; then we draw a new set of $\beta_l \sim \Pi_r(\beta)$ in the next step. Effectively, this is equivalent to the stochastic gradient descent on an infinitely large training sample, since we can draw infinitely many samples from $\Pi_r(\beta)$. Such a method is routinely used in the variational inference \citep{kingma2013auto}, and prevents overfitting to the finite number of training samples. We provide more details on the optimization in the next subsection.

{\noindent \bf Drawing $\theta$ via Random Transport:} After the optimization converges, the samples of $\theta$ can be obtained via
\be
& \beta \stackrel{iid}\sim \Pi_r, \\
& c\sim \text{Categorical}\{v_1(\beta),\ldots, v_K(\beta)\},\\
& \theta =T_c(\beta).
\ee
   Strictly speaking, the samples of $\theta$ generated from above are approximation  to $\theta\sim\Pi(\theta;y)$, since we substitute $\tilde \Pi(\beta)$ by $\Pi_r$ in the first line. As shown in all of our cases,
 we found the approximations indistinguishable from the ones obtained from a long-time run of MCMC.

\subsection{Details on the Optimization}

We now provide more details on the optimization stage on: (i) how to effectively optimize all $K$ components, especially in a high dimensional setting; (ii) how to diagnose if the selected $K$ is sufficiently large; and (iii) how to initialize the working parameters.

\subsubsection{Component-wise Optimization}
Since we approximate the infinite mixture components at a truncation $K$, it would be desirable that those $K$ components contain most of the ``effective'' transforms for minimizing the divergence. To quantify the effectiveness, note that \eqref{eq:loss} contains a LogSumExp function 
\[
\t{LSE}(l_1,\ldots,l_K)= \log\{\sum_{k=1}^K \exp(l_k)\}\]
where $l_k=\log [ w_k\{T_k(\beta)\}{L\{y; T_k(\beta)\}\Pi_0\{T_k(\beta) \}\ \prod_{j=1}^p s_{k,j} }] -\chi_\Theta\{ T_k(\beta) \}$, with $\chi_\Theta(t)=0$ if $t\in\Theta$ and $\chi_\Theta(t)=\infty$ otherwise. If we reorder the $(l_1,\ldots,l_K)$, $l_{(1)}\le l_{(2)} \le \ldots \le  l_{(K-h)} \le l_{(K-h+1)} \le \ldots\le l_{(K)}$, then we can bound this function from both sides:
\[ l_{(K)}\le \t{LSE}(l_1,\ldots,l_K) \le l_{(K)} + \log h +  \exp(l_{(K-h)}-l_{(K)})(K-h)/h ,
\]
where the lower bound is due to $\exp(l_{(K)})\le \sum_{k=1}^K \exp(l_k)$, and the upper bound is due to $\sum_{k=1}^{K-h} \exp(l_k) \le (K-h) \exp(l_{(K-h)})$ and $ \sum_{k=K-h+1}^{K} \le  h \exp(l_{(K)})$, in addition to $\log(a+b)\le \log a + b/a$ for $a>0,b>0$. Now, note that if $l_{(K-h)} \ll l_{(K)}$, then the last term is close to $0$, hence the LSE function is almost fully determined by the top $h$ components that are close to $l_{(K)}$. 
Therefore,  with  $l_k$ and $l_{(K)}$ dependent on $\beta$, a useful score measuring the effectiveness of the component $k$   is $\xi_k=\mathbb{E}_{\beta\sim \Pi_r}\exp(l_{k}-l_{(K)})$. 

To see how this score impacts the gradient descent algorithm, we can compute the magnitude of the gradient with respect to the working parameters in the $k$th component,
\[
\|\nabla_k {\t{LSE}(l_1,\ldots,l_K)} \|= \frac{\exp(l_k)}{\sum_{k'=1}^K \exp(l_{k'})} \|\nabla_k l_k
\|\le \exp(l_k-l_{(K)}) \|\nabla_k l_k\|.
\]
Now, if a component is initialized at $\exp(l_k-l_{(K)})\approx0$, then the gradient descent would almost not update the working parameters.
This is very common when the parameters are in high dimension and randomly initialized.

Therefore, a component-wise optimization with a good initialization,
is more useful than simultaneously updating all components. Specifically, for $k=1,\ldots,K$,
\begin{enumerate}
        \item Compute the scores:   $\xi_j=\mathbb{E}_{\beta\sim \Pi_r}\exp( l_j- l_{(K)})$ using   the samples of $\beta$, for all $j=1,\ldots,K$. Find the set of components with scores $H=\{j: \xi_j > \tau\}$.
        \item Re-initialize the weak component: if the current $\xi_k<\tau$,  draw an index $j$ from $H$, and      set the parameters in $(T_k, w_k)$ to be equal to $(T_{j}, w_{j})$ plus a small perturbation.
        \item Optimize the $k$th component: optimize $(T_k, w_k)$ using the gradient descent until the empirical KL divergence  converges, while keeping the other $(T_{k'},w_{k'})$ fixed.
\end{enumerate}
In the implementation, we use a threshold $\tau=0.01$ and perturbation $N(0,0.01/p)$ to yield a good initialization for each component.
During this process, we use the PyTorch framework for auto-differentiation and ADAM optimizer \citep{kingma2014adam} for gradient descent. 
 We consider each optimization converged if the change in the empirical KL divergence is smaller than a threshold of over $100$ iterations.

\subsubsection{Diagnostics on $K$ and Convergence}

At the same time, this strategy enables us to easily diagnose if the chosen $K$ is large enough. Note that the KL divergence is always greater or equal to zero, hence in the loss function \eqref{eq:loss}, the function $- \mathbb{E}_{\beta \sim \Pi_r (\beta)}   \t{LSE}(l_1,\ldots,l_K)$ is bounded from below at a constant [to be exact, the log of the normalizing constant $\log z(y)$]. Therefore, we can collect the optimized value every time we finish updating a component, creating a curve over $k=1,\ldots,K$. If we see the curve flattening well before $K$, then the selected $K$ is very likely to be sufficient; otherwise, we should increase $K$. We provide an example diagnostic plot in the supplementary materials.

\begin{remark}
One could extend our algorithm by indefinitely adding and optimizing a new component, until the KL divergence does not decrease any further. This could prevent the need to specify an upper bound $K$. See \cite{miller2017variational} for a similar algorithm.
\end{remark}

\subsubsection{Initialization}

Since the loss function \eqref{eq:loss} is usually non-convex (for example, when the posterior density is non-convex), it is helpful to use a good initialization on the working parameters. With $\Pi_r$ chosen as the $\t{Uniform}\{(0,1)^p\}$, we set  all $a_k$ and $b_k$ to be zero in the logistic function so that the initial weights are all equal, and the shift parameters $m_k$ to be in the high posterior density region of $\theta$. For the target distribution with log-concave density, we can first calculate the posterior mode $\hat\theta$ using optimization, then generate $m_k$'s near $\hat\theta$. On the other hand, this is more difficult for the multi-modal target distribution, especially when the modes are far apart; in such a case, we randomly generate $m_k$'s uniformly in an estimated range of $\theta$, then rely on the mixture framework for the parallel searches for all the modes. We found this strategy yield good empirical performance, as demonstrated in the example of sampling a target posterior with $25$ modes in the supplementary materials --- although in more sophisticated cases, some customized initialization should be used instead.

\subsection{Combining with Independence Hastings Algorithm}

As in the other popular approaches, there are approximation errors incurred in the TMC algorithm, for example, due to the use of finite $K$, the $\epsilon$-suboptimal convergence of the optimization, etc. Since such errors are often intractable, some control methods are needed.

For this purpose, we develop an extension to combine TMC with the independence Hastings algorithm. For conciseness of presentation, we defer the method to the supplementary materials.

\section{Theoretic Study}
In this section, we give a more theoretical exposition on the TMC method. For the mathematic rigor, we will focus on $\Pi(\theta;y)$ being the posterior density corresponding to $\{\Theta,\mathcal B(\Theta),\mu\}$, with $ \Theta\subseteq \mathbb{R}^p$, $\mathcal B$ the Borel $\sigma$-algebra and $\mu$ absolutely continuous with respect to the Lebesgue measure.

\begin{figure}[h]
     \centering
         \includegraphics[ width=.5\textwidth]{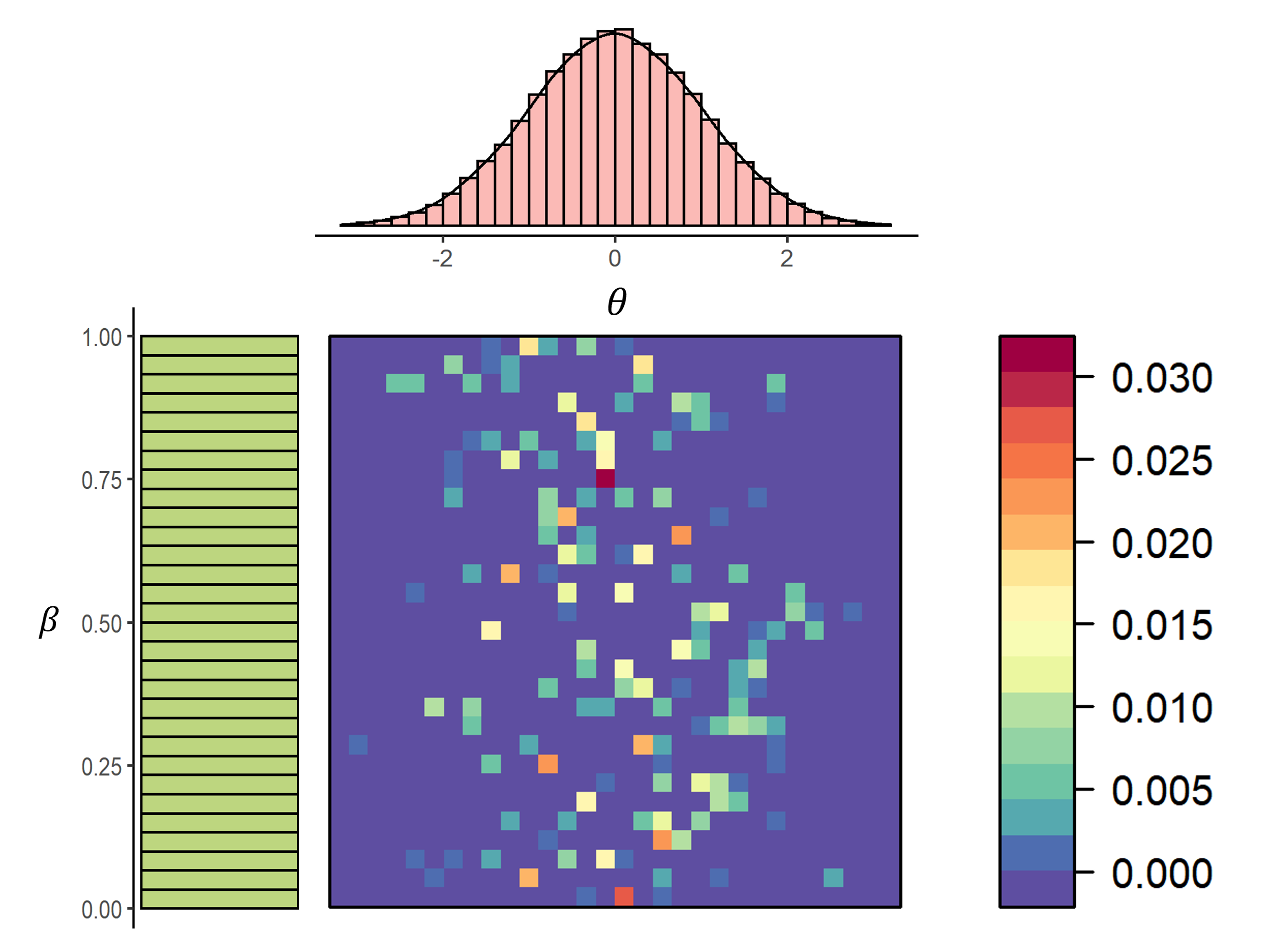}
         \caption{Visualization of an approximate transport plan --- a joint distribution (with probabilities $\t{pr}_{\theta,\beta}(\Theta_j,\Beta_k)$ shown in the heatmap) having one marginal equal to the histogram  of $\beta\sim \t{Uniform}(0,1)$ (left), and the other equal to the histogram   (a piece-wise uniform approximation) of $\theta\sim N(0,1)$ (top). In each cell of the heatmap, a simple location-scale change is carried out that changes the set $\Beta_k$ to $\Theta_j$. As the number of bins increases (as $K\to \infty$), the top histogram converges to the target distribution $N(0,1)$.
\label{fig:transport_plan}}
\end{figure}

Before presenting  results, we use a simple example to show the intuition of ``changing one histogram into another'', behind the choice of uniform distribution for $\Pi_r$ and location-scale transformation for $T_k$'s. Suppose that we want to find an approximate transport plan between $\beta\sim U(0,1)$ and $\theta\sim N(0,1)$, a standard normal distribution: we first find the high density region of $\theta$, divide it into disjoint bins (denoted by $\Theta_j$'s),  compute the probability within each bin [denoted by $\t{pr}_\theta(\Theta_j)$] and re-normalize them so that $ \sum_j \t{pr}_\theta(\Theta_j)=1$.
As a result, this produces the commonly used histogram, where we replace the density within each bin by a flat constant (therefore, leading to an approximation).
Similarly, we
 divide the  support of the uniform $(0,1)$ into multiple bins (denoted by $\Beta_k$'s), and obtain corresponding probabilities $\t{pr}_\beta(\Beta_k)$'s.

With these two marginal histograms, we can find a joint distribution $\t{pr}_{\theta,\beta}(\Theta_j,\Beta_k)$, such that $\sum_k \t{pr}_{\theta,\beta}(\Theta_j,\Beta_k)=\t{pr}_{\theta}(\Theta_j)$  and $\sum_j \t{pr}_{\theta,\beta}(\Theta_j,\Beta_k)=\t{pr}_{\beta}(\Beta_k)$ --- that is, solving for a {\em contingency table} with known marginal values. Obviously, there are more than one solutions.
  We plot a solved table of $\t{pr}_{\theta,\beta}(\Theta_j,\Beta_k)$ in Figure~\ref{fig:transport_plan} with $30$ bins for  $\theta$ and $30$ bins for  $\beta$. To show more details in each table cell, we also solved for a smaller table (Table~1 in the supplementary materials) with $6$ bins for $\theta$ and $2$ bins for $\beta$.

Note that these solutions are sparse with some $pr_{\theta,\beta} (\Theta_j,\Beta_k)=0$. And in each $(\Theta_j,\Beta_k)$, we can impose a one-to-one transform $\theta=T_l(\beta)$ that changes a uniform distribution in $\Beta_k$ to a uniform distribution in $\Theta_j$ --- clearly, the simple scale-location change $T_l$  is adequate for this task.
As the number of bins  $\Theta_j$'s increases, we can expect the histogram to converge to the target density of $\theta$. We now formalize this intuition.


Without loss of generality, we use  $\beta\sim \text{Uniform}(0,1)^p$;  after the location-scale change, the transformed follows   $T_k(\beta) \sim \text{Uniform}\{\times_{j=1}^{p}(m_{k,j},s_{k,j}+m_{k,j})\}$.
The density of \eqref{eq:generator} at a specific value $\theta_0$ is:
\be
\tilde\Pi(\theta_0)& =\int\Pi_r(\beta) \sum_{k=1}^Kv_k(\beta)\delta\{\theta_0-T_k(\beta)\} \textup{d}\beta\\
&=  \sum_{k=1}^K v_k\{T^{-1}_k(\theta_0)\}
\frac{1}{\prod_{j=1}^p s_{k,j}}
 1\{T^{-1}_k(\theta_0)\in (0,1)^p \},
\ee
where we exchange the summation and integral, and use $\int _{X}f(x) \delta(x-y) \textup{d} x=  f(y)1(y\in X)$. Rewriting this as
\bel{eq:simple_func_den}
&\tilde\Pi_K(\theta_0)= \sum_{k=1}^K \tilde a_k (\theta_0) 1(\theta_0 \in C_k), \\
& \tilde a_k(\theta_0) =\frac{v^*_k(\theta_0)}{ \prod_{j=1}^{ p}s_{k,j}}, C_k= \times_{j=1}^{p}(m_{k,j},s_{k,j}+m_{k,j}),
\eel
where $v^*_k(\theta_0)=v_k\{s_k^{-1} \odot (\theta_0 - m_k)\}$. Note that if we had $\Pi^*_K(\theta_0)=\sum_{k=1}^K \tilde a_k 1(\theta_0 \in C_k)$ with $\tilde a_k$ does not depend on $\theta_0$, then $\Pi^*_K(\theta_0)$ would be a ``simple function''. The simple function is routinely used for approximating any Lebesgue-measurable function, as stated in the following theorem.

\begin{theorem}\citep{schilling2017measures}
Let $\{\Theta,\mathcal B(\Theta)\}$ be a measurable space, and $f$ be a measurable function.
Then there exists a sequence $\left({f_K}\right)_{K \in \mathbb{Z}_+}$, with each $f_K$ a
simple function, such that,
$\forall \theta_0 \in \Theta: f \left({\theta_0}\right) = \displaystyle \lim_{K \to \infty} f_K \left({\theta_0}\right)$.
\end{theorem}

On the other hand, now since each $\tilde a_k(\theta_0)$  does depend on  $\theta_0$, with $\tilde a_k(\theta)$  set by a specific form of $v_k(\theta)$ as in \eqref{eq:rev_conditional}. Therefore, some additional work is needed to show a similar asymptotic result. We state the result as followed and provide a construction in the proof.
For conciseness, we provide all the proofs in the supplementary materials.

\begin{theorem}
Let $\Theta\subseteq \mathbb{R}^p$ be the set for $0<\Pi(\theta;y)<\infty$, if $\Pi(\theta;y)$ is a continuous function in $\Theta$ except for finite number of points,  then there exists a sequence   $({\tilde\Pi_K })_{K\in \mathbb Z_+}$, with each ${\tilde\Pi_K }$\ in the form of \eqref{eq:simple_func_den}, such that,
$ \Pi(\theta; y) = \displaystyle \lim_{K \to \infty}   \tilde\Pi_K(\theta)$ almost everywhere in $\Theta$.
\end{theorem}

Next, we focus on the random transport plan \eqref{eq:random_transport}, as the joint probability betweena $\theta$ and $\beta$. Recall that we obtain its estimate via  discretizing the conditional $\Pi(\beta\mid \theta)$ using a mixture and matching its marginal $\tilde\Pi(\beta)$ to a chosen $\Pi_r(\beta)$. We want to show that there exists  solution for $\{(w_k, T_k)\}_k$ with negligibly small difference.
To formalize, let $P_0(\theta, \beta)$ be a joint density of $\theta$ and $\beta$, that has the marginals exactly as $\beta\sim \text{Uniform} \{(0,1)^{p}\}$ and $\theta\sim\Pi(\theta;y)$. For a measurable $\mathcal A\in \mathcal B(\Beta)$ with $\Beta= (0,1)^{p}$, denote the conditional measure by
\be
\nu_{\beta\mid\theta}(\mathcal A)=\int_\mathcal{A}\frac{P_0(\theta, \beta)}{\Pi(\theta; y)} \textup{d}\beta \in (0,1].
\ee
Since there exists more than one $P_0(\theta, \beta)$, we can focus on those with $\nu_{\beta\mid\theta}$ corresponding to a measure absolutely continuous with respect to a Lebesgue measure.

\begin{theorem}
Denote the measures corresponding to $\Pi_r(\beta)$ and $\tilde\Pi(\beta)$ by $\pi_{r}$ and $\tilde\pi_{\beta}$, respectively.
 If $P_0(\theta, \beta)$ satisfies that   $\nu_{\beta\mid\theta}$ is  absolutely continuous with respect to the Lebesgue measure  for any $\|\theta\|_2<\infty$,
 then there exists a sequence $\{w_k,T_k\}_{k=1}^K$
with $T_k$ parameterized by location-scale transform \eqref{eq:affine_transform} and $w_k: w_k(\theta)\ge 0,\sum_{k=1}^K w_k(\theta)=1$,
 such that the total variation distance
\[
\lim_{K\to \infty}\sup_{\mathcal A\in \mathcal B(\Beta)}|\pi_{r}(\mathcal A) - \tilde\pi_{\beta}(\mathcal A)|= 0.
  \]
\end{theorem}
 Strictly speaking, in the above we consider a broad class of  functions $w:\Theta\to \Delta^{K-1}$, with $ \Delta^{K-1}$ the probability simplex.

 We now turn to the algorithmic details of the TMC. As suggested in the last theorem,  under a large $K$, we can expect $\tilde \pi(\beta_l)$ to be close to $\pi_r(\beta_l)$ for most of the samples $\beta_l$'s generated during the optimization. On the other hand, since $(0,1)^p$ is a continuous space, there are always points that we have not trained on --- if we generate a new $\beta^*$ and sample $\theta^*$ through \eqref{eq:generator}, how can we guarantee that it still has a low approximation error?

Since we use stochastic gradient descent with batch size $n_b$, by the time when we stop after $t$ iterations, the optimization is effectively based on $n= n_b t$ training samples.
Intuitively, if the training $\{\beta_l\}_{l=1}^n$ are ``dense'' enough to cover most of $(0,1)^p$  --- that is, the maximal spacing $\max_i \min_j \|\beta_i -\beta_j\|$ is small, any new sample $\beta^*$ drawn in the sampling stage will  be near a certain training $\beta_l$; hence, the associated $\tilde\Pi(\beta^*)$ should be very close to $\Pi(\beta_l)$ (on the logarithmic scale, as in the optimization).
The following theorem formalizes this intuition and quantifies how the error vanishes in terms of $n$.
\begin{theorem}
If $\Pi(\theta; y)$ is absolutely continuous, then

 \be
\inf_{l\in\{1\ldots n\}}\|\log\tilde\Pi(\beta^*) - \log\tilde\Pi(\beta_l)\|
=\mathcal{O}\{\ \frac{ p( 2\log\log n + \log n) }{n}\}.
\ee
\end{theorem}

 This above rate is due to the uniform reference $\Pi_r(\beta)$ having a compact support; hence the maximal spacing drops to zero rapidly in a roughly $\mathcal O (1/n)$ rate. For the other reference with unbounded support, such as multivariate normal, we would not have such a guarantee. In fact, the rate based on a normal $\Pi_r(\beta)$ would be  approximately $\mathcal O( 1/\sqrt{\log n })$ \citep{deheuvels1986influence}, substantially slower than uniform.


\section{Comparison with the Hamiltonian Monte Carlo}
With the augmented random variable $\beta\in\mathbb{R}^p$ and the deterministic transforms, the TMC may appear similar to the popular Hamiltonian Monte Carlo (HMC) algorithm. Therefore, it is interesting to compare those two methods.

To provide some background, the HMC uses an augmented ``momentum'' variable $v\in \mathbb{R}^p$, with $v\sim \Pi(v)$ (independent from $\theta$ in the original HMC algorithm \citep{neal2011mcmc}, or dependent on $\theta$ in the Riemannian manifold HMC, RMHMC \citep{girolami2011riemann}).  With $\Pi(\theta,v) = \exp\{-H (\theta,v)\}$, $H(\theta,v)$ is referred to as the Hamiltonian. For multiple copies of $(\theta,v)$  indexed over time $t\in [0,\infty)$, denoted by $\{(\theta^t,v^t)\}_t$, they change smoothly according to the Hamilton's equations:
\bel{eq:hamilton}
\frac{\partial \theta^t}{\partial t}= \frac{\partial H(\theta^t,v^t)}{\partial v^t},\qquad
\frac{\partial v^t}{\partial t}= -\frac{\partial H(\theta^t,v^t)}{\partial \theta^t}.
\eel
Using one Markov chain sample of $(\theta, v)$ as  the initial $(\theta^0,v^0)$, one could obtain another sample $(\theta^T,v^T)$ at time $T$, deterministically using the exact solution to  \eqref{eq:hamilton}. On the other hand, for most of the posterior densities, the equation \eqref{eq:hamilton} cannot 
be solved in closed form; therefore, some approximation is often used, such as the leapfrog scheme: $\theta^{t+\varepsilon/2} = \theta^t+ (\varepsilon/2){\partial H(\theta^t,v^t)}/{\partial v^t}$, and $v^{t+\varepsilon/2} = v^t -(\varepsilon/2){\partial H(\theta^t,v^t)}/{\partial \theta^t}$ for $t=0,\varepsilon/2,\ldots, \varepsilon L$, and then $(\theta^T, v^T)$ at $T=\varepsilon L$ is accepted/rejected using the Metropolis-Hastings step. When $L=1$, the above is equivalent to the Metropolis-adjusted Langevin Algorithm (MALA). Now, to compare them with the TMC:

\begin{enumerate}
\item From the perspective of transport, both the exact and approximate solutions to
the Hamilton's equations  (before the Metropolis-Hastings correction) can be viewed as a transport plan from $(\theta^0, v^0)$ to $(\theta^T, v^T)$. That is, it forms a joint distribution among $(\theta^0, v^0, \theta^T,v^T)$ --- importantly, although $\theta^0$ and $ v^{0}$ can be independent, across time, $\theta^T$ is dependent on $v^0$, and $v^T$ is dependent on $\theta^0$. In comparison, the TMC focus on one copy of $(\theta,\beta)$, with a dependency in between; across two copies, $(\theta^i,\beta^i)$ are independent from $(\theta^{i'},\beta^{i'})$ for $i'\neq i$.
\item 
In terms of the computation burden, in the HMC, the transformation from $(\theta^0,v^0)\to (\theta^T,v^T)$ is pre-determined as the solution to \eqref{eq:hamilton}, but due to the often lack of closed-form, some intensive and iterative algorithm (such as the leapfrog) is needed to produce a new $\theta^T$ relatively far away from the current $\theta^0$. In the TMC, the transformation $\beta\to \theta$ is not known beforehand and needs to be estimated, but after the optimization, the transformation is simple to compute.
\item In both approaches,  a Metropolis-Hastings step would be needed to correct the numeric errors, so that the sample collected can converge to the target posterior distribution as the number of samples diverges.
\end{enumerate}

\begin{figure}[H]
     \centering
         \begin{subfigure}[t]{.32\textwidth}
         \centering
         \includegraphics[ width=1\textwidth, height = 1.2in]{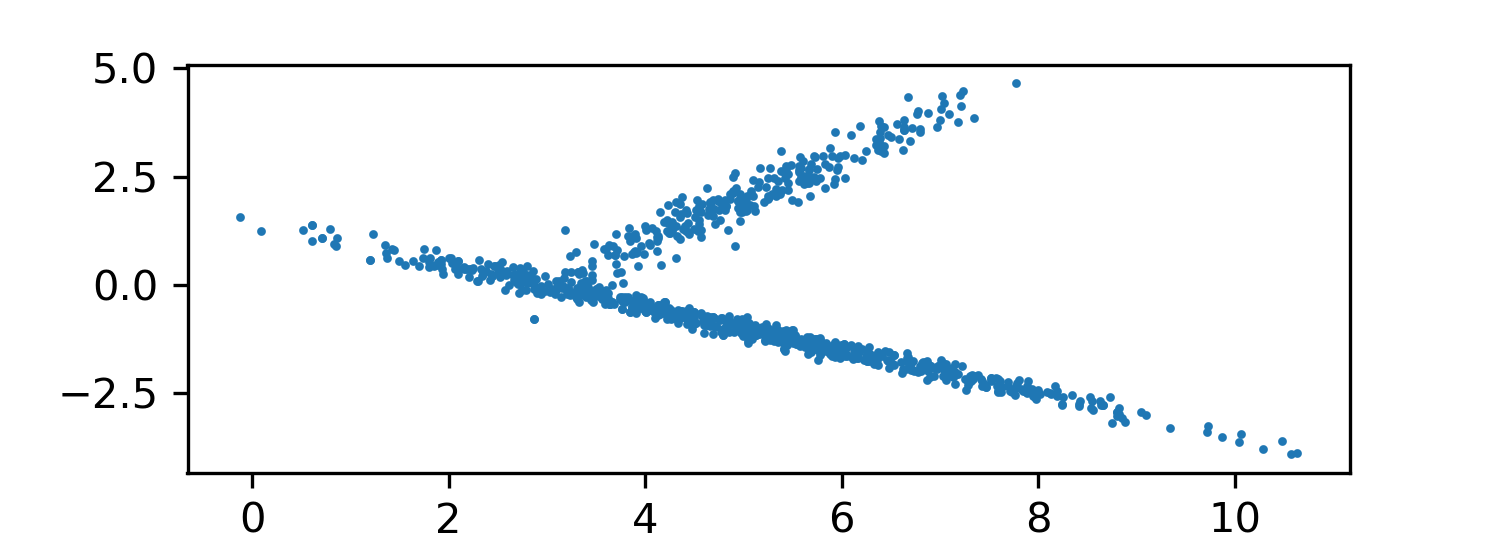}
         \caption{Scatterplot of the sample from the Gaussian mixture.}
     \end{subfigure}
       \begin{subfigure}[t]{0.32\textwidth}
         \centering
         \includegraphics[ width=\textwidth, height = 1.1in]{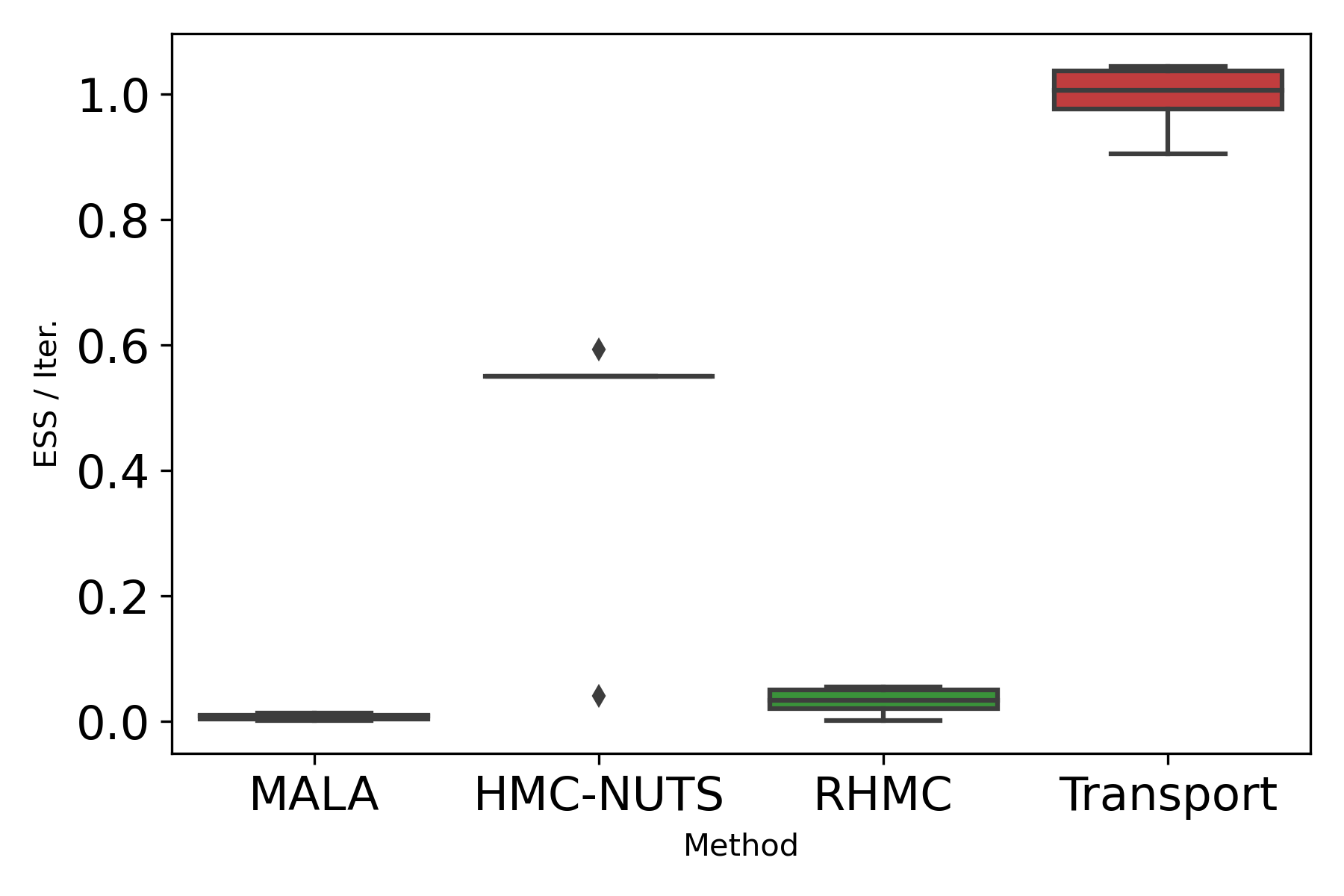}
         \caption{Boxplots of the effective sample size (ESS) per iteration.}
     \end{subfigure}\;
           \begin{subfigure}[t]{.32\textwidth}
         \centering
         \includegraphics[ width=1\textwidth, height = 1.2in]{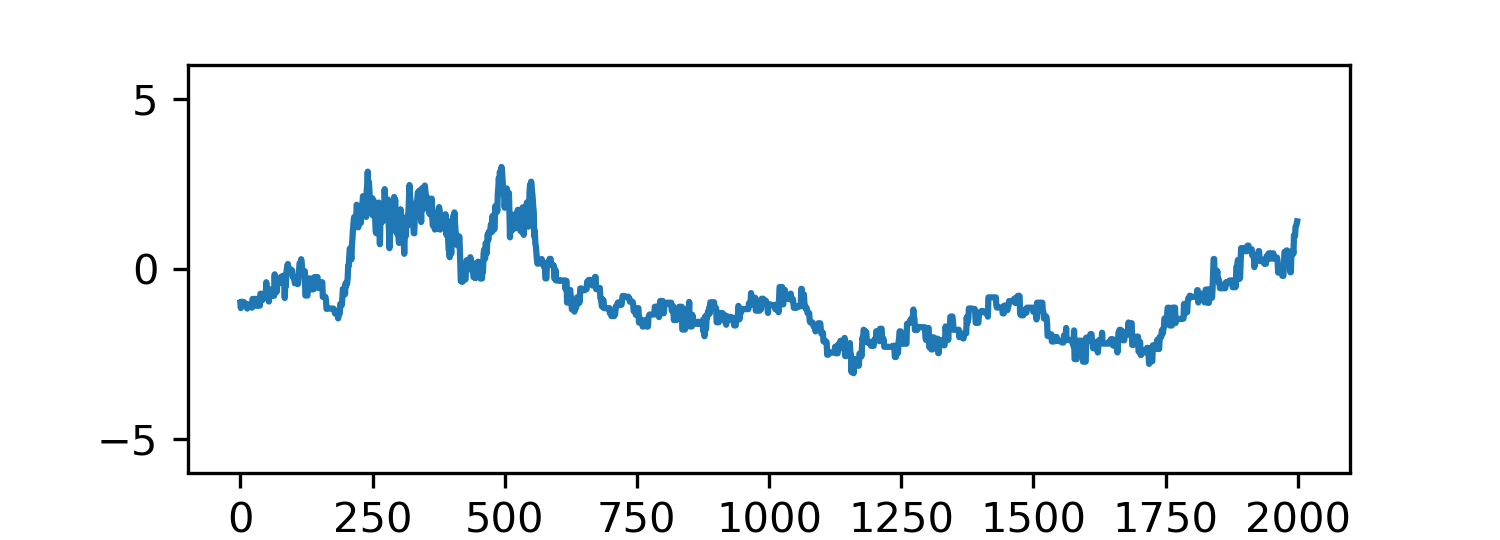}
         \caption{Traceplot of $\theta_2$ produced by the Metropolis-adjusted Langevin algorithm (MALA).}
     \end{subfigure}
            \begin{subfigure}[t]{.32\textwidth}
         \centering
         \includegraphics[ width=1\textwidth, height = 1.2in]{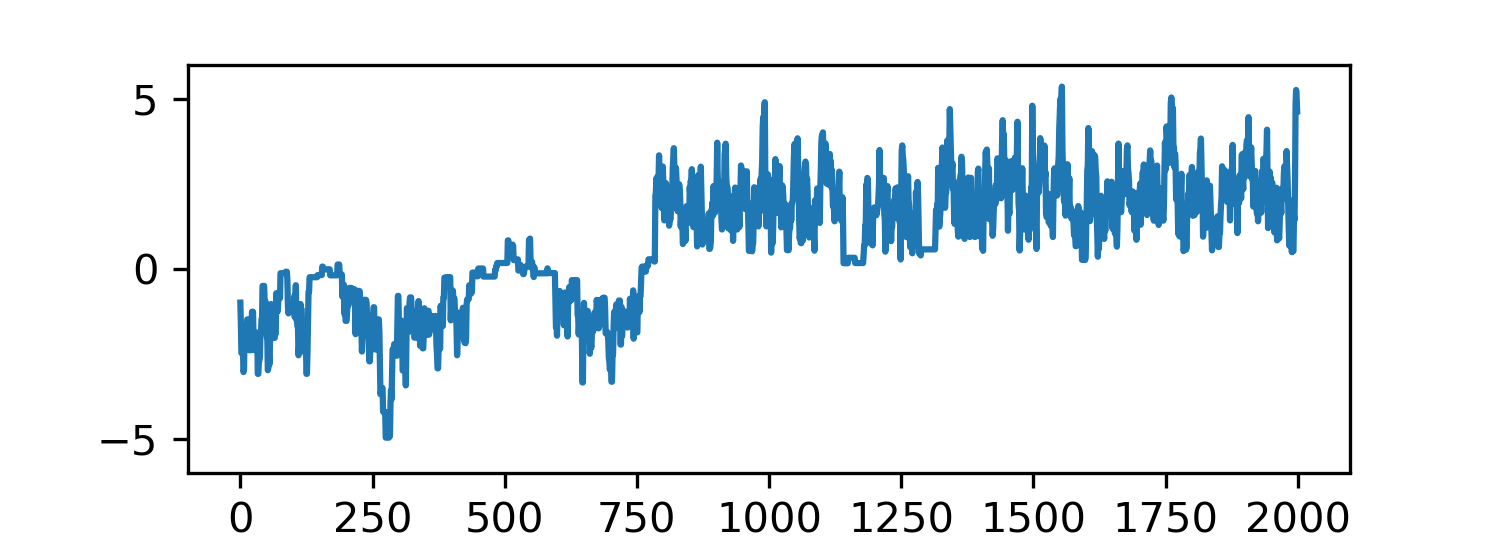}
         \caption{Traceplot of $\theta_2$ produced by the Riemannian manifold Hamiltonian Monte Carlo (RMHMC).}
     \end{subfigure}
           \begin{subfigure}[t]{.32\textwidth}
         \centering
         \includegraphics[ width=1\textwidth, height = 1.2in]{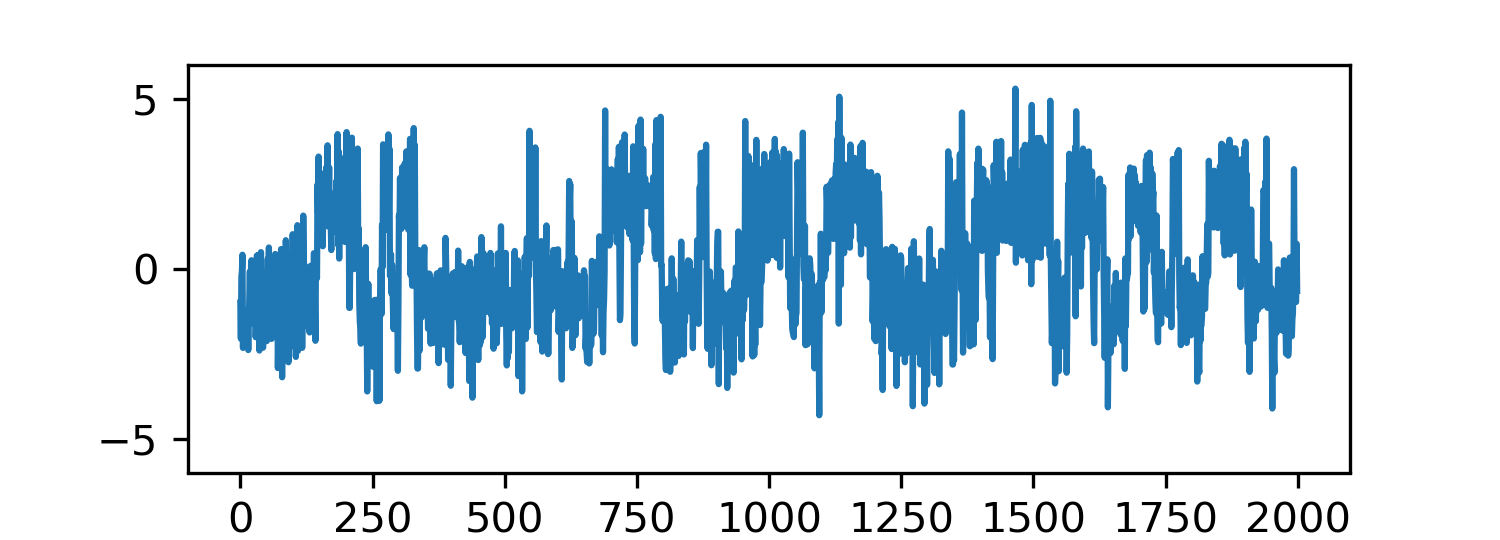}
         \caption{Traceplot of $\theta_2$ produced by the Hamiltonian Monte Carlo (No-U-Turn Sampler, HMC-NUTS).}
     \end{subfigure}
           \begin{subfigure}[t]{.32\textwidth}
         \centering
         \includegraphics[ width=1\textwidth, height = 1.2in]{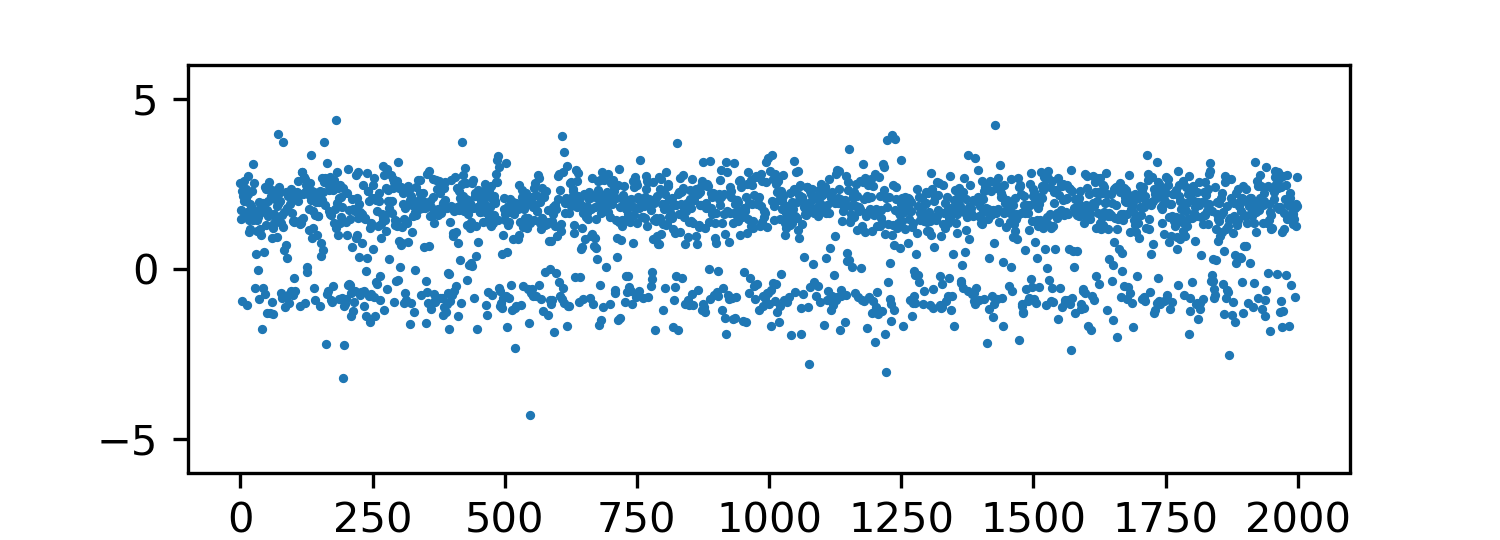}
         \caption{Traceplot of $\theta_2$ produced by the Transport Monte Carlo, corrected using the independence Hastings algorithm.}
     \end{subfigure} 
        \caption{Experiments of sampling from a multivariate Gaussian mixture distribution, comparing the performances between  the Transport Monte Carlo and various Hamiltonian Monte Carlo algorithms. \label{fig:compare_w_hmc}
        }
\end{figure}

To illustrate the above points, we use the two-component normal mixture in $\mathbb R^2$ as the target distribution [Figure~\ref{fig:compare_w_hmc}(a)]: \(
\theta  \sim N[ \begin{psmallmatrix}
  5 \\-1
\end{psmallmatrix},
\begin{psmallmatrix}
    1 & -0.9 \\-0.9 & 1
  \end{psmallmatrix}
)+
0.5  \; N( \begin{psmallmatrix}
    5 \\ 2
  \end{psmallmatrix},
  \begin{psmallmatrix}
      1 & 0.9 \\0.9 & 1
    \end{psmallmatrix}].
\)
This modifies the previous example by bringing the two components means closer to each other, so that the HMC can visit both components more easily. We initialized each algorithm at one of the means $\hat\theta=[5,2]$, and tuned each algorithm to have the acceptance rate close to $70\%$. We ran each HMC algorithm for $20,000$ steps.

Since $\theta_2$ has two local optima at $-1$ and $2$, we can use its traceplot to visualize how quickly each algorithm can jump from one normal component to another. Among the HMC algorithms, the MALA was the fastest to run (due to $L=1$ in the numeric approximation, it took $36$ seconds) but suffered from high autocorrelation, as each new proposal is very close to the current one (panel c); the HMC with No-U-Turn sampler (HMC-NUTS) had a much lower autocorrelation (panel e), although at a much higher computational cost (it took $1260$ seconds to run, with each new proposal taking on average $L=53$ steps to generate). In addition, we tested the Riemannian manifold HMC (RMHMC), which makes the covariance matrix of $v$ depend on the current state of $\theta$ via the observed Fisher information \citep{girolami2011riemann}, hence potentially giving better adaptation to the local geometry of the high posterior density region. We set $L=1$ in RMHMC and found  a much better performance than the MALA. However, since each step was computationally intensive [we used the state-of-art explicit-scheme integrator \citep{cobb2019introducing} that improves upon the implicit-scheme one \citep{girolami2011riemann} in speed], it took a longer time (1800 seconds)  and was less efficient in exploring two components (panel d), compared to the HMC-NUTS. To summarize, the HMC-NUTS algorithm was the most efficient among all the HMC algorithms in this experiment.

To compare, the TMC took 18 seconds to optimize and less than 1 second to generate $20,000$ samples. We further applied the Metropolis-Hastings step on the generated samples, which lead to an acceptance rate of $90\%$. As a result, the produced samples were almost independent.

In addition, we carried out experiments on: (i) the estimation of high-dimensional regression using the shrinkage prior, (ii) sampling from multi-modal distribution,  (iii) comparing the performance with various normalizing flow neural networks. The details are provided in the supplementary materials.

\section{Application: Graph Estimation under Degree Regularization}
We now illustrate the performance of discrete parameter estimation using a   data application. The data are the multivariate electroencephalogram (EEG)
time series collected over $V=128$ electrodes when the human subject is performing a working memory task. Our goal is to estimate an undirected graph $G=\{\mathcal V,\mathcal E\}$ with $\mathcal V=\{1,\ldots, V\}$ the nodes and
$E=\{e_{i,j}\}$ the edges, based on the temporal correlation among those time series. The parameter of the interest is a binary adjacency matrix $A=\{A_{i,j}\}$, with $A_{i,j}=1$ if $e_{i,j}\in E$, $0$ otherwise for $j<i$; $A_{j,i}=A_{i,j}$ and we fix $A_{i,i}=0$. In particular, we are interested in finding a subset of nodes that are well connected during this memory task, while excluding the remaining as isolated singletons. Therefore, it is useful to consider a prior shrinkage
on the graph degree $D_i= \sum_{j\neq i} A_{i,j}$ for $i=1
\ldots, V$.

To prescribe a likelihood for graph estimation, we are motivated by the popularity of the simple hard-thresholding on the empirical correlation matrix $A_{i,j}= 1( |\rho_{i,j}| > \tau)$ with some $\tau\in(0,1)$. Although appearing heuristic, it was recently shown to have an equivalence to the more sophisticated graphical lasso \citep{sojoudi2016equivalence}. Therefore, it is interesting to develop a generalized Bayes extension that allows prior regularization.  Assuming $|\rho_{i,j}|\neq 0$ or $1$, we assign a Beta pseudo-likelihood for each $|\rho_{i,j}|$ and a degree shrinkage-prior,
\be
& L(\rho_{i,j}; A_{i,j}) \propto |\rho_{i,j}|^{A_{i,j}} (1-|\rho_{i,j}|)^{(1-A_{i,j})} \text{ for } j<i, \\
& \Pi_{0,A}(A) \propto \prod_{i=1}^V (\phi_i\tau)^{-1}\exp( -  \frac{D_i}{\phi_i \tau}),\\
& \Pi_{0,\phi}(\phi) \propto \prod_{i=1}^V\phi^{\alpha-1}_i, \qquad \Pi_{0,\tau}(\tau)\propto \exp(-\frac{\tau}{V}).
\ee
Each $A_{i,j}$ can be viewed as if a Bernoulli random variable and therefore a  ``soft'' thresholding.
Note that although it ignores the positive definite constraint for the correlation matrix, this generalized Bayes posterior still enjoys coherence in decision theory, as studied by \cite{bissiri2016general}. For the prior, we use the Dirichlet-Laplace shrinkage prior \citep{bhattacharya2015dirichlet} for the degrees $(D_1,\ldots,D_V)$ , with $\phi\sim \text{Dirichlet}(\alpha,\ldots,\alpha)$, $\tau\sim \text{Exp}(V)$ with a weakly-informative mean at $V$. We use $\alpha=0.01$ to encourage sparsity in $(\phi_1,\ldots,\phi_V)$.

 In this case, the parameter is in high dimension   $p=8,193$, and we have both continuous and discrete elements. To accommodate this,
we separate the output of each $T_k(\beta)$ into three parts $(\gamma^k_1, \gamma^k_2, \gamma^k_3)$, corresponding to $(A, \phi, \tau)$, and use
\[
\tilde\Pi(\beta) \propto \sum_{k=1}^K w_k   \big\{T_k(\beta) \big\}    {|\textup{det} \nabla T_k(\beta)|}  \Pi_{0,\phi}(\gamma ^k_2) \Pi_{0,\tau}(\gamma ^k_3)
\prod_{j<i} L\{ \rho_{i,j}; R(\gamma ^k_1)\} \Pi_{0,A}\{R(\gamma^k_1)\},
 \]
where $\Pi_{0,\phi}(\gamma ^k_2)$ is the Dirichlet density re-parameterized as the transform from gamma random variables Gamma$(\alpha,1)$, multiplied to the associated Jacobian.

\begin{figure}[H]
     \centering
      \begin{subfigure}[t]{0.32\textwidth}
         \centering
         \includegraphics[ width=\textwidth]{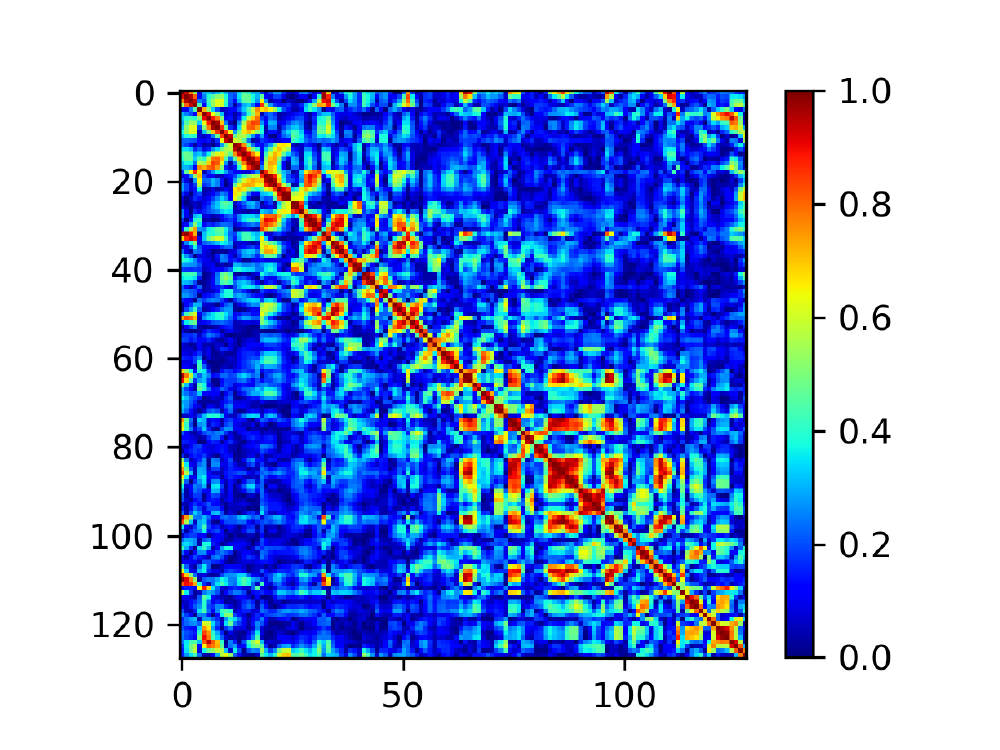}
         \caption{Empirical correlation matrix, shown in absolute values.}
     \end{subfigure}
               \hfill
     \begin{subfigure}[t]{0.32\textwidth}
         \centering
          \includegraphics[width=\textwidth]{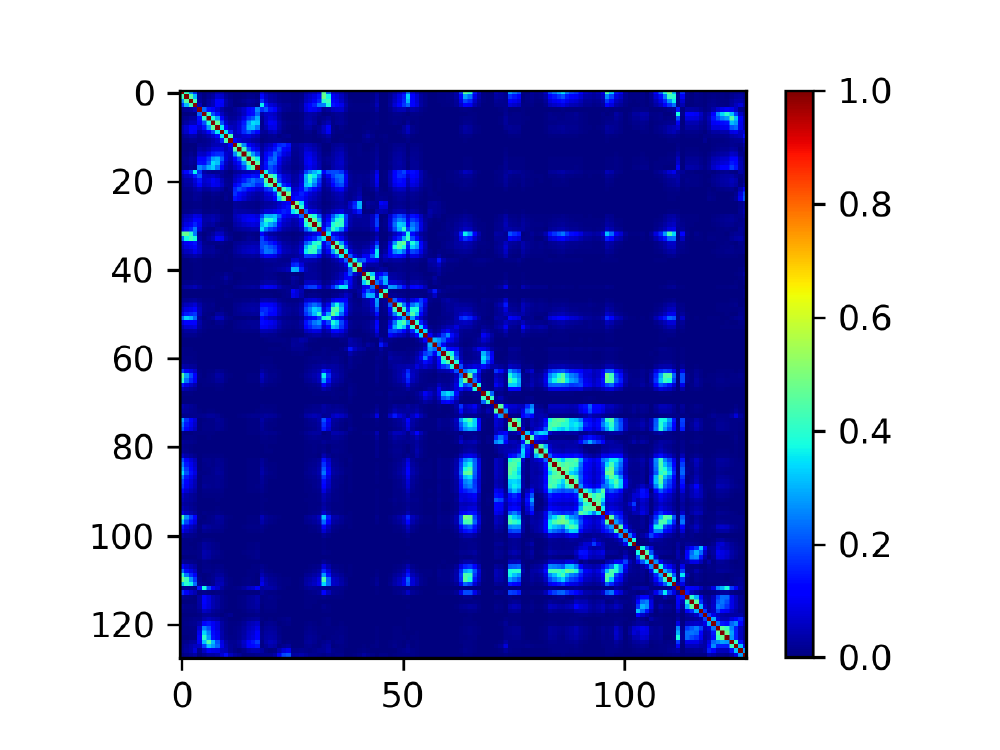}
         \caption{Estimated correlation via graphical lasso, shown in absolute values.}
     \end{subfigure}
     \hfill
     \begin{subfigure}[t]{0.32\textwidth}
         \centering
          \includegraphics[width=\textwidth]{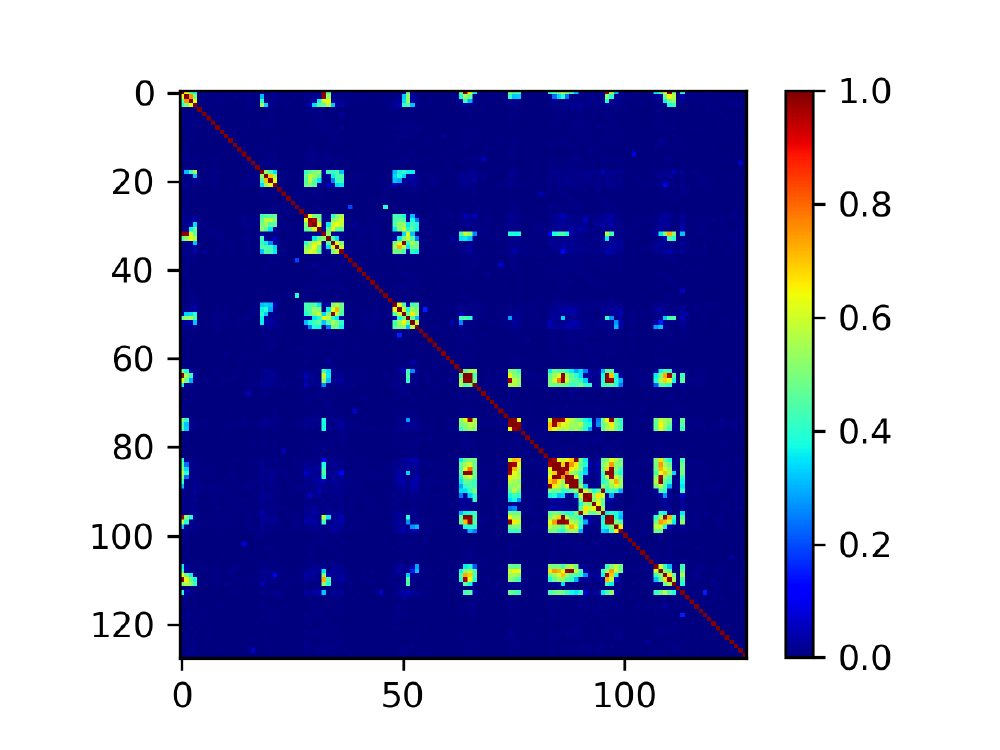}
         \caption{Posterior mean of $A_{i,j}$ using the Beta likelihood and the degree shrinkage prior, estimated using Transport Monte Carlo.}
     \end{subfigure}
          \hfill
     \begin{subfigure}[t]{0.45\textwidth}
         \centering
          \includegraphics[width=\textwidth,height=1.5in]{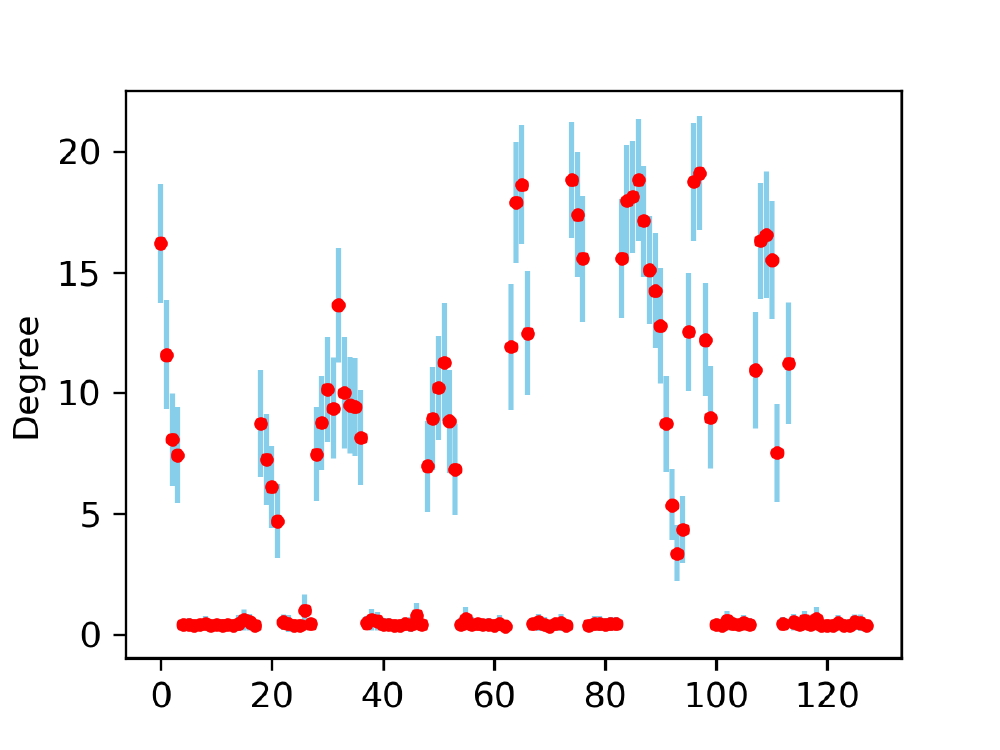}
         \caption{Estimated degree $D_{i}$ with the posterior mean (red) and $95\%$ point-wise credible interval (blue), estimated using Transport Monte Carlo.}
     \end{subfigure}\hfil
          \begin{subfigure}[t]{0.45\textwidth}
         \centering
          \includegraphics[width=1\textwidth,height=1.5in]{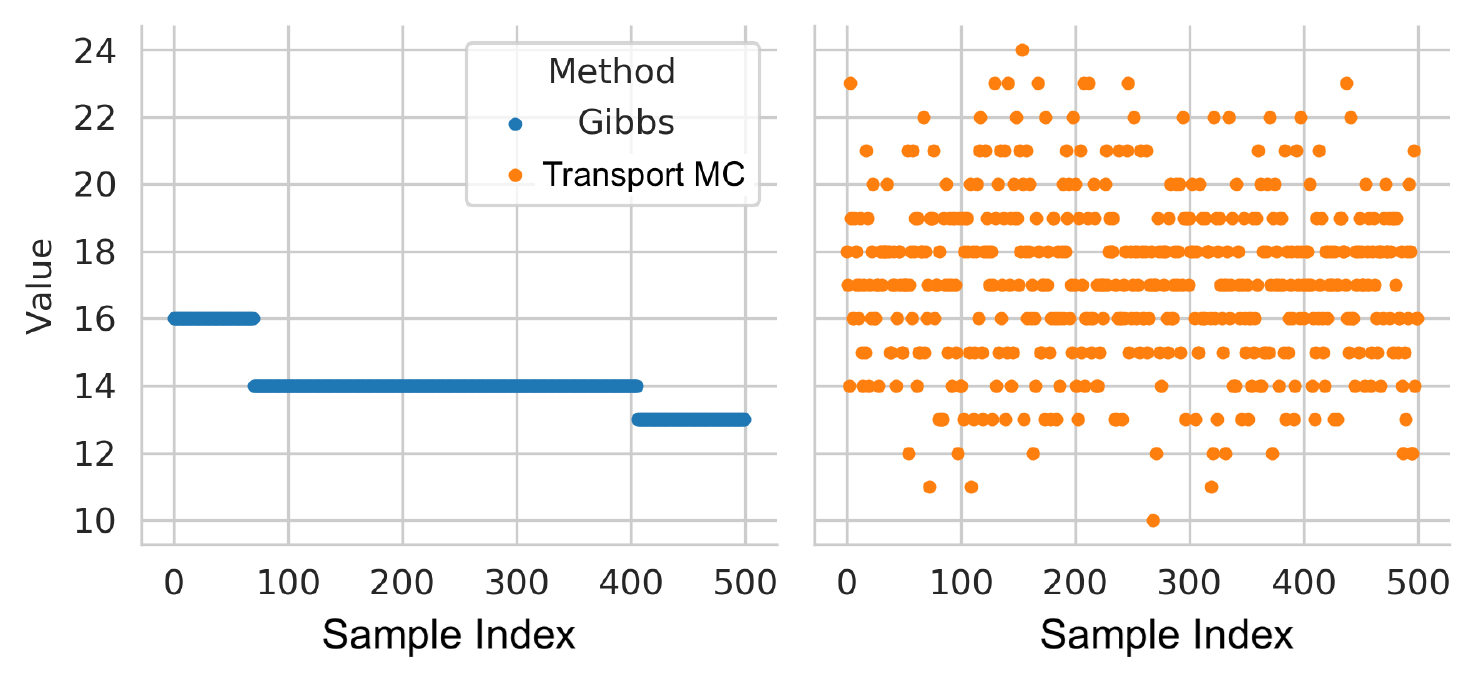}
         \caption{Traceplots of $D_1$ from Gibbs sampling and Transport Monte Carlo algorithms.}
     \end{subfigure}
        \caption{Data application using the Transport Monte Carlo to estimate a binary adjacency matrix $A$, based on the empirical correlation (panel a) using  Dirichlet-Laplace shrinkage on the degrees. By shrinking the degrees of some nodes to zero (panel d), we found well-connected sub-graphs (panel c). The graphical lasso found a similar structure  (panel b), except that it did not have degree-sparsity and it under-estimated the large signals. \label{fig:graph_est}}
\end{figure}

Figure~\ref{fig:graph_est} shows the result of posterior estimation. We successfully shrunk the degrees of some nodes to zero (panel d). The remaining nodes correspond to well-connected sub-graphs (panel c). To compare, we also ran graphical lasso, and it discovered a similar structure  (panel b), except that it did not have degree-sparsity and under-estimated the large signals (as a known side-effect of the $l_1$-regularization). For comparison, we ran the Gibbs sampling algorithm that updated one $A_{i,j}$ at a time. The mixing was extremely slow, as shown in Figure~\ref{fig:graph_est}(e). The TMC was free from this issue as the samples were independent.

\section{Discussion}
In recent years, it has become increasingly easier to develop and apply Bayesian models, thanks to the new tools capable of handling posterior without closed-form conditional. A primary factor that contributes to their success is arguably the reduction of efforts and time needed for deriving and implementing an algorithm; as a result, statisticians can focus more on model design and calibration.

We believe our proposed method is another advance in this direction. In particular, our method substantially reduces the time needed from writing a Bayesian model to collecting posterior samples in the high probability region. In all of our experiments, including the high dimensional ones, the computation took at most a few minutes. Such a close to real-time feedback is very beneficial, and it encourages statisticians to explore new forms of likelihood and prior through rapid experiments. For example, our new regularized graph estimation was made possible because of the ability to avoid sequential search in the graph space, which would be a highly combinatorial and challenging problem.

Compared to the other optimization-based approaches, the main distinction of our proposal is the infinite mixture of simple transforms. This not only enables a ``shotgun'' algorithm that can handle multi-modality in the posterior, but also leads to a tractable theoretic analysis via piecewise probability approximation. 

Lastly, this framework can be extended for general statistical inferences, such as the conditional density estimation. For example, one could estimate a transport plan between the empirical distributions of some predictors and an outcome of interest. This is similar to the estimation of Wasserstein distances \citep{solomon2015convolutional,NEURIPS2019_f0935e4c}; nevertheless, to prevent overfitting under a finite sample size, it would be important to choose an appropriate cost function or regularization to yield a parsimonious transport plan. The methodology, as well as the signal recovery theory, is still an underexplored but interesting topic.

\section*{Acknowledgement}
The author would like to thank James Hobert, David Dunson, and Yun Yang for useful discussions.
\footnotesize

{\centering  \LARGE \bf Supplementary Materials}

\section*{Extension: Independence Hastings Algorithm}

A unique advantage of Markov chain Monte Carlo (MCMC)   is the ``asymptotic exactness'': as the number of collected samples increases to infinity, under some conditions, the empirical distribution of the  Markov chain samples will converge to the posterior distribution \citep{roberts1996exponential}.
Since the optimized random transport can generate samples with small approximation errors, we can use it to build a proposal-generating distribution. In the Markov chain, we denote a given state by $(\beta^t,\theta^t)$  with
$\theta^t=T_{c^t}(\beta^t)$,
 and the new proposal by $(\beta^*,\theta^*)$.
To compare, the transport map-based MCMC algorithm \citep{parno2018transport} transforms a simple  Metropolis proposal  $Q(\beta^{*}\mid\beta^{t})$ on the reference into a sophisticated one $Q(\theta^{*}\mid\theta^{t}) $ for the target, hence the proposal is dependent on the current state; whereas in our extension, the proposal is independent $Q(\beta^*,\theta^* \mid\beta^{t},\theta^{t} ) = Q(\beta^*,\theta^*  )$ hence potentially more efficient in exploring the parameter space.

To formalize, consider the  target distribution as the augmented
 $\Pi(\theta;y)\tilde\Pi(\beta\mid\theta)$, with the later  a categorical distribution $\text{Pr}\{\beta=T^{-1}_k(\theta)\}=w_k(\theta)$ as defined in the main text (except that we use the truncation at $K$ and treat the $\{T_k,w_k\}_{k}$ as fixed).
Clearly, the $\theta$-marginal distribution is  still the posterior $\Pi(\theta;y)$.

When devising the proposal kernel, we recognize that if the target state space $\Theta$ is unbounded, such as $\mathbb{R}^p$,  there will be a small discrepancy from the image of $T_k$'s from a uniform  reference sample $\beta\sim \Pi_r(\beta)$. To correct this, we now generate the $\beta$ from a two-component mixture, with one component from uniform $\Pi_r(\beta)$ and the other from  distribution $\Pi_a(\beta)$ with the unbounded support. This leads to a proposal kernel:
\be
& Q(\beta^*,\theta^*) =
\{\rho\Pi_r(\beta^{*})+ (1-\rho)\Pi_a(\beta^*)\}\tilde \Pi(\theta^*\mid \beta^*)\\
&=
\{\rho\Pi_r(\beta^{*})+ (1-\rho)\Pi_a(\beta^*)\}\\
&\;\;\;\;\;\;\times\frac{ w_c\{ T_c(\beta^*)\} \Pi\{T_c(\beta^{*}); y\}
{|\textup{det} \nabla T_c(\beta^{*})|1\{\theta^*=T_c(\beta^*), T_c(\beta^*)\in \Theta \}} }{
\sum_{k=1}^{K}   w_k \{\ T_k(\beta^{*})\}\ \Pi\{ T_k(\beta^{*}); y\}
{|\textup{det} \nabla T_k(\beta^{*})| 1\{ T_k(\beta^*)\in \Theta \}}},
\ee
with $\rho\in(0,1]$ and chosen to be a  value close to $1$. 
Using the Hastings algorithm, we accept $(\beta^*,\theta^*)$ with probability
\be
\min
\left\{ 1,
\frac{\Pi(\theta^*;y)\tilde \Pi(\beta^{*}\mid\theta^{*})}{\Pi(\theta^t;y)\tilde\Pi(\beta^t\mid\theta^t)}
\frac{
  Q(\beta^{t},\theta^{t})
}
{
  Q(\beta^*,\theta^*)
}
\right\},
\ee
Applying change of variable $\theta=T_c(\beta)$ and some cancellations (detail provided later), the above acceptance rate becomes

\be
\min\left[ 1,\frac{
  \{\rho\Pi_r(\beta^{t})+ (1-\rho)\Pi_a(\beta^t)\}\tilde\Pi(\beta^*)|\textup{det} \nabla T_{c^t}(\beta^{t})|
}
{
  \{\rho\Pi_r(\beta^{*})+ (1-\rho)\Pi_a(\beta^*)\}\tilde\Pi(\beta^t)|\textup{det} \nabla T_{c^*}(\beta^{*})|
}1\{ T_{c^*}(\beta^*)\in \Theta\}\right].
\ee

Recall that $\tilde\Pi(\beta)$ is an approximation to $\Pi_r(\beta)$--- a uniform. Therefore, with $\rho\approx 1$ and $\{T_k,w_k\}_k$ optimized,
 the acceptance rate will be close to a constant. Further, if we can ensure the  $\textup{det}
\nabla T_{k}(\beta)=\prod_{j=1}^p s_{k,j}$ is similar for all $k$, then the acceptance rate will be close to one.

\begin{remark}
       Note that the proposal $(\theta^*,\beta^*)$ is independent of the current state $(\theta^t,\beta^t)$, making this an independence Hastings algorithm \citep{tierney1994markov}.
\end{remark}

As shown in early work [\cite{tierney1994markov,mengersen1996rates} among others], a sufficient condition to ensure asymptotic exactness of MCMC, is when the ratio between the proposal and target is bounded from below. In our case, this can be achieved with $\Pi_a(\beta)/\Pi[ T_k(\beta); y]>\lambda$ for all $k=1,\ldots,K$. To see this,
\begin{equation}
\begin{aligned}
 \frac{
Q(\beta,\theta)
} { \Pi(\theta;y)\tilde\Pi(\beta\mid\theta)}&=
\frac{
 \{\rho\Pi_r(\beta^{})+ (1-\rho)\Pi_a(\beta)\}|\textup{det} \nabla T_{c}(\beta^{})|\   }
  {
\sum_{k=1}^{K}   w_k \{T_k(\beta^{})\} \Pi\{ T_k(\beta^{}); y\}
{|\textup{det} \nabla T_k(\beta^{})| 1\{ T_k(\beta)\in \Theta\} }}
 \\
  &  \ge (1-\rho)\lambda (\min_{l=1\ldots,K}\prod_{j=1}^ps_{l,j})/( \sum_{k=1}^{K}\prod_{j=1}^ps_{k,j}  ),
\end{aligned}
\end{equation}
due to the cancellation $\tilde\Pi\{\beta= T_c(\beta)\mid\theta\}= w_c\{ T_c(\beta)\}$, and $\Pi(\theta;y)= \Pi\{T_c(\beta); y\}$, $\rho\Pi_r(\beta^{})>0$ and each $w_k \{T_k(\beta^{})\}\le 1$.
In practice, a common choice for $\Pi_a(\beta)$ is a heavy-tail distribution, such as multivariate $t$-distribution (provided it can satisfy the above condition). Another potential issue is that as the dimension $p\to\infty$, the independence Hastings algorithm could suffer from the curse of dimensionality, with the acceptance rate approaching $0$. A common remedy is to use block-wise updating, that each time proposes change to only one part of the parameter.

\subsection*{Details of Hastings Acceptance Rate}

At the current state, $T_{c^t}(\beta^t)\in \Theta$; if $T_{c^*}(\beta)\not\in \Theta$, we will reject it; therefore, we focus on $T_{c^*}(\beta^*)\in \Theta$
as well.
\be
&\frac{\cancel{\Pi(\theta^*;y)}\tilde \Pi(\beta^{*}\mid\theta^{*}) }
{\cancel{\Pi(\theta^t;y)}\tilde\Pi(\beta^t\mid\theta^t) }
\frac{
  \{\rho\Pi_r(\beta^{t})+ (1-\rho)\Pi_a(\beta^t)\}\frac{ w_{c^t}\{ T_{c^t}(\beta^t)\}\cancel{
 \Pi\{T_{c^t}(\beta^{t});
y\}}{
{
|\textup{det} \nabla T_{c^t}(\beta^{t})|}}1\{\theta^t=T_{c^t}(\beta^t), T_{c^t}(\beta^t)\in
\Theta \}
 }{
\sum_{k=1}^{\infty}   w_k \{ T_k(\beta^{t})\} \Pi\{T_k(\beta^{t}); y\}
{|\textup{det} \nabla T_k(\beta^{t})| 1\{ T_k(\beta^t)\in
\Theta\}}}
}
{
  \{\rho\Pi_r(\beta^{*})+ (1-\rho)\Pi_a(\beta^*)\}\frac{ w_{c^*}\{ T_{c^*}(\beta^*)\}\cancel{
\Pi\{T_{c^*}(\beta^{*});
y\}}
{{ |\textup{det} \nabla T_{c^*}(\beta^{*})|}} 1\{\theta^*=T_{c^*}(\beta^*), T_{c^*}(\beta^*)\in
\Theta \}}{
\sum_{k=1}^{\infty}   w_k \{ T_k(\beta^{*})\} \Pi\{T_k(\beta^{*}); y\}
{|\textup{det} \nabla T_k(\beta^{*})| 1\{ T_{c^*}(\beta^*)\in
\Theta\}}}
}
\\
&=\frac{  \cancel{\tilde \Pi(\beta^{*}\mid\theta^{*})} }
{\cancel{\tilde\Pi(\beta^t\mid\theta^t)} }
\frac{
  \{\rho\Pi_r(\beta^{t})+ (1-\rho)\Pi_a(\beta^t)\}\frac{\cancel{ w_{c^t}\{ T_{c^t}(\beta^t)\}}
{
{
|\textup{det} \nabla T_{c^t}(\beta^{t})|}}1\{\theta^t=T_{c^t}(\beta^t) \}
 }{
\sum_{k=1}^{\infty}   w_k \{ T_k(\beta^{t})\} \Pi\{T_k(\beta^{t}); y\}
{|\textup{det} \nabla T_k(\beta^{t})| 1\{ T_k(\beta^t)\in
\Theta\}}}
}
{
  \{\rho\Pi_r(\beta^{*})+ (1-\rho)\Pi_a(\beta^*)\}\frac{\cancel{ w_{c^*}\{ T_{c^*}(\beta^*)\}}
{{ |\textup{det} \nabla T_{c^*}(\beta^{*})|}} 1\{\theta^*=T_{c^*}(\beta^*) \}}{
\sum_{k=1}^{\infty}   w_k \{ T_k(\beta^{*})\} \Pi\{T_k(\beta^{*}); y\}
{|\textup{det} \nabla T_k(\beta^{*})| 1\{ T_{c^*}(\beta^*)\in
\Theta\}}}
}\\
& =
\frac{
  \{\rho\Pi_r(\beta^{t})+ (1-\rho)\Pi_a(\beta^t)\}|\textup{det} \nabla T_{c^t}(\beta^{t})|{ \sum_{k=1}^{\infty}   w_k
\{ T_k(\beta^{*})\} \Pi\{T_k(\beta^{*}); y\}
{|\textup{det} \nabla T_k(\beta^{*})| 1\{ T_{c^*}(\beta^*)\in
\Theta\}}
}
}
{
  \{\rho\Pi_r(\beta^{*})+ (1-\rho)\Pi_a(\beta^*)\}|\textup{det} \nabla T_{c^*}(\beta^{*})|
  {
\sum_{k=1}^{\infty}   w_k \{ T_k(\beta^{t})\} \Pi\{T_k(\beta^{t}); y\}
{|\textup{det} \nabla T_k(\beta^{t})| 1\{ T_{c^t}(\beta^t)\in
\Theta\}}
}
}\\
& =
\frac{
  \{\rho\Pi_r(\beta^{t})+ (1-\rho)\Pi_a(\beta^t)\}|\textup{det} \nabla T_{c^t}(\beta^{t})|\tilde\Pi(\beta^*)
}
{
  \{\rho\Pi_r(\beta^{*})+ (1-\rho)\Pi_a(\beta^*)\}|\textup{det} \nabla T_{c^*}(\beta^{*})|\tilde\Pi(\beta^t)
}.
\ee

\subsection*{Table of an Approximate Transport Plan}

\begin{table}[h]\scriptsize
\begin{tabular}{ p{4mm} | p{19mm}  ||  p{20mm}|  p{20mm} | p{17mm} | p{17mm} | p{15mm} | p{15mm} | }
& $pr_{\theta}$ & $\bf 0.02$ & $\bf 0.14$ & $\bf 0.34$ & $\bf 0.34$ & $\bf 0.14$ & $\bf 0.02$ \\
\hline
$pr_{\beta}$ &  $pr_{\theta,\beta}$, \newline $\theta=T_k(\beta)$ 
& \(\Theta_1 = (-3,-2)\)
& \(\Theta_2 = (-2,-1)\) & \(\Theta_3 =(-1,0)\) & \(\Theta_4 =(0,1)\) & \(\Theta_5 =(1,2)\) 
& \(\Theta_6 = (2,3)\)
\\
\hline
\hline
$\bf 0.5$ & \(\Beta_1=(0,0.5)\)
 & $\bf 0.02$ \newline  \(T_1(\beta)= \newline  2\beta -3\)
 & $\bf 0$ \newline  \(\)
&\(  \bf 0.24\) \newline \(T_3(\beta) =\newline 2\beta -1\) &
\(  \bf 0.1\)\newline \(T_4(\beta)=\newline 2\beta\)   &
 \( \bf  0.14\)  \newline \(T_5(\beta)=\newline 2\beta+1\)
  & $\bf 0$ \newline  \(\)
 \\
\hline
$\bf 0.5$ & \(\Beta_2=(0.5,1)\) & \bf 0  \newline  \(\) &
 \( \bf   0.14\)\newline \(T_1(\beta)=\newline 2\beta -3\)  &
\(\bf 0.1\) \newline \(T_2(\beta) =\newline 2\beta-2\) &
\(\bf 0.24\)\newline \(T_3(\beta)=\newline 2\beta -1\) &
\(\bf 0\)\newline \(\) 
& $\bf 0.02$ \newline  \(T_5(\beta)= \newline  2\beta +1\) 
\\
\hline
\end{tabular}
\caption{Table of an approxiamte transport plan that change the histogram of $\beta\sim U(0,1)$ into a histogram of $\theta\sim N(0,1)$ (using fewer bins than the heatmap for conciseness). The marginal probability (bold) for each histogram bin is shown in the row/column heading, while the joint probability (bold) is shown in each cell, along with the used location-scale change.  \label{tb:transport_plan}}
\end{table}

\subsection*{Proof of Theorem 1}

Proof of Theorem 1 can be found in \cite{schilling2017measures}.

\subsection*{Proof of Theorem 2}
\begin{proof}

We first focus on $\Theta=\mathbb{R}^p$ and $\Pi(\theta;y)>0$ for any $\theta\in \Theta$. For simplicity, we denote $\theta_{0kl}= T_l\{T_k^{-1}(\theta_0)\}$.
\be
\tilde a_k(\theta_0) & = \frac{ w_{k}(\theta_0)\Pi\{T_k(T^{-1}_k(\theta_0)); y\}
{|\textup{det} \nabla T_k|} }{
\sum_{l=1}^{K}   w_{k}(T_l(T^{-1}_k(\theta_0)) \Pi\{T_l(T^{-1}_k(\theta_0)); y\}
{|\textup{det} \nabla T_l|  }
}
\frac{1}{\prod_{j=1}^p s_{k,j}}
\\
& = \frac{  w_{k}(\theta_0)\Pi(\theta_0; y)
{|\textup{det} \nabla T_k|} }{
\sum_{l=1}^{K}    w_{l}(\theta_{0kl})\Pi(\theta_{0kl}; y)
{|\textup{det} \nabla T_l| }
}
\frac{1}{\prod_{j=1}^p s_{k,j}}\\
&=
\frac{  \Pi(\theta_0; y) }{
\sum_{l=1}^{K}  \{w_{l}(\theta_{0kl})/w_{k}(\theta_0)\}  \Pi\{\theta_{0k_{}l}; y\}\prod_{j=1}^p s_{l,j}
}
\ee


{\bf a) Making $C_k$'s pairwise disjoint}.

For any given $K$, we can select $\{m_k,s_{k}\}_{k=1}^K$ with $m_k\neq m_{k'}$ if $k\neq k'$, and $s_{0,j}$ sufficiently small, so that all $C_k$'s are pairwise disjoint.

 Further, if $\Pi(\theta;y)$ contains points of discontinuity at set $\{\theta^\dagger_i\}_i$, we can partition the rest $\Theta \setminus \{\theta^\dagger_i\}_i=\Theta_1 \cup\ldots\cup\Theta_H$, with $\Pi(\theta;y)$ continuous in each $\Theta_h$. Then we can choose suitable $\{m_k,s_{k}\}_{k=1}^K$, so that $C_k$'s do not contain any $\theta^\dagger_i$.

{\bf b) Piecewise approximation}.

For any $\theta_0\in \setminus \{\theta^\dagger_i\}_i$, we can have $\theta_0\in C_{k_0}$ in a set $C_{k_0}$. Since $C_k$'s are pairwise disjoint:

\be
\sum_{k=1}^K \tilde a_k (\theta_0)1(\theta_0 \in C_k)=\frac{  \Pi(\theta_0; y) }{
\sum_{l=1}^{K}  \{w_{l}(\theta_{0k_{0}l})/w_{k_{0}}(\theta_0)\}  \Pi(\theta_{0k_{0}l}; y)(\prod_{j=1}^p s_{0,j})
},
\ee
We denote the denominator on the right-hand side by $G$.

For each $l$, by the continuity of $\Pi(\theta;y)$, and each $\theta_{0 k_0l}\in C_l$ (a compact set), there exists a pair of constants $(q_{K,l},r_{K,l})$ such that $q_{K,l}\ge \Pi(\theta_{0 k_0l}; y) \ge r_{K,l}>0$, and $q_{K,l}/r_{K,l}\to 1$ as $\|s_{l}\|\to 0$.

We now choose $w_l(\theta)$ to be a constant-output function (that is, with the output invariant to the input $\theta$) therefore, we will use short notation $w_l(\theta)=w_l$ from now on.
We choose $ w_{l} \propto 1/(q_{K,l} \prod_{j=1}^p s_{l,j})$, subject to $\sum_{l=1}^K w_l=1$.

(i) If  $\Pi(\theta_0; y)>\epsilon$, we will show that $G$ can go to $1$ as $K\to\infty$. We choose $\prod_{j=1}^p s_{k_0,j}=1/(K q_{K,k_0})$. Therefore,
\be
G\le & \sum_{l=1}^{K}  ( w_{l}/w_{k_{0}}  )q_{K,l}\prod_{j=1}^p s_{l,j}\\
 = & K( q_{K,k_0}\prod_{j=1}^p s_{{k_{0}}  ,j}) \\
= & 1,
\ee

On the other hand,
\be
G\ge & \sum_{l=1}^{K}  ( w_{l}/w_{k_{0}}  )r_{K,l}(\prod_{j=1}^p s_{l,j})\\
 =&(\prod_{j=1}^p s_{{k_{0}}  ,j})q_{K,k_0} \sum_{l=1}^K \frac{r_{K,l}}{q_{K,l}}\\
\ge & \frac{1}{K} K \inf_l \frac{r_{K,l}}{q_{K,l}} ,
\ee
which goes to $1$ as $K\to \infty$.

(ii) If  $\Pi(\theta_0; y)\le \epsilon$, we choose $\prod_{j=1}^p s_{k_0,j}=1/(K \epsilon)$

\be
G\ge & (1/K\epsilon)q_{K,k_0} \sum_{l=1}^K \frac{r_{K,l}}{q_{K,l}}
\ee

Therefore,
\be
\frac{\Pi(\theta_0; y)}{G}\le \frac{q_{K,k_0}}{(1/K\epsilon)q_{K,k_0} \sum_{l=1}^K \frac{r_{K,l}}{q_{K,l}}} \le \epsilon \frac{1}{ \inf_l \frac{r_{K,l}}{q_{K,l}}},
\ee
which goes to $\epsilon$ as $K\to \infty$.

Lastly, it is easy to verify that there exists $\{s_{l}\}_{l=1}^K$ so that $\|s_l\|\to 0$ for all $l=1,\ldots, K$ as $K\to\infty$.

 Therefore, for any $\theta_0\in \Theta\setminus \{\theta^\dagger_i\}_i$, there exists a sequence of $\{\sum_{k=1}^K \tilde a_k (\theta_0)1(\theta_0 \in C_k)\}_K$, such that,
\be
\lim_{K\to\infty}|\sum_{k=1}^K \tilde a_k (\theta_0)1(\theta_0 \in C_k)-\Pi(\theta_0;y)| \to 0.
\ee

To see how the above extends to $\Theta$ as a subset of  $\mathbb{R}^p$, as a regularity, we define   $\Pi\{\theta_{0k_{}l}; y\}=0$ if $\theta_{0k_{}l}\not \in \Theta$.
\be
\sum_{k=1}^{K}  \tilde a_k (\theta_0)1(\theta_0 \in C_k)=
\frac{ \Pi(\theta_0; y) 1(\theta_{0}\in \Theta) }{
\sum_{l=1}^{K}  \{w_{l}/w_{k}\}   \Pi(\theta_{0k_{}l}; y)\prod_{j=1}^p s_{l,j} 1(\theta_{0k_0l}\in \Theta)
}
\ee

For each $l$,  if $T_l\{ (0,1)^{p}\}\subseteq\Theta_h$ for any $h=1\ldots H$, then $(q_{K,l},r_{K,l})$ as mentioned before still exist, and set $w_l \propto 1/(q_{K,l} \prod_{j=1}^p s_{l,j})$. If if $T_l\{ (0,1)^{p}\}\not\subseteq\Theta_h$ for any $h$, we set $w_l=0$. Record $K^*= \sum_{l=1}^K 1(w_{l}>0)$.
Since each $\Theta_h$ is a continuous set, it is not hard to see $K^*$ can go to infinity, with appropriate $\{m_k,s_k\}_{k=1}^K$ and $\|s_k\|_1\to 0$.

If $\Pi(\theta_0; y)=0$, we have $\sum_{k=1}^{K}  \tilde a_k (\theta_0)1(\theta_0 \in C_k)=0$ for any $K^*\ge 1$.

If   $\Pi(\theta_0; y)>\epsilon,$  we set $\prod_{j=1}^p s_{k_0,j}=1/(K^{*} q_{K,k_0})$, we have the denominator:
\be
G= &\sum_{l=1}^{K}  (w_{l}/w_{k})   \Pi(\theta_{0k_{}l}; y)\prod_{j=1}^p s_{l,j} 1(\theta_{0k_0l}\in \Theta)\to 1,
\ee
If  $0<\Pi(\theta_0; y)\le \epsilon$, we choose $\prod_{j=1}^p s_{k_0,j}=1/(K^* \epsilon)$, then
 the upper bound on $\Pi(\theta_0; y)/{G}$ goes to $\epsilon$ as well, when $K^*\to \infty$.
\end{proof}

\subsection*{Proof of Theorem 3}
 \begin{proof}

We show the existence via one (among many) constructions.

The total variational distance is,

\be
\|\pi_r-\tilde\pi\|_{TV}=
\sup_{\mathcal A\subseteq\in \mathcal B(\Beta)}|\int_{\mathcal A} \int_\Theta & \{\tilde \Pi( \beta\mid \theta) - \Pi( \beta\mid \theta)\} \Pi(\theta; y) \textup{d} \theta |,
\ee

For a measurable $\mathcal A\in \mathcal B(\Beta)$, denote the conditional probability by
\be
\nu_{\beta\mid\theta}(\mathcal A)=\int_\mathcal{A}\frac{P(\theta, \beta)}{\Pi(\theta\mid y)} d\beta \in (0,1\}.
\ee
%

We will divide the $\Theta$ into a two sets: a bounded subset $\Theta^*$: $\int_{\Theta^*}  \Pi(\theta; y) \textup{d} \theta =1-\epsilon/2$, and the rest $\Theta\setminus \Theta^*$ with negligibly small measure $\epsilon/2$ w.r.t. $\Pi(\theta; y)$.

{\bf a) When $\theta\in\Theta^*$:}

Let $E_{\{n\}1},\ldots, E_{\{n\}2^{pn}} $ be the partitioning cubes for $(0,1)^p$,
define as $E_{\{n\}k}=\times_{j=1}^p \{ k^*_j/2^{n}, (k^*_j+1)/2^{n})$ for $k^{*}_j=0,\ldots,(2^n-1)$, and $k=(k^*_1,\ldots,k^*_p)$.
Because $\nu_{\beta\mid\theta}$ is a Lebesgue measurable for any bounded $\theta$, it is not hard to see
that for any $\mathcal A\in(0,1)^p$
\be
&\nu_{\beta\mid\theta} ( \mathcal A ) = \lim_{n\to \infty}\sum_{k=1}^{2^{np}} 1(E_{\{n\}k}\subseteq \mathcal A )\nu_{\beta\mid\theta} (  E_{\{n\}k}),\\
&\nu_{\beta\mid\theta} ( \mathcal A ) = \lim_{n\to \infty}\sum_{k=1}^{2^{np}} \{1(E_{\{n\}k}\subseteq \mathcal A)+1(E_{\{n\}k}\not\subseteq \mathcal A,E_{\{n\}k}\cap \mathcal A \neq \varnothing)\}\nu_{\beta\mid\theta} (  E_{\{n\}k}),\\
\ee
That is, the limit measures of the maximum packing cubes, and the minimum covering cubes.

For a sufficiently large $n\ge N(\mathcal A,\epsilon)$, we have
\be
&\nu_{\beta\mid\theta} ( \mathcal A ) - \sum_{k=1}^{2^{np}} 1(E_{\{n\}k}\subseteq \mathcal A )\nu_{\beta\mid\theta} (  E_{\{n\}k}) \le\epsilon/4,\\
&\sum_{k=1}^{2^{np}} \{1(E_{\{n\}k}\not\subseteq \mathcal A,E_{\{n\}k}\cap \mathcal A \neq \varnothing)\}\nu_{\beta\mid\theta} (  E_{\{n\}k})\le \epsilon/4.
\ee

On the other hand, for the mixture distribution:
$$
\phi(\beta\mid \theta)= \sum_{k=1}^{K}w_k(\theta)\delta\{\beta
-T^{-1}_k(\theta)\},
$$
we can find $s_k, m_k$ such that $T^{-1}_k(\theta) ={\theta} \odot s_k^{-1} - m_k\odot s_k^{-1} \in  E_{\{n\}k}\;\; \forall\theta\in\Theta^*$ (that is, reducing the scale and shifting the location, so that all the  $T^{-1}_k(\Theta^*)$  falls inside the cube). Let $w_k(\theta)=\nu_{\beta\mid\theta}( E_{\{n\}k})$ and integrate over $\mathcal A,$
\be
\Phi_{\beta\mid\theta} ( \mathcal A ) &=\int_{\mathcal A} \phi(\beta\mid \theta) d\beta \\ &= \sum_{k=1}^{K} \nu_{\beta\mid\theta}( E_{\{n\}k}) \int_{\mathcal A}\delta\{\beta
- T^{-1}_k(\theta)\} \textup{d}\beta \\
&\stackrel{(i)}= \sum_{k=1}^{K} \nu_{\beta\mid\theta}( E_{\{n\}k}) 1\{T^{-1}_k(\theta)\in \mathcal A\} \\
&= \sum_{k=1}^{K} \nu_{\beta\mid\theta}( E_{\{n\}k})\{ 1(E_{\{n\}k}\subseteq \mathcal A,T^{-1}_k(\theta)\in \mathcal A)+ 1(E_{\{n\}k}\not\subseteq \mathcal A,T^{-1}_k(\theta)\in \mathcal A)\},\\
&\stackrel{(ii)}= \sum_{k=1}^{K} \nu_{\beta\mid\theta}( E_{\{n\}k})\{ 1(E_{\{n\}k}\subseteq \mathcal A)+ 1(E_{\{n\}k}\not\subseteq \mathcal A,T^{-1}_k(\theta)\in \mathcal A)\},
\ee
where $(i)$ is due to $\int_{\mathcal A} \delta(x-a)\textup{d}x= 1(a\in \mathcal A)$, and $(ii)$ is due to $E_{\{n\}k}\subseteq \mathcal A$ guarantees $T^{-1}_k(\theta)\in \mathcal A$.

Letting $K=2^{np}$, we have
\be
|\nu_{\beta\mid\theta} ( \mathcal A )-\Phi_{\beta\mid\theta} ( \mathcal A )|&\stackrel{(i)}\le \epsilon/4+ \sum_{k=1}^{2^{np}} \nu_{\beta\mid\theta}( E_{\{n\}k})1(E_{\{n\}k}\not\subseteq \mathcal A,T^{-1}_k(\theta)\in \mathcal A) \\
&\stackrel{(ii)} \le \epsilon/4+ \sum_{k=1}^{2^{np}} \nu_{\beta\mid\theta}( E_{\{n\}k})1(E_{\{n\}k}\not\subseteq \mathcal A,E_{\{n\}k}\cap \mathcal A \neq \varnothing) \\
&\le  \epsilon/2,
\ee
where $(i)$ uses triangle inequality, and (ii) is due to $T^{-1}_k(\theta)\in \mathcal A$ implies $E_{\{n\}k}\cap \mathcal A \neq \varnothing$.

{\bf b) When $\theta\in \Theta\setminus \Theta^*$}:

\be
   |\int_{\Theta\setminus\Theta^*} \int_{\mathcal A}  \{\tilde \Pi( \beta\mid \theta) - \Pi( \beta\mid \theta)\} \textup{d} \beta\Pi(\theta; y) \textup{d} \theta |
\le    |\int_{\Theta\setminus\Theta^*} 1\; \Pi(\theta; y) \textup{d} \theta |= \epsilon/2,
\ee
due to $|\int_{\mathcal A}  \{\tilde \Pi( \beta\mid \theta) - \Pi( \beta\mid \theta)\} \textup{d} \beta| \le\|\Phi_{\beta\mid\theta} - \nu_{\beta\mid\theta}\|_{TV}\le 1$.

Combining a) and b) we have.
  \be
  |\int_{\mathcal A} \int_\Theta & \{\tilde \Pi( \beta\mid \theta) - \Pi( \beta\mid \theta)\} \Pi(\theta; y) \textup{d} \theta \textup{d} \beta |\\
& \stackrel{(i)}=
  |\int_\Theta \int_{\mathcal A}  \{\tilde \Pi( \beta\mid \theta) - \Pi( \beta\mid \theta)\}\textup{d} \beta \Pi(\theta; y) \textup{d} \theta |\\
  & \stackrel{(ii)}\le
  |\int_{\Theta^*} \int_{\mathcal A}  \{\tilde \Pi( \beta\mid \theta) - \Pi( \beta\mid \theta)\}\textup{d} \beta \Pi(\theta; y) \textup{d} \theta |
  +   |\int_{\Theta\setminus\Theta^*} \int_{\mathcal A}  \{\tilde \Pi( \beta\mid \theta) - \Pi( \beta\mid \theta)\} \textup{d} \beta\Pi(\theta; y) \textup{d} \theta |\\
& \le
  \epsilon/2  \int_{\Theta^*} \Pi(\theta; y) \textup{d} \theta
  +   |\int_{\Theta\setminus\Theta^*} 1\; \Pi(\theta; y) \textup{d} \theta |
\\
&\stackrel{(iii)}\le
  \epsilon/2+
  \epsilon/2
\\
&= \epsilon
  \ee
  where $(i)$ uses Fubini, $(ii)$ uses triangle inequality and $(iii)$ uses $\int_{\Theta^*} \Pi(\theta; y) \textup{d} \theta=1-\epsilon/2\le 1$.

 \end{proof}

\subsection*{Proof of Theorem 4}

\begin{proof}

We first quantify the maximal uniform spacing in $(0,1)^p$ as
$$
\Delta_n = \sup_{i\in \{1\ldots n\}} \inf_{j:j\neq i, j\in \{1 \ldots n\}} \|\beta_i-\beta_j\|.
$$

\cite{devroye1982log} showed in one dimension $(0,1)$ the uniform spacing $\Delta^*_n$ has
$$
 \lim\sup (n \Delta^*_n - \log n)/(2\log\log n) =1 \quad a.s.,
$$
which means for $n$ large enough
$$
\Delta^*_n \le \frac{  2 \log\log n+\log n}{n}
$$
As $x_i \sim \text{Uniform}(0,1)^p$ is equivalent to combining $p$ independent $\text{Uniform}(0,1)$'s, by the triangle inequality
$$
\Delta_n \le p  \frac{  2\log\log n+\log n}{n} .
$$

This means a new $\beta^*$ will be within $\Delta_n$ of an existing $\beta_l$. Our next task is equivalent to showing $g(\beta)=\log\tilde \Pi(\beta)$ has a bounded derivative almost everywhere. Rewriting
\be
 &g(\beta) = \log     \sum_{k=1}^{K}\exp h_k(\beta),\\
          &h_k(\beta)=\log w_k   \{T_k(\beta)\}+\log
\Pi
\{T_k(\beta);y \}+\log
        {|\prod_{j=1}^p s_{k,j}|},
       \ee
       and taking derivative with respect to the $j$th sub-coordinate of $\beta$,
denoted by $\beta_{\{j\}}$, its magnitude satisfies
       \be
    \bigg | \frac{\partial g(\beta)}{\partial \beta_{\{j\}}} \bigg
| = &\bigg |\sum_{k=1}^{K}\frac{\exp h_k(\beta)\partial h_k(\beta)/\partial
\beta_{\{j\}}}
        {\sum_{l=1}^{K}\exp h_l(\beta)} \bigg |\\
        \le &\sum_{k=1}^{K}\frac{\exp h_k(\beta)}
        {\sum_{l=1}^{K}\exp h_l(\beta)} \bigg |\frac{\partial
h_k(\beta)}
        {\partial \beta_{\{j\}}} \bigg |\\
        \le & \max_{k\in\{1\ldots K\}} \bigg |\frac{\partial
h_k(\beta)}
        {\partial \beta_{\{j\}}} \bigg |.
        \ee
Examining the derivative yields
      \be
    \bigg| \frac{\partial
h_k(\beta)}
        {\partial \beta_{\{j\}}} \bigg| \le
 \bigg|\frac{\partial\log
w_k
  (\theta)}{
  \partial \theta
  }\vert_{\theta=T_k(\beta)}
+ \frac{\partial \log\ \Pi
( \theta;y) }{\partial \theta}\vert_{\theta=T_k(\beta)}
\bigg|\bigg|\frac{\partial T_k(\beta)}{\partial \beta_{\{j\}}}\bigg|.
      \ee
      Since $w_k(\theta)$ as logistic function is continuous, and $\Pi
( \theta;y)$ is  absolutely continuous,
then first absolute value is finite almost
everywhere, and ${\partial T_k(\beta)}/{\partial \beta_{\{j\}}}=s_{k,j}$

Denote the index that achieves the minimum distance as $l_0=\arg\inf_{l\in\{1\ldots
n\}}\|\beta^* -\beta_i\|$, then
\be
\inf_{l\in\{1\ldots n\}}\|g(\beta^*) - g(\beta_l)\|
\le\|g(\beta^*) - g(\beta_{l_0})\|
=\mathcal{O}( p  \frac{  2\log\log n+\log n}{n}).
\ee
\end{proof}
\newpage

\subsection*{Simulation: High Dimensional Regression using the Shrinkage Prior}

We  experiment with a sparse linear regression problem using the shrinkage prior. As the original horseshoe prior \citep{carvalho2010horseshoe} can be estimated with the fast Gibbs sampler \citep{bhattacharya2016fast}, we focus on a variant called the ``regularized horseshoe''  \citep{piironen2017sparsity}. For the data index $i=1,\ldots,n$ and covariate index $j=1,\ldots,p,$
\be
& y_i\sim N( x_i'b , \sigma^2),\\
& b_j\sim N(0,\tilde\lambda_j^2 \tau^2),
\; \tilde\lambda^2_j=\frac{\tilde c^2 \lambda_j^2}{\tilde c^2+\tau^2\lambda_j^2},
 \;\lambda_j\sim \text{C}^{+}(0,1),\; \\
& \tilde c^{2}\sim \text{Inverse-Gamma}(\xi_1/2,\xi_1\xi_2^{2}/2).
\ee
where $x_i\in \mathbb{R}^p$ is the predictor; $\text{C}^+$ denotes the half-Cauchy distribution. The difference from the original horseshoe prior \citep{carvalho2010horseshoe} is that, as $\lambda_j$ increases, the prior for $b_j$ will approximately follow a normal $N(0, \tilde c^2)$. This property can be useful when one needs to specify a minimum level of regularization to the largest signals. Due to the unique form of $\tilde\lambda^2_j$, Gibbs sampler is no longer suitable, \cite{piironen2017sparsity} used the Hamiltonian Monte Carlo.

To simulate the data, we followed \cite{bhadra2019lasso} and chose a correlated predictor $x_i\sim N(0,\Sigma)$, with $\Sigma_{j,k}=\rho^{|j-k|}$. We used a moderately high correlation $\rho=0.7$, as it posed some challenge for the posterior computation, while still retained identifiabiltiy for $b$ (see \cite{castillo2015bayesian} on the mutual coherenece condition). To induce a $p\gg n$ setting, we used $p=1,000$ and $n=200$. We specified the ground-truth $b_j$'s as $(b_1,b_2,\ldots,b_5)=(5,3.5,5.5,5,4.5)$ and used $b_j=0$ for $j=6\ldots 1000$, based on which we simulated the outcome  $y_i\sim N( x_i'b , \sigma^2)$ with $\sigma^2=0.1$.

To choose the hyper-priors and hyper-parameters,  for both $\sigma^2$ and $\tau^2$, we used the informative prior $\text{Exp}(0.01)$ to favor a low noise and a small global scale (to induce a strong shrinakge); for $\tilde c^2$, we set $\xi_1=5$, $\xi_2=10$, as suggested by  \cite{piironen2017sparsity}.

We compare the computing performances between the Metropolis-adjusted Langevin algorithm (MALA), the Hamiltonian Monte Carlo with the No-U-Turn Sampler (HMC-NUTS) and the Transport Monte Carlo (TMC) (The Riemannian manifold Hamiltonian Monte Carlo is not suitable in this case due to the unscalability of the large Fisher information matrix).

For the MALA and HMC-NUTS algorithms, we used the ``hamiltorch'' python package \citep{cobb2020scaling} to tune the step size automatically.
Due to the high dimensionality of the parameters, both algorithms require some additional tuning to yield satisfactory mixing --- most importantly, the working parameter $M$ known as the ``mass'' in the Hamiltonian needs to adapt to the width of the high posterior density region for each model parameter. For example, the width for $b_1$ (non-zero signal) would be much larger than $b_6$ (concentrated at zero). In order to obtain a good tuning, we first initialized the Markov chain at the maximum-a-posteriori $\hat\theta$ (using the ADAM optimizer), and then set the mass to the diagonal matrix $M=\diag{\tilde M_{j}}$  with $\tilde M_{j}=|\hat\theta_j|^{-2}$. This resulted in a much better mixing compared to using simple identity matrix for $M$.

\begin{figure}[H]
     \centering
                     \begin{subfigure}[t]{0.32\textwidth}
         \centering
         \includegraphics[ width=\textwidth]{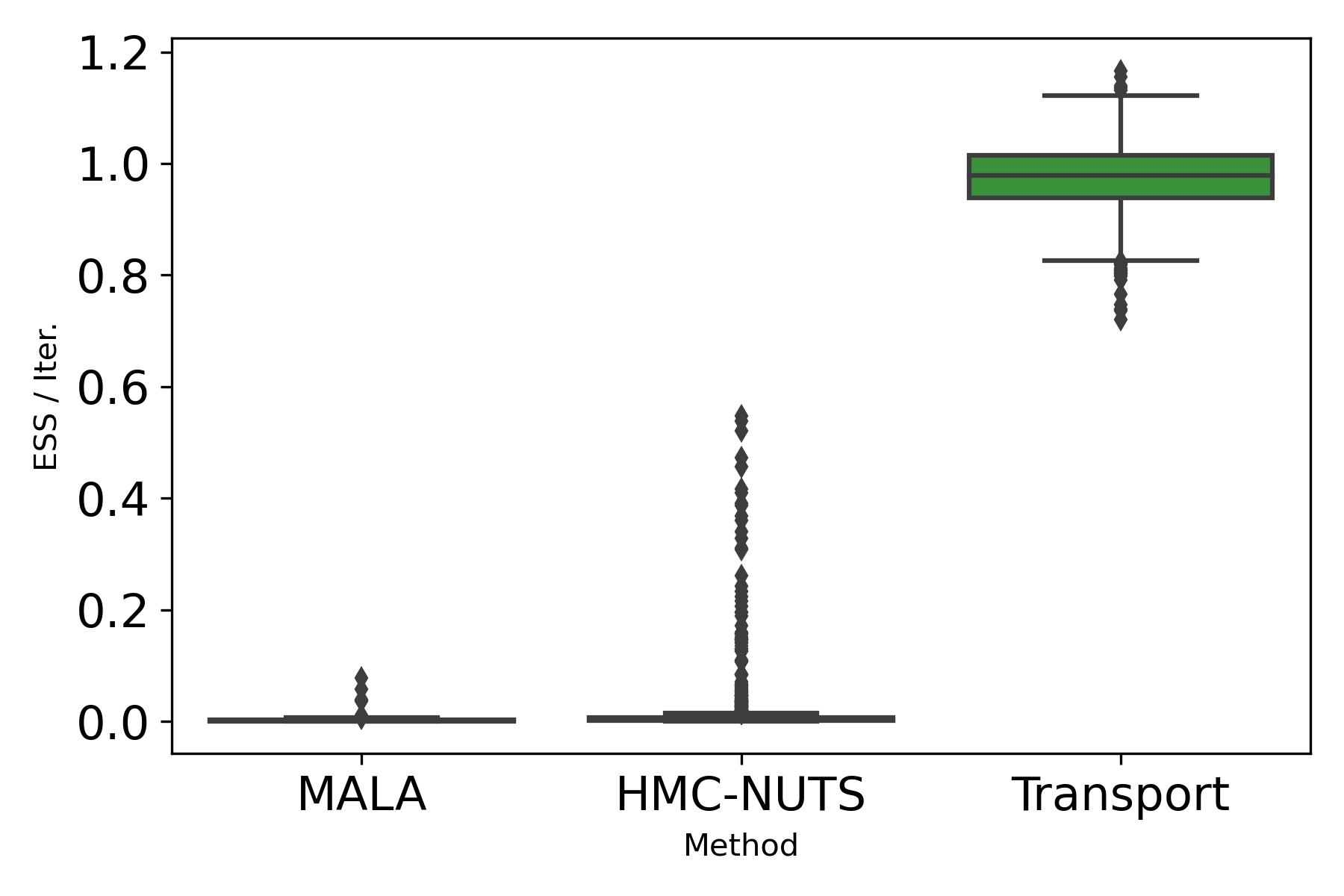}
         \caption{Boxplots of the effective sample size (ESS) per iteration.}
     \end{subfigure}\;
           \begin{subfigure}[t]{0.6\textwidth}
         \centering
         \includegraphics[ width=\textwidth]{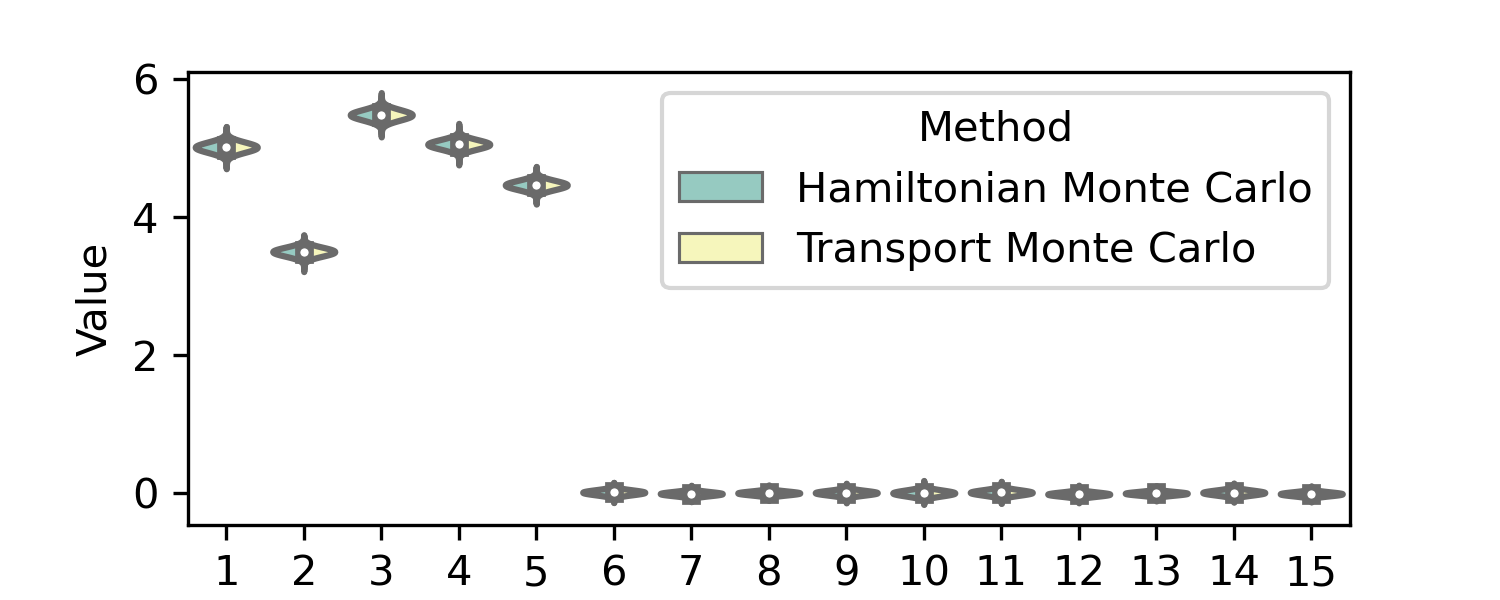}
         \caption{The samples of $b_j$'s produced by the Transport Monte Carlo are almost indistinguishable from the ones produced by the Hamiltonian Monte Carlo.}
     \end{subfigure}
           \begin{subfigure}[t]{1\textwidth}
         \centering
         \includegraphics[ width=.8\textwidth, height = 1.2in]{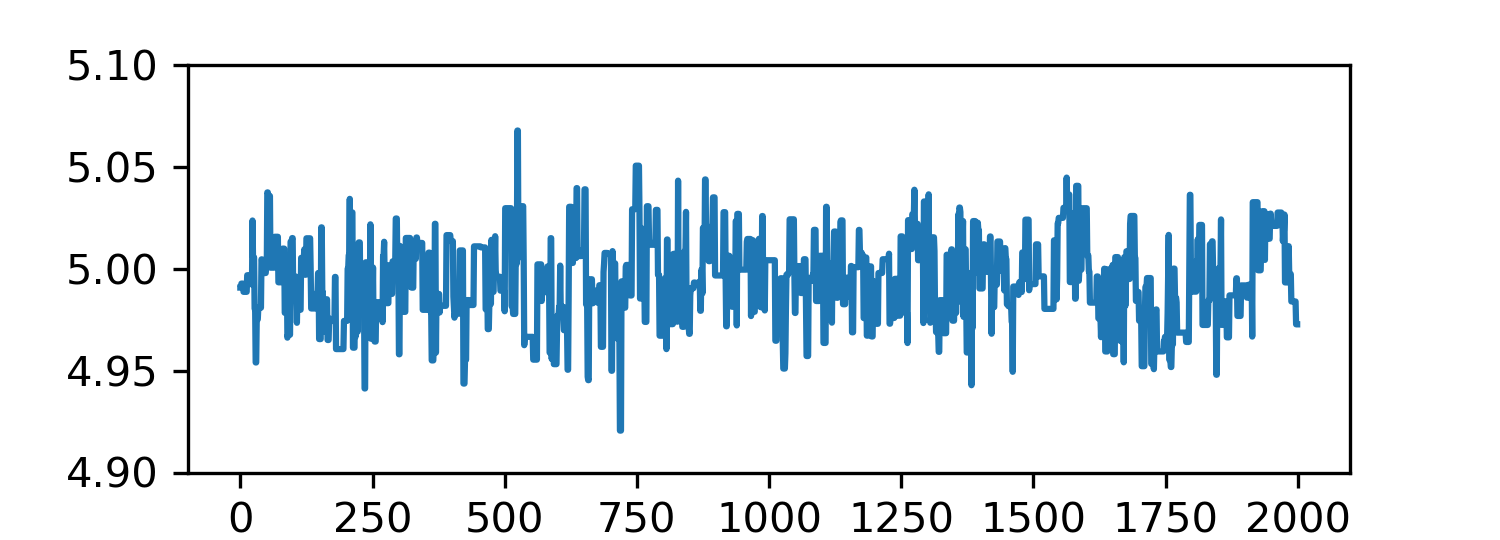}
         \caption{Traceplot of $b_1$ produced by the Metropolis-adjusted Langevin algorithm (MALA).}
     \end{subfigure}
           \begin{subfigure}[t]{1\textwidth}
         \centering
         \includegraphics[ width=.8\textwidth, height = 1.2in]{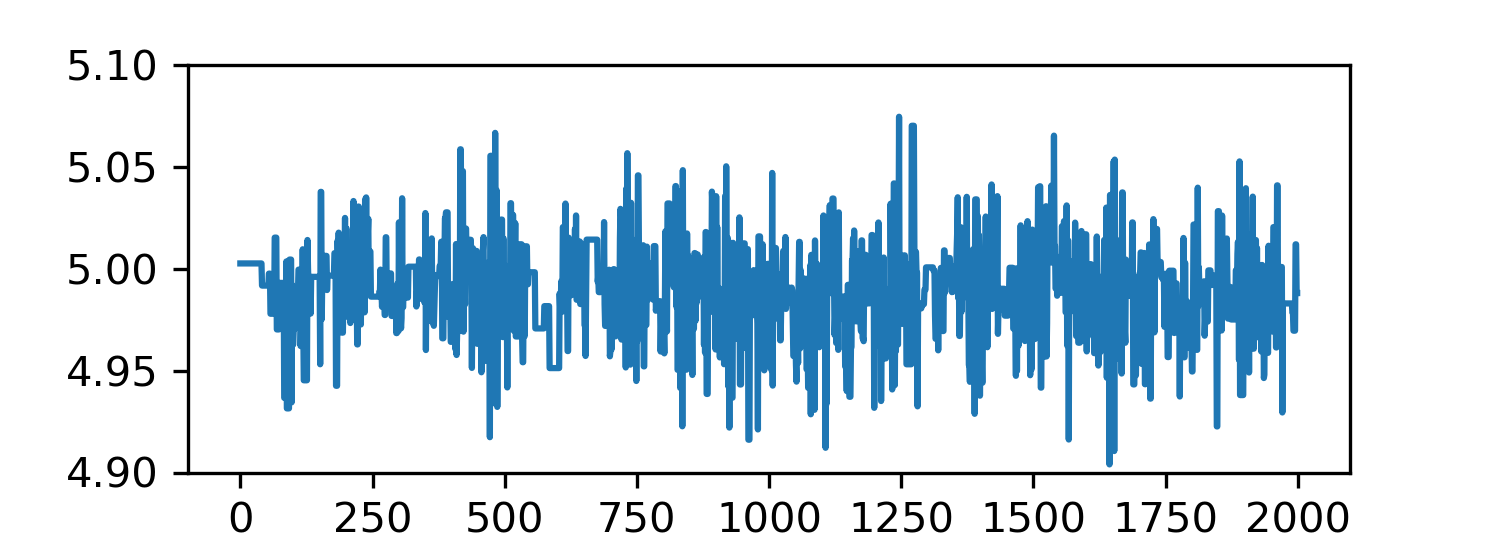}
         \caption{Traceplot of $b_1$ produced by the Hamiltonian Monte Carlo (No-U-Turn Sampler, HMC-NUTS).}
     \end{subfigure}
           \begin{subfigure}[t]{1\textwidth}
         \centering
         \includegraphics[ width=.8\textwidth, height = 1.2in]{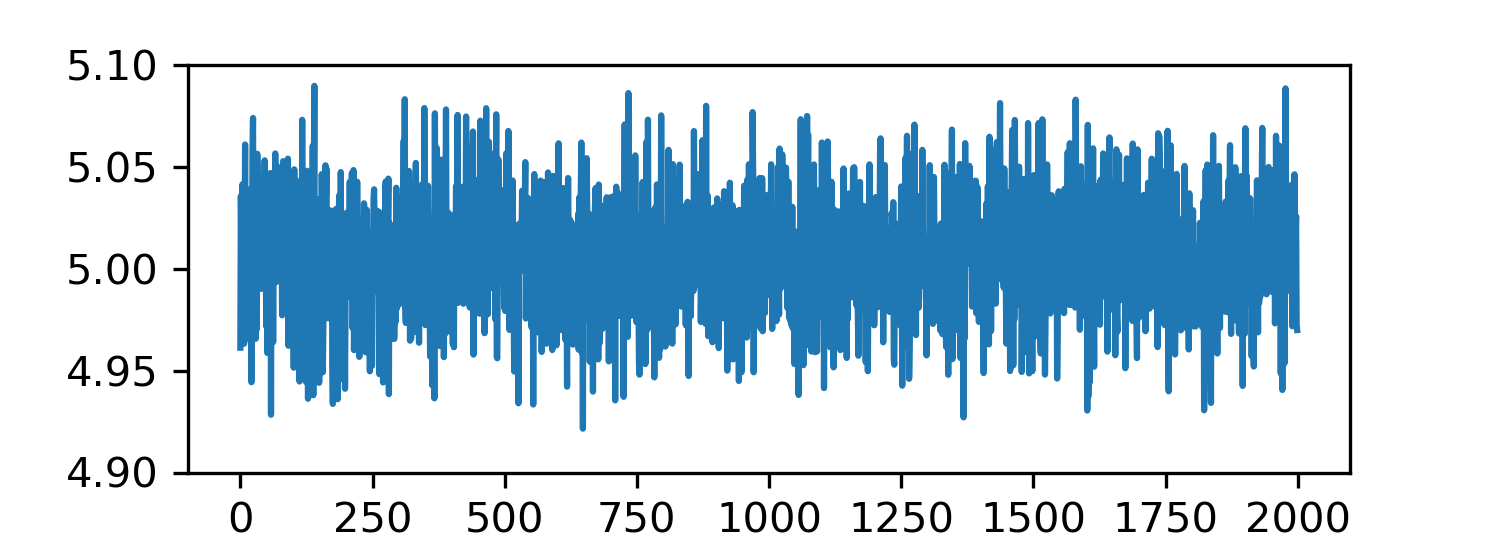}
         \caption{Traceplot of $b_1$ produced by the Transport Monte Carlo.}
     \end{subfigure} 
        \caption{Estimation of sparse regression signals using the regularized horseshoe prior: the posterior samples of $b_j$'s produced by the Transport Monte Carlo are almost indistinguishable from the ones produced by the Hamiltonian Monte Carlo. \label{fig:sparse_regression}
        }
\end{figure}

We show the traceplots of the samples in Figure~\ref{fig:sparse_regression} (panel c-e). Both the MALA and HMC-NUTS algorithms showed high autocorrelations in the Markov chains:  for most of  $b_j$'s, the effective sample size (ESS) per sampling iteration was less than $10\%$ (panel a) [we also experimented these two algorithms using an identity mass matrix (as the default option in most of the HMC software), and the ESSs per iteration got worse and were less than $0.1\%$ in both]. Between the two, the HMC-NUTS algorithm showed a slightly higher ESS; therefore, we ran the HMC-NUTS for an extended period of 2,000,000 iterations, and used thinning at every $100$th sample.  This process took about 11 hours on a 12-core Intel computer.

As shown in the violin plot (panel b), the samples collected from the TMC were almost identical in distribution to the ones from HMC-NUTS (with thinning). On the other hand, due to the independence, the ESS's per iteration were close to $1$ for almost all the samples. This process took about 2 minutes on an NVIDIA GTX 1080TI GPU.

\thispagestyle{empty}

\subsection*{Benchmark: Assessing Approximation Error}

To assess the approximation error, we compare with three alternative approximations: (i) the deterministic transport using normalizing flow neural network [RealNVP  with $6$ hidden layer, with each containing $256$ latent dimensions \citep{dinh2016density}]; (ii) the variational approximation using normal mixture, with each component having a diagonal covariance $\sum  _{k}v_{k}^* N_k(\mu_{k}, \diag {\sigma^2_{k1}, \sigma^2_{k2}} )$; (iii) the variational approximation using simple uniform mixture   $\sum  v_{k}^* {U}_k(.,.)$, with constant $\sum_k v_{k}^*=1$.

We first revisit the bivariate normal mixture example as in the main text. Since the target density $\Pi(\theta)$ is fully known including the normalizing constant $z(y)$, we can compute   \( \log \{\tilde\Pi(\theta)  / \Pi(\theta; y) \} \) directly and compare  the mean log-ratio (empirical KL divergence) against the ideal $\mathbb{E}_{\theta\sim \tilde\Pi(\theta)} \log \{\tilde\Pi(\theta)/ \Pi(\theta; y) \}=0$.

\begin{figure}[H]
      \begin{subfigure}[t]{.40\textwidth}
         \includegraphics[ width=\textwidth,height= 1.6in]{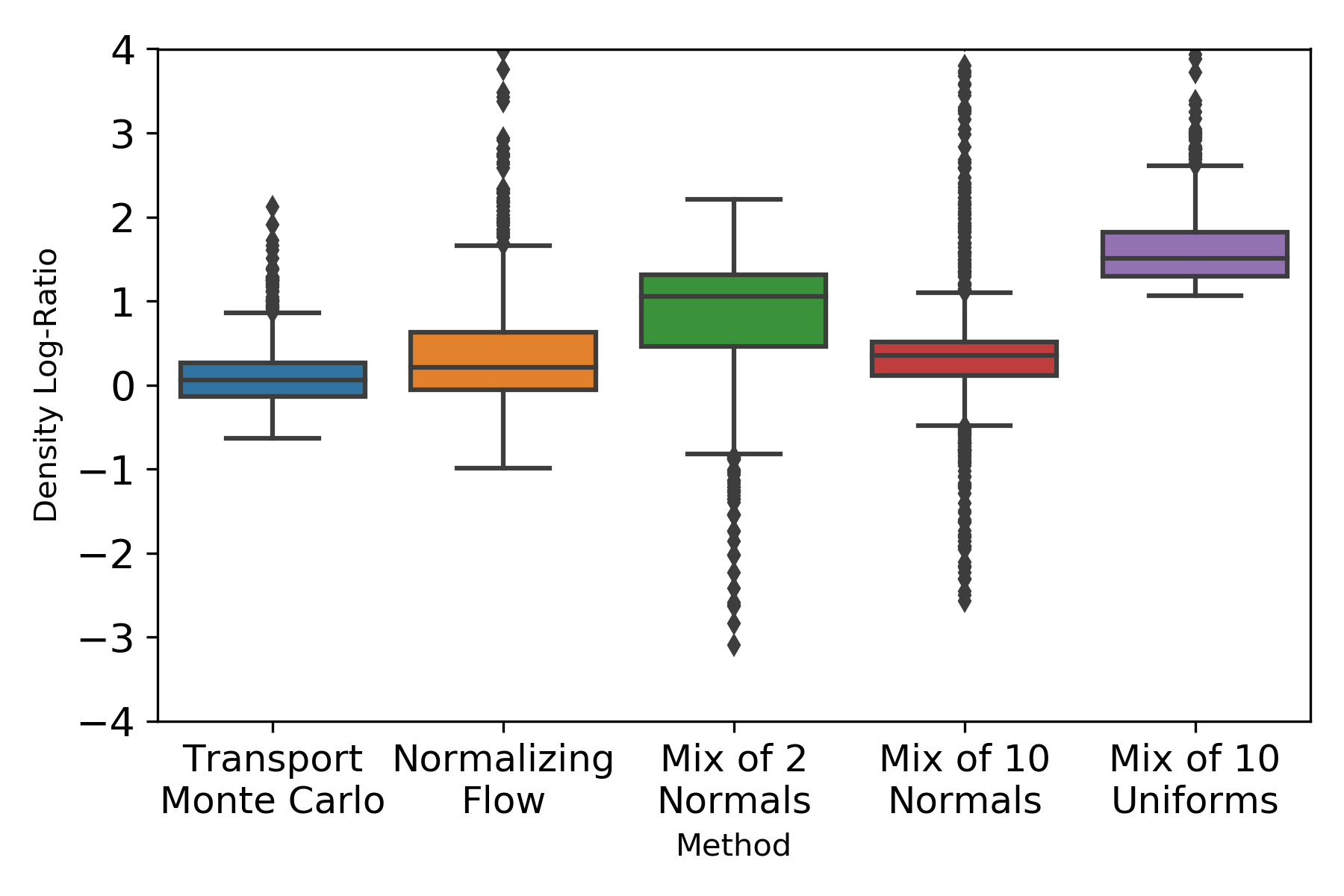}
         \caption{Approximation error: log-density ratio between approximate and true target $\log\{\tilde\Pi(\theta)/ \Pi(\theta)\}$}
     \end{subfigure}
     \quad
           \begin{subfigure}[t]{.28\textwidth}
         \centering
         \includegraphics[width=1.6in, height= 1.7in]{MixoNormal_target}
         \caption{Transport Monte Carlo}
     \end{subfigure}
      \begin{subfigure}[t]{.28\textwidth}
         \centering
         \includegraphics[ width=1.6in, height= 1.7in]{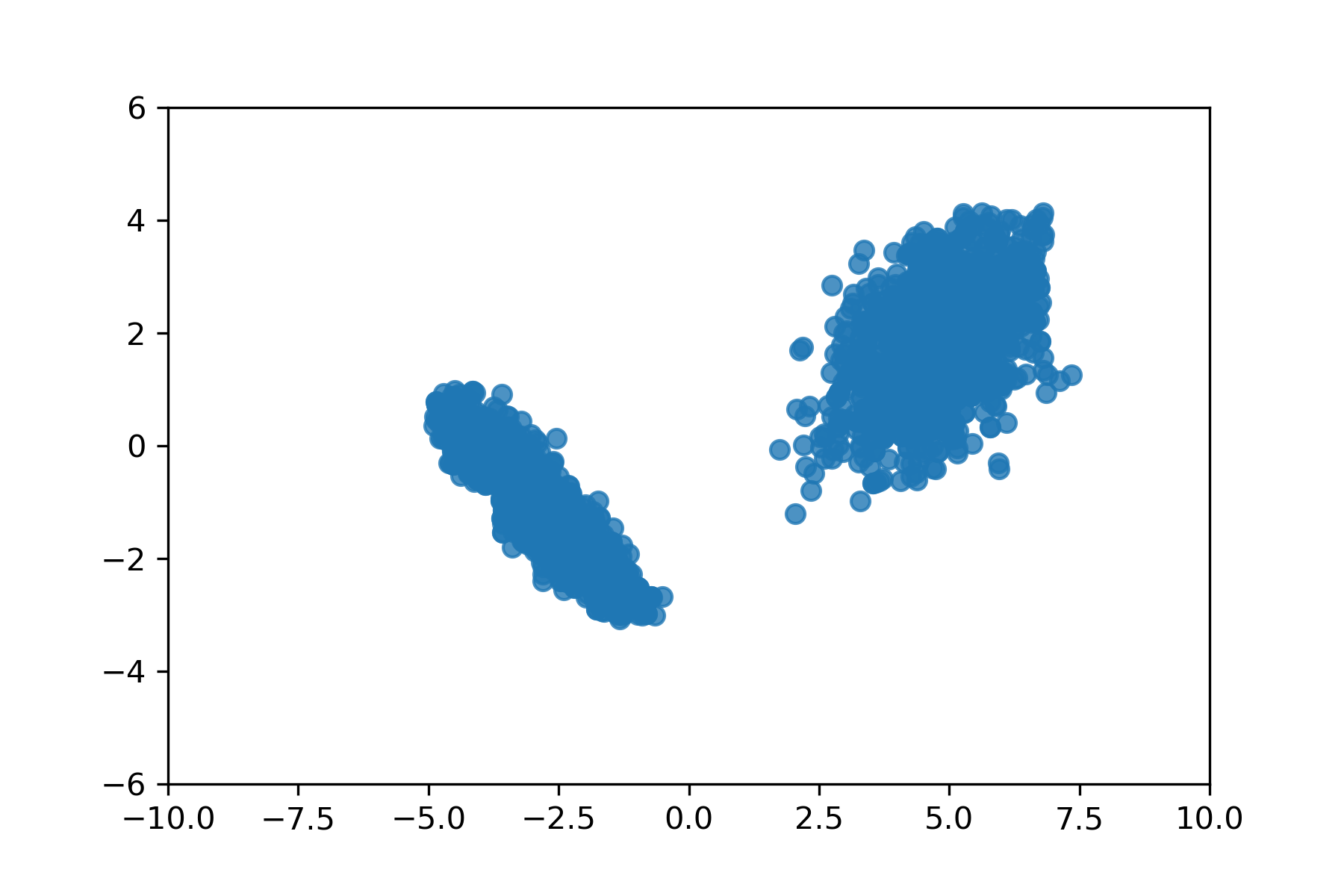}
         \caption{Normalizing flow.}
     \end{subfigure}\\
     \begin{subfigure}[t]{.28\textwidth}
         \centering
          \includegraphics[ width=1.6in, height= 1.7in]{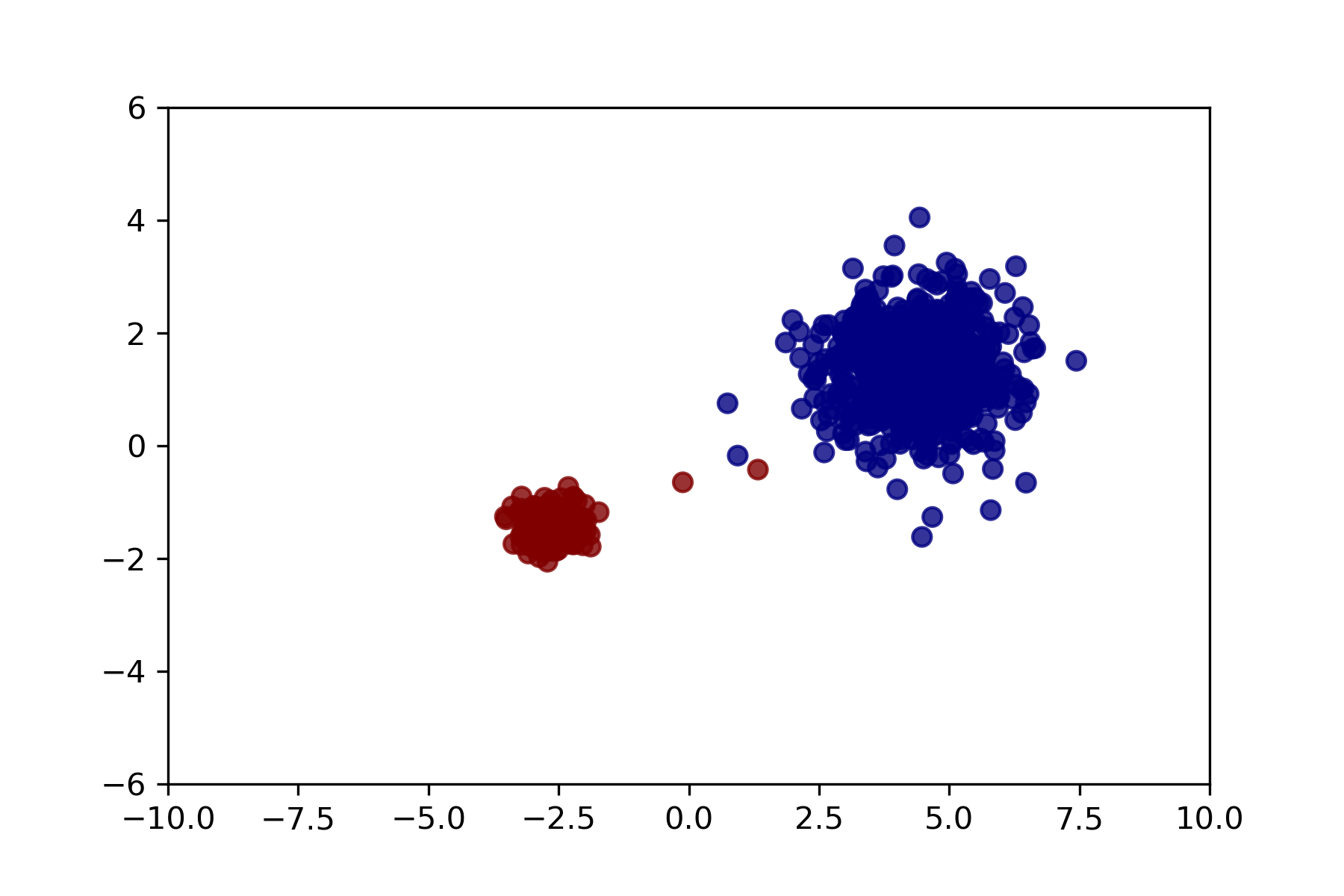}
\caption{Variational approximation using 2-component normal mixture with diagonal covariance.}
     \end{subfigure}
          \hfill
 \begin{subfigure}[t]{.28\textwidth}
         \centering
         \includegraphics[ width=1.6in, height= 1.7in]{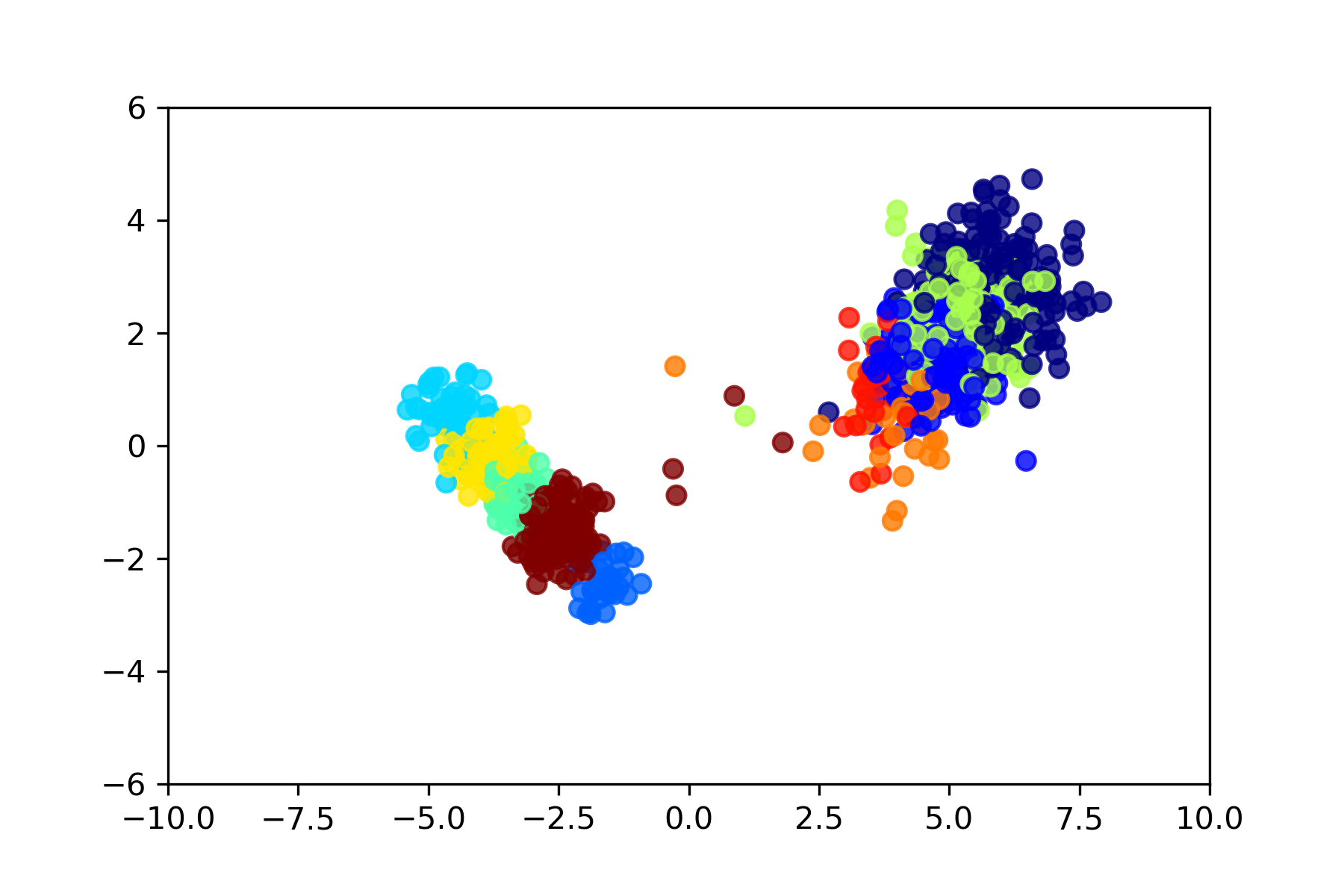}
         \caption{Variational approximation using 10-component normal mixture with diagonal covariance.}
     \end{subfigure}
     \hfill
     \begin{subfigure}[t]{.28\textwidth}
         \centering
          \includegraphics[ width=1.6in, height= 1.7in]{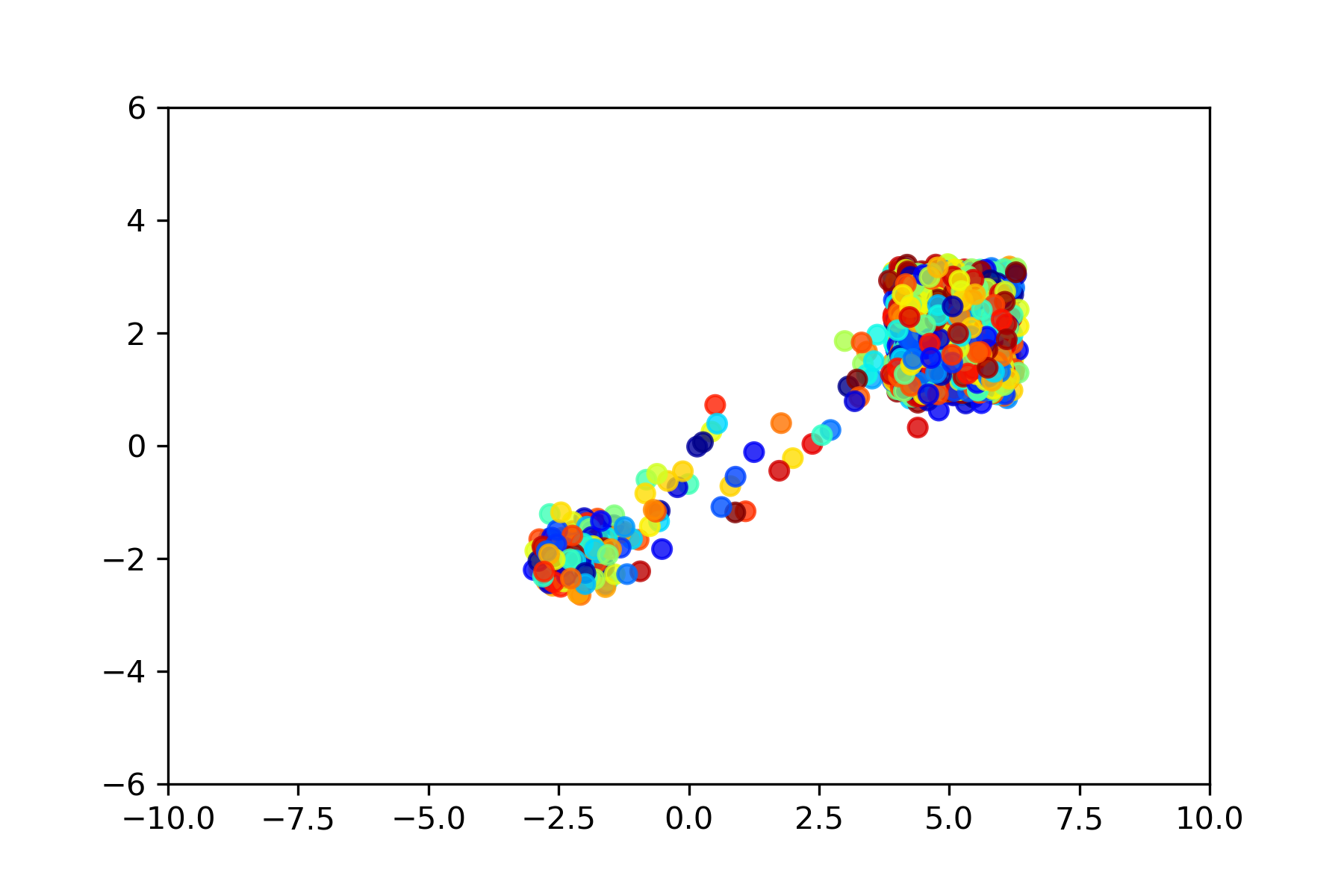}
\caption{Variational approximation using 10-component uniform mixture.}
     \end{subfigure}
     \caption{Approximation of a normal mixture using various methods. \label{fig:approximation_error}}
\end{figure}

Figure~\ref{fig:approximation_error}(a) plots the the log-density ratio $\log\{\tilde\Pi(\theta)/ \Pi(\theta; y)\}$ based on the samples collected using various method. The TMC (panel b) showed very high accuracy, with the mean log-ratio $0.10$. The deterministic transport using normalizing flow (panel c) also showed high accuracy (mean log-ratio $0.22$), although it used a large number of working parameters in the transform (total $804,900$, versus $80$  used in the TMC). On the other hand, for the variational approximations, due to the diagonal covariance, the one using a 2-component normal mixture (panel d) gave a poor result (mean log-ratio $0.78$); and increasing the number of components to $10$  (panel e) reduced it to $0.60$. The simple mixture of 10-component uniforms had the worst result  (panel f) with the mean log-ratio  $1.97$ --- clearly, the dramatic difference between the simple uniform mixture and the TMC was due to the varying mixture weight $v_k(\beta)$ in the latter.

We next  sample from a more challenging density that contains multiple local maxima:
\be
\Pi(\theta ; y) = & z(\lambda)^{-1}\exp\{  \lambda H(\theta_1,\theta_2)\},\\
 H( \theta_1, \theta_2) = &\{\theta_1 \sin (20 \theta_2)+y \sin (20 \theta_1)\}^{2} \cosh \{\sin (10 \theta_1) \theta_1\}\ \\
& +\{\theta_1 \cos (10 \theta_2)-\theta_2 \sin (10 \theta_1)\}^{2} \cosh\{\cos (20 \theta_2) \theta_2\}.
\ee
with support in $(-1.1,1.1)^2$. 

\begin{figure}[H]
      \begin{subfigure}[t]{.35\textwidth}
         \includegraphics[ width=\textwidth,height= 1.9in]{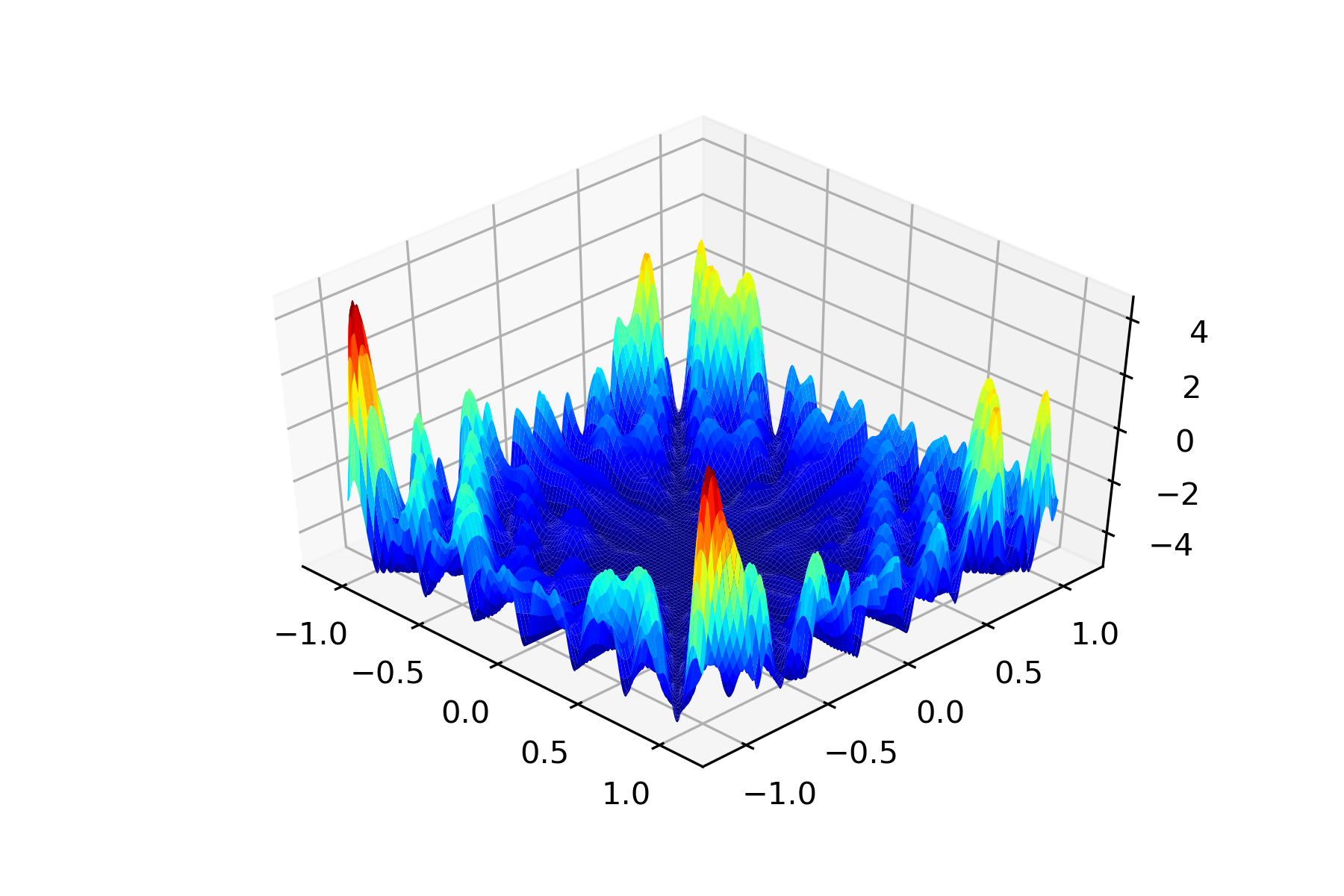}
         \caption{Target log-density $\log \Pi(\theta;y)$.}
     \end{subfigure}
 \hfill
       \begin{subfigure}[t]{.30\textwidth}
        \centering
        \includegraphics[width=\textwidth, height= 1.7in]{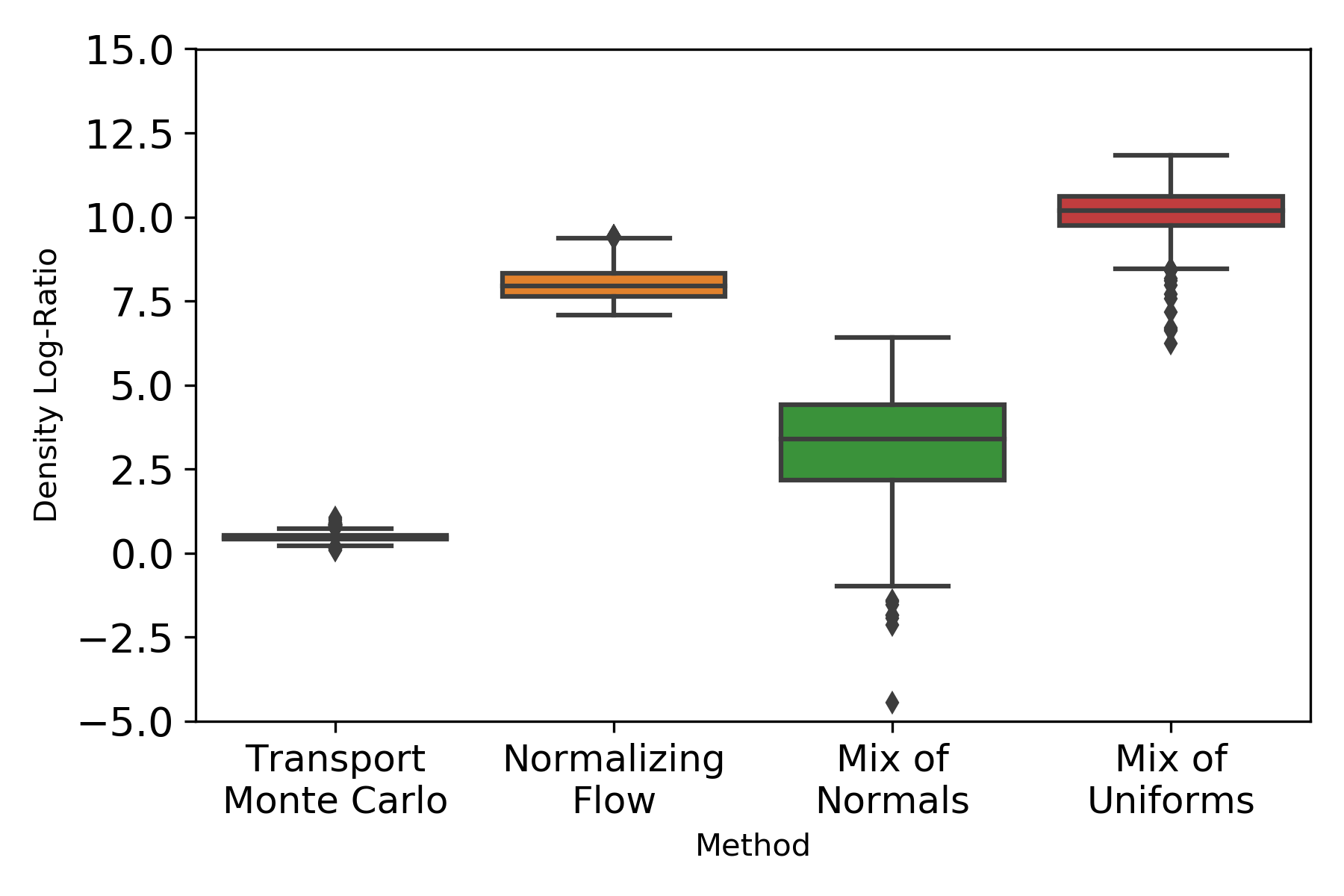}
        \caption{Approximation error: $\log [\tilde\Pi(\theta)/ \Pi(\theta; y) ]$ between approximate and true target.}
 \end{subfigure}
\hfill
           \begin{subfigure}[t]{.30\textwidth}
         \centering
         \includegraphics[width=\textwidth, height= 1.85in]{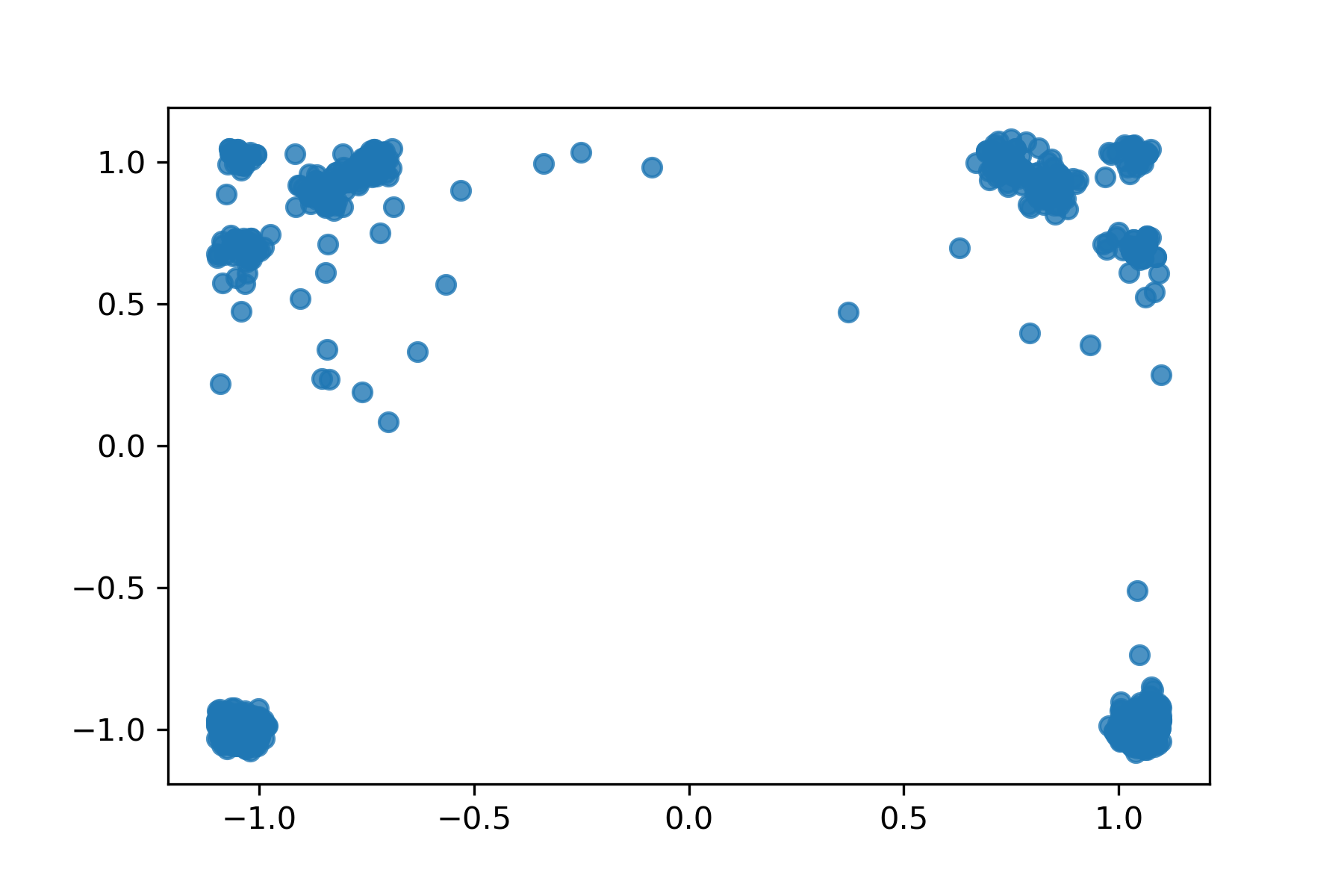}
         \caption{Samples from Transport Monte Carlo}
     \end{subfigure}
     \caption{Appproximated samples from a density that contains multiple local maxima. \label{fig:multi_modes}}
\end{figure}

This example was originally proposed by \cite{casella1999monte} and later  modified by \cite{liang2005generalized}. We plot the $H(\theta_1,\theta_2)$ function in Figure~\ref{fig:multi_modes}(a). And we chose $\lambda=1.2$, so that the high probability region is dominated by  $8$ major peaks, located near the four corners of the support. Using numerical integration, we have the normalizing constant $z(\lambda)\approx 173.1$. We used $K=100$ in the TMC and $100$ components in all the mixture-based methods.

Figure~\ref{fig:multi_modes}(b) shows the log-density ratios. The Transport Monte Carlo showed a very low approximation error with the mean log-ratio $0.47$, and the generated samples indeed recovered the $8$ density peaks (panel c). On the other hand, since the target distribution was no longer normal, the variational inference with normal mixture performed much worse this time, with the mean log-ratio $3.18$; the one with the uniform mixture had a mean log-ratio $10.01$.  The normalizing flow neural network had a surprisingly poor mean log-ratio $8.00$, despite the large number of working parameters it used --- we found out that all of the produced samples were trapped near one local density maximum.

In both the simulated examples above, it is worth noting that TMC also had the smallest standard deviation in the log-ratios. This can be particularly advantageous if we use the generated samples in the independence Hastings algorithm. In our experiments, The acceptance rates were $87\%$ in the first and  $93\%$ in the second example.

\subsection*{Comparison with Normalizing Flow Neural Networks}
As discussed in the introduction, the normalizing flow neural networks are a popular class of transport-based methods. They have demonstrated very good empirical performance, especially when the target density is log-concave.

On the other hand, the normalizing flow is known to have difficulties in handling a density with multiple local maxima. To demonstrate this, we experiment with 
a case of sampling from a $25$-modal distribution:
\[
\theta\sim \sum_{l=1}^{25} 1/25 \; \text{N}(\tilde\mu_l, 0.1^2),
\]
where $\tilde\mu_l$ is from a two dimensional lattice ranging from $(-2,2)$ to $(2,2)$. We used a small variance $0.1^2$, so that the modes were well separated.

\begin{figure}[H]
      \begin{subfigure}[t]{.32\textwidth}
         \centering
         \includegraphics[width=\textwidth]{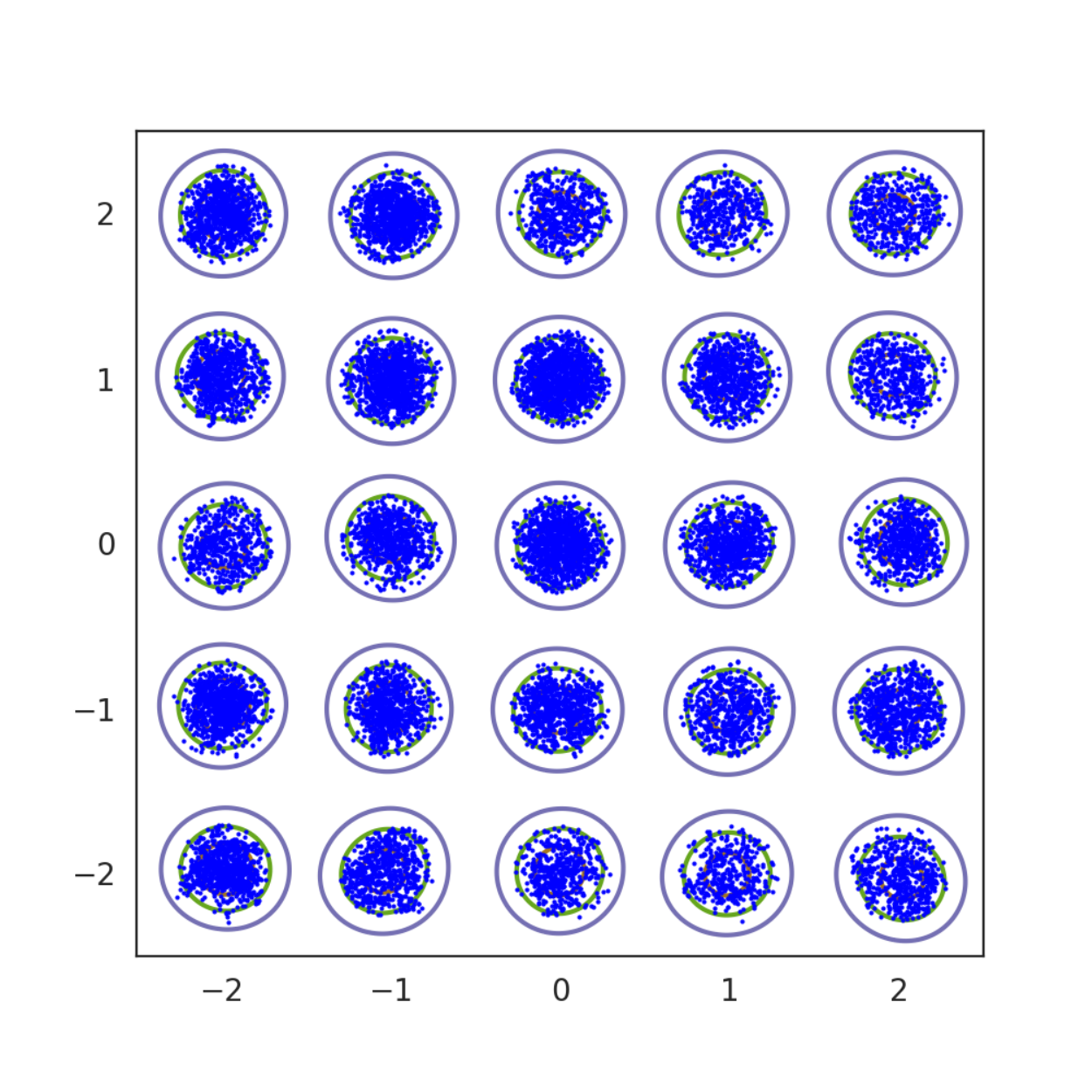}
       \caption{Samples from the Transport Monte Carlo.}
     \end{subfigure}
       \centering
      \begin{subfigure}[t]{.32\textwidth}
         \centering
         \includegraphics[width=\textwidth]{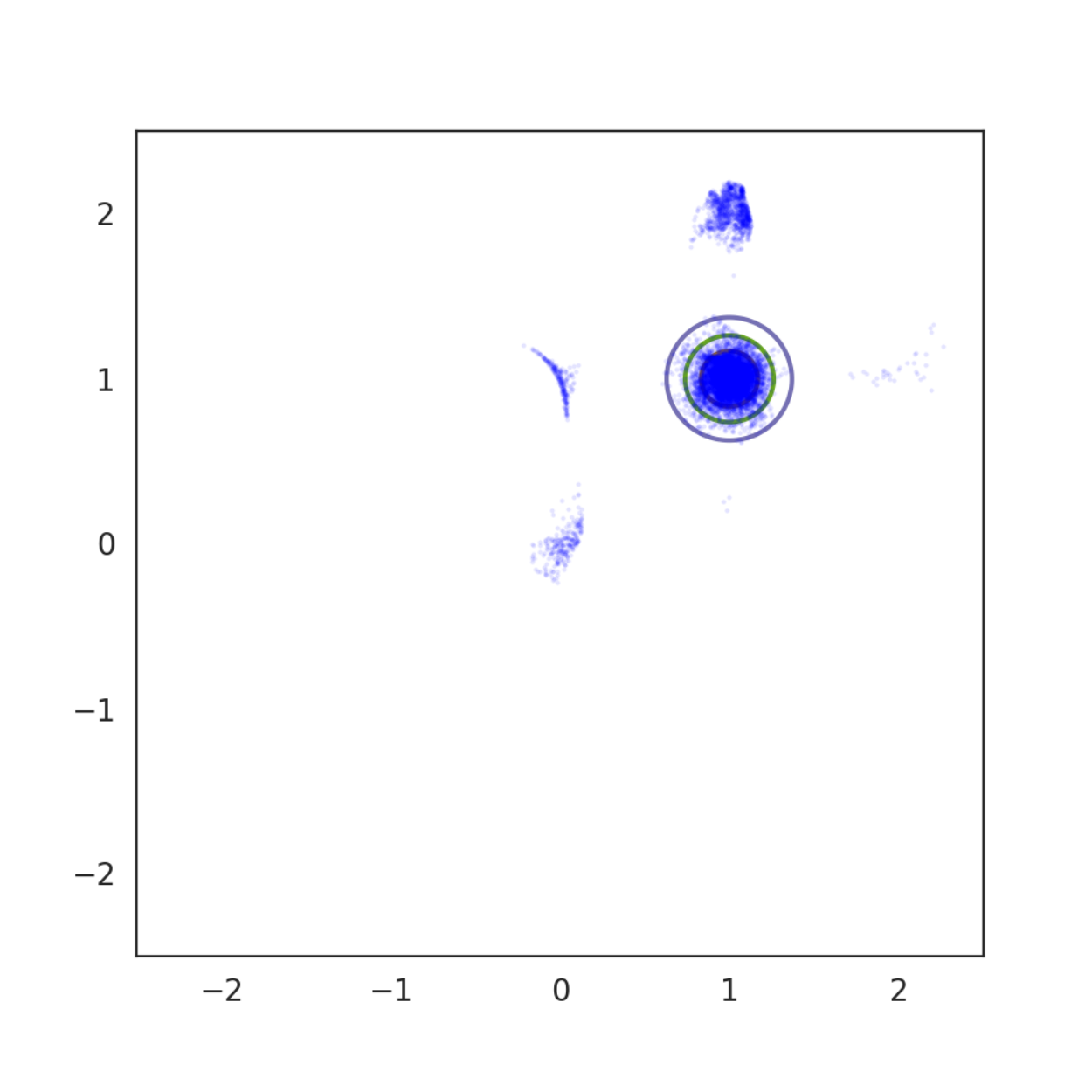}
         \caption{Samples from the RealNVP.}
     \end{subfigure}
      \begin{subfigure}[t]{.32\textwidth}
         \centering
         \includegraphics[height=2in,width=2in]{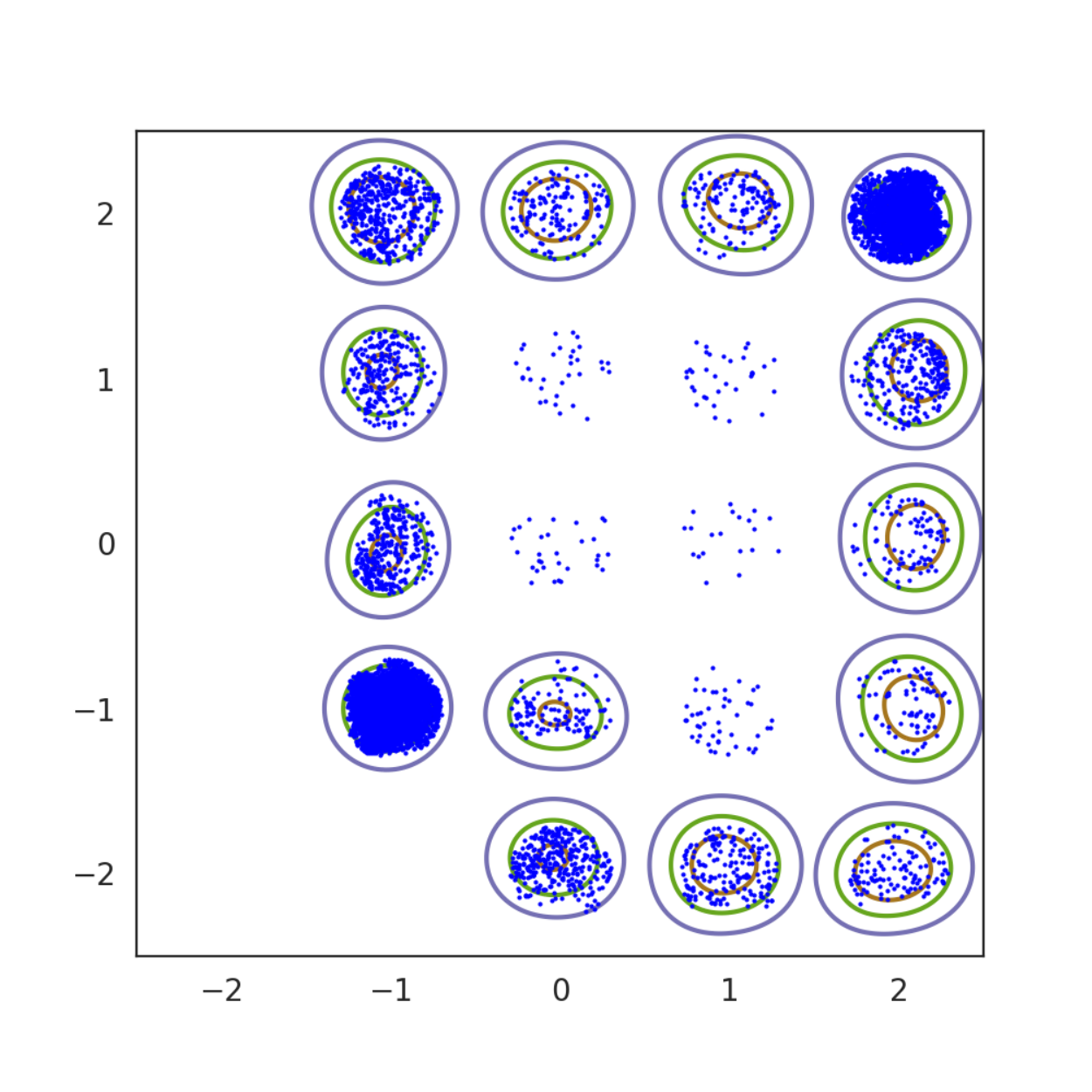}
         \caption{Samples from the Glow.}
     \end{subfigure}
     \hfill
      \begin{subfigure}[t]{.32\textwidth}
         \centering
         \includegraphics[height=2in,width=2in]{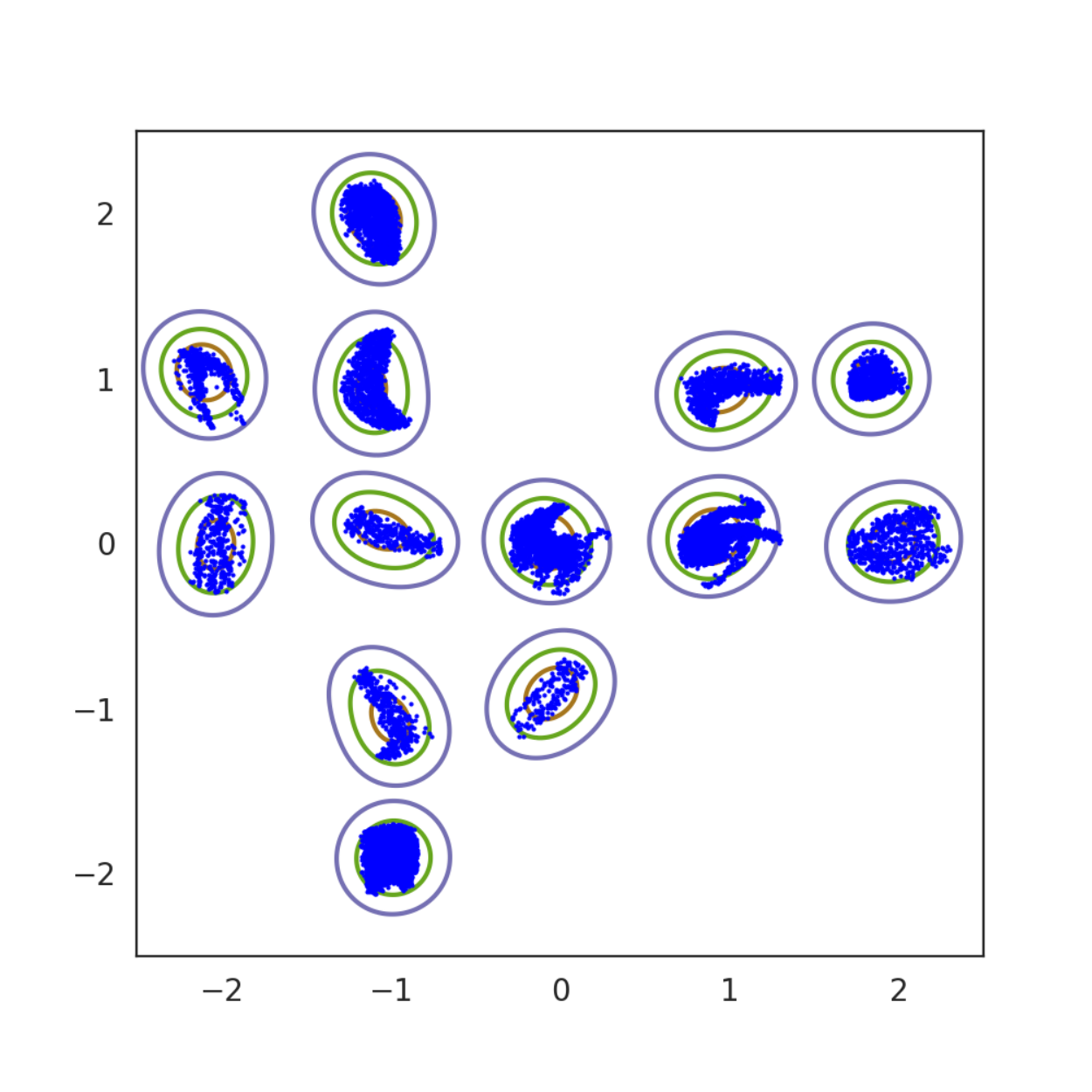}
         \caption{Samples from the Masked Autoregressive Flow.}
     \end{subfigure}
      \begin{subfigure}[t]{.32\textwidth}
         \centering
         \includegraphics[height=2in,width=2in]{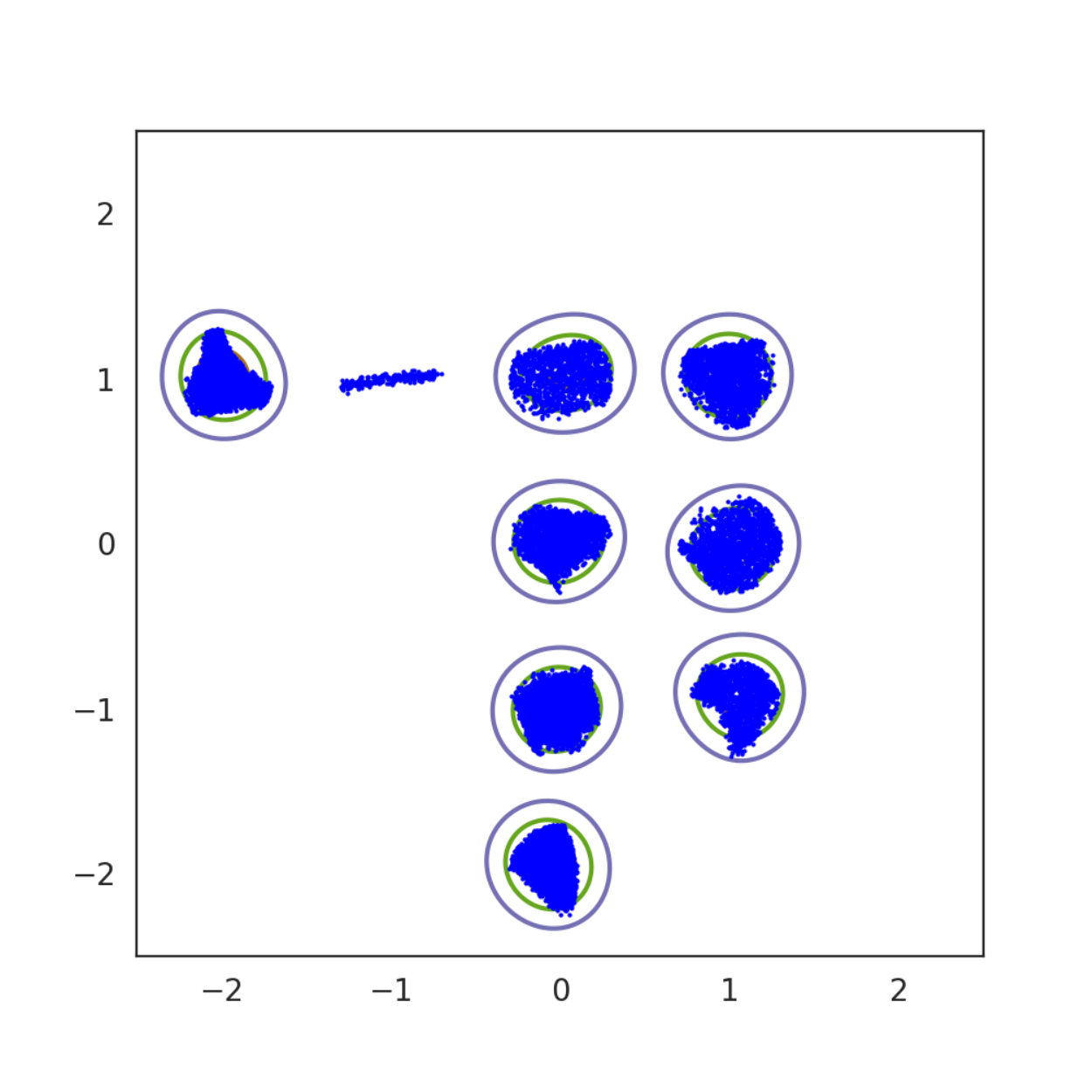}
         \caption{Samples from the Inverse Autoregressive Flow.}
     \end{subfigure}
          \hfill
      \begin{subfigure}[t]{.32\textwidth}
         \centering
         \includegraphics[height=2in,width=2in]{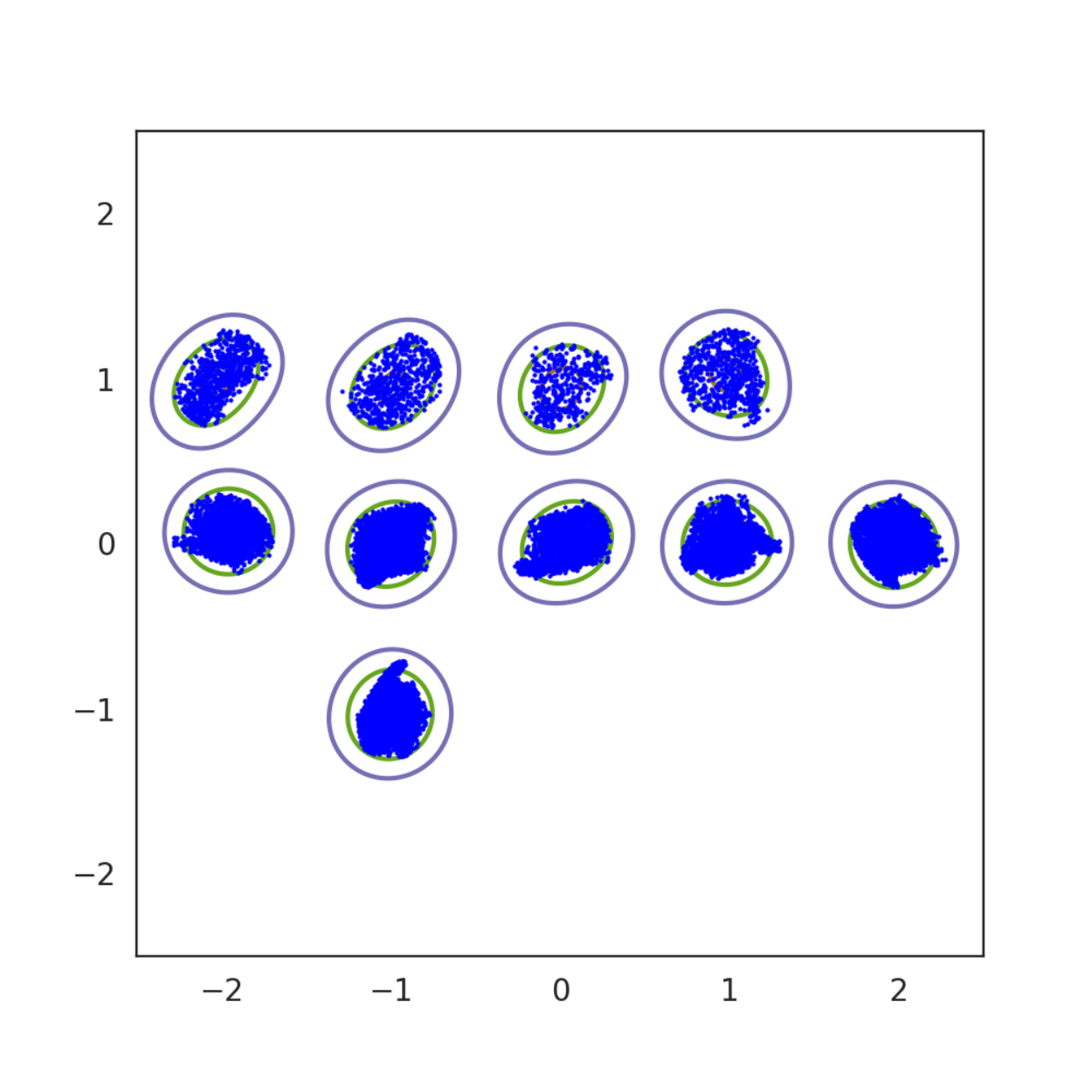}
         \caption{Samples from the Masked Autoregressive Flow/Glow.}
     \end{subfigure}
         \caption{Sampling from a distribution with $25$ modes. Due to the high complexity, the normalizing flow networks fail to discover all the modes (panels b-f). Using multiple maps, the Transport Monte Carlo can explore larger high posterior probability region (panel a).
         }
\end{figure}
 
As shown, using the RealNVP normalizing flow \citep{dinh2016density} (with $5$ layers, each with $256$ hidden units) resulted in a severe underestimation of the modes. Empirically, we found almost no difference when doubling the depth and/or width. We also experimented with other normalizing flow neural networks \citep{kingma2016improved,papamakarios2017masked,kingma2018glow}. Although they improved the performance; however, none of them recovered all 25 modes.

In comparison, due to the use of multiple maps, the TMC is much less sensitive to this issue, and discovered all the modes in this case. As an alternative, one could use the normalizing flow as the mixture component transform in the TMC framework. We could not experiment with this extension, since each normalizing flow involved about $800,000$ to $1,200,000$ working parameters, which exceeded our memory capacity at $K\ge 3$. 
Although at the larger computing system, we can expect to see an improved performance.

\subsection*{Example of Diagnostic Plot on $K$}

\begin{figure}[H]
         \centering
         \includegraphics[width=.5\textwidth]{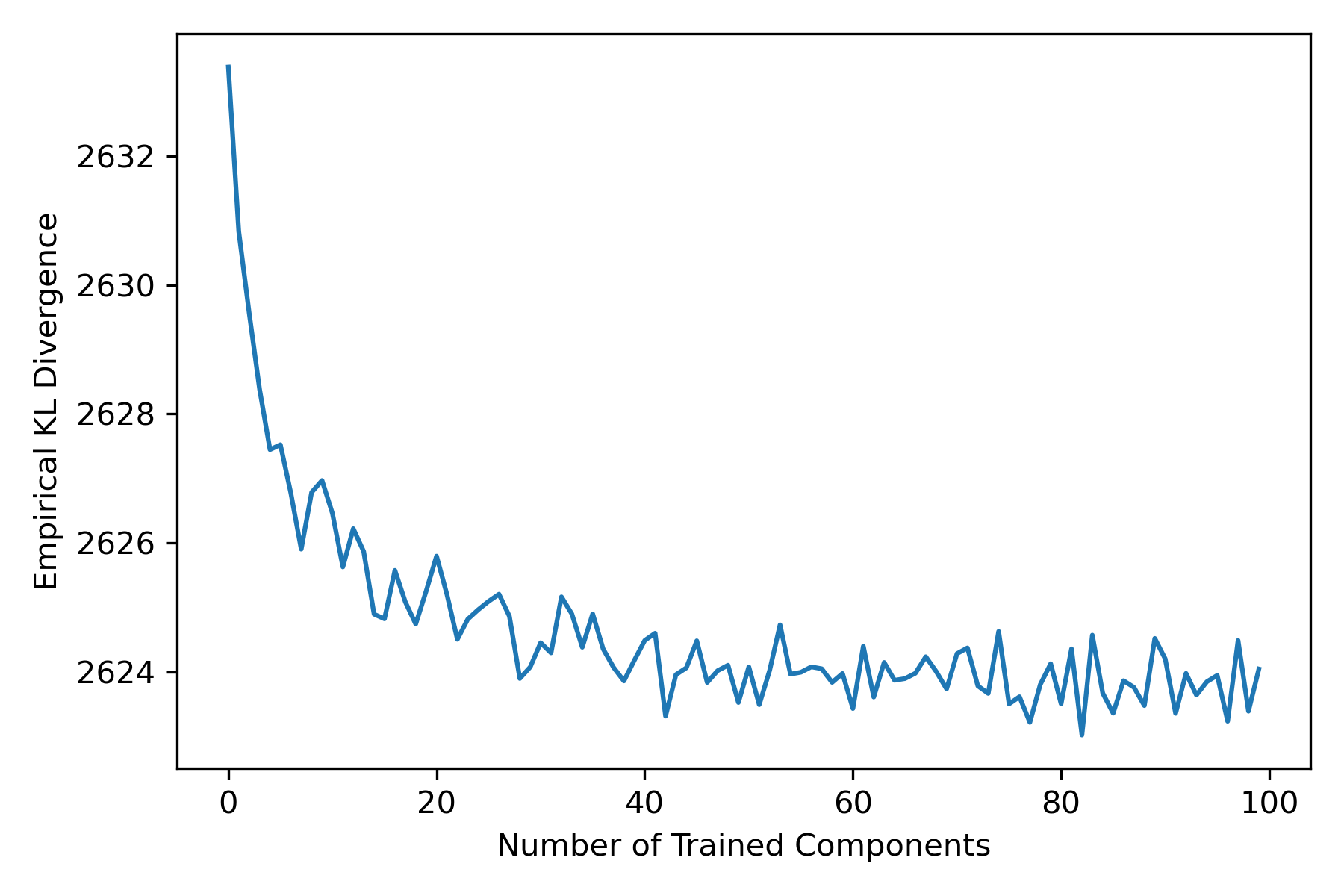}
         \caption{The minimized empirical KL divergence (omitting constant) until the $k$th component is optimized, collected from the high dimensional regression using the shrinkage prior. The flattening of the curve suggests the specified $K=100$ is sufficient.
         }
\end{figure}

\bibliographystyle{chicago}
\bibliography{reference}

\begin{thebibliography}{}

\bibitem[\protect\citeauthoryear{Ambrosio, Gigli, and Savar{\'e}}{Ambrosio
  et~al.}{2008}]{ambrosio2008gradient}
Ambrosio, L., N.~Gigli, and G.~Savar{\'e} (2008).
\newblock {\em {Gradient Flows: In Metric Spaces and in the Space of
  Probability Measures}}.
\newblock Springer Science \& Business Media.

\bibitem[\protect\citeauthoryear{Beaumont, Cornuet, Marin, and Robert}{Beaumont
  et~al.}{2009}]{beaumont2009adaptive}
Beaumont, M.~A., J.-M. Cornuet, J.-M. Marin, and C.~P. Robert (2009).
\newblock {Adaptive Approximate Bayesian Computation}.
\newblock {\em Biometrika\/}~{\em 96\/}(4), 983--990.

\bibitem[\protect\citeauthoryear{Bhadra, Datta, Polson, and Willard}{Bhadra
  et~al.}{2019}]{bhadra2019lasso}
Bhadra, A., J.~Datta, N.~G. Polson, and B.~Willard (2019).
\newblock {Lasso Meets Horseshoe: A Survey}.
\newblock {\em Statistical Science\/}~{\em 34\/}(3), 405--427.

\bibitem[\protect\citeauthoryear{Bhattacharya, Chakraborty, and
  Mallick}{Bhattacharya et~al.}{2016}]{bhattacharya2016fast}
Bhattacharya, A., A.~Chakraborty, and B.~K. Mallick (2016).
\newblock {Fast Sampling With Gaussian Scale Mixture Priors in High-Dimensional
  Regression}.
\newblock {\em Biometrika\/}, 985--991.

\bibitem[\protect\citeauthoryear{Bhattacharya, Pati, Pillai, and
  Dunson}{Bhattacharya et~al.}{2015}]{bhattacharya2015dirichlet}
Bhattacharya, A., D.~Pati, N.~S. Pillai, and D.~B. Dunson (2015).
\newblock {Dirichlet--Laplace Priors for Optimal Shrinkage}.
\newblock {\em Journal of the American Statistical Association\/}~{\em
  110\/}(512), 1479--1490.

\bibitem[\protect\citeauthoryear{Bierkens, Fearnhead, and Roberts}{Bierkens
  et~al.}{2019}]{bierkens2019zig}
Bierkens, J., P.~Fearnhead, and G.~Roberts (2019).
\newblock {The Zig-Zag Process and Super-efficient Sampling for Bayesian
  Analysis of Big Data}.
\newblock {\em Annals of Statistics\/}~{\em 47\/}(3), 1288--1320.

\bibitem[\protect\citeauthoryear{Bissiri, Holmes, and Walker}{Bissiri
  et~al.}{2016}]{bissiri2016general}
Bissiri, P.~G., C.~C. Holmes, and S.~G. Walker (2016).
\newblock {A General Framework for Updating Belief Distributions}.
\newblock {\em Journal of the Royal Statistical Society: Series B (Statistical
  Methodology)\/}~{\em 78\/}(5), 1103--1130.

\bibitem[\protect\citeauthoryear{Blei, Kucukelbir, and McAuliffe}{Blei
  et~al.}{2017}]{blei2017variational}
Blei, D.~M., A.~Kucukelbir, and J.~D. McAuliffe (2017).
\newblock {Variational Inference: a Review for Statisticians}.
\newblock {\em Journal of the American Statistical Association\/}~{\em
  112\/}(518), 859--877.

\bibitem[\protect\citeauthoryear{Carvalho, Polson, and Scott}{Carvalho
  et~al.}{2010}]{carvalho2010horseshoe}
Carvalho, C.~M., N.~G. Polson, and J.~G. Scott (2010).
\newblock {The Horseshoe Estimator for Sparse Signals}.
\newblock {\em Biometrika\/}~{\em 97\/}(2), 465--480.

\bibitem[\protect\citeauthoryear{Castillo, Schmidt-Hieber, and Van~der
  Vaart}{Castillo et~al.}{2015}]{castillo2015bayesian}
Castillo, I., J.~Schmidt-Hieber, and A.~Van~der Vaart (2015).
\newblock {Bayesian Linear Regression with Sparse Priors}.
\newblock {\em The Annals of Statistics\/}~{\em 43\/}(5), 1986--2018.

\bibitem[\protect\citeauthoryear{Chen, Rubanova, Bettencourt, and
  Duvenaud}{Chen et~al.}{2018}]{chen2018neural}
Chen, R.~T., Y.~Rubanova, J.~Bettencourt, and D.~Duvenaud (2018).
\newblock {Neural Ordinary Differential Equations}.
\newblock In {\em Advances in Neural Information Processing Systems}, pp.\
  261--272. Curran Associates, Inc.

\bibitem[\protect\citeauthoryear{Cobb, Baydin, Markham, and Roberts}{Cobb
  et~al.}{2019}]{cobb2019introducing}
Cobb, A.~D., A.~G. Baydin, A.~Markham, and S.~J. Roberts (2019).
\newblock {Introducing an Explicit Symplectic Integration Scheme for Riemannian
  Manifold Hamiltonian Monte Carlo}.
\newblock {\em arXiv preprint arXiv:1910.06243\/}.

\bibitem[\protect\citeauthoryear{Cobb and Jalaian}{Cobb and
  Jalaian}{2020}]{cobb2020scaling}
Cobb, A.~D. and B.~Jalaian (2020).
\newblock {Scaling Hamiltonian Monte Carlo Inference for Bayesian Neural
  Networks With Symmetric Splitting}.
\newblock {\em arXiv preprint arXiv:2010.06772\/}.

\bibitem[\protect\citeauthoryear{Cuturi}{Cuturi}{2013}]{cuturi2013sinkhorn}
Cuturi, M. (2013).
\newblock {Sinkhorn Distances: Lightspeed Computation of Optimal Transport}.
\newblock {\em Advances in Neural Information Processing Systems\/}~{\em 26},
  2292--2300.

\bibitem[\protect\citeauthoryear{Deheuvels et~al.}{Deheuvels
  et~al.}{1986}]{deheuvels1986influence}
Deheuvels, P. et~al. (1986).
\newblock {On the Influence of the Extremes of an IID Sequence on the Maximal
  Spacings}.
\newblock {\em The Annals of Probability\/}~{\em 14\/}(1), 194--208.

\bibitem[\protect\citeauthoryear{Devroye}{Devroye}{1982}]{devroye1982log}
Devroye, L. (1982).
\newblock {A Log Log Law for Maximal Uniform Spacings}.
\newblock {\em Annals of Probability\/}~{\em 10\/}(3), 863--868.

\bibitem[\protect\citeauthoryear{Dinh, Sohl-Dickstein, and Bengio}{Dinh
  et~al.}{2017}]{dinh2016density}
Dinh, L., J.~Sohl-Dickstein, and S.~Bengio (2017).
\newblock {Density Estimation using Real NVP}.
\newblock In {\em International Conference on Learning Representations}.

\bibitem[\protect\citeauthoryear{Doucet, Heng, and Pokern}{Doucet
  et~al.}{2021}]{doucetgibbs2021}
Doucet, A., J.~Heng, and Y.~Pokern (2021).
\newblock {Gibbs Flow for Approximate Transport With Applications to Bayesian
  Computation}.
\newblock {\em Journal of the Royal Statistical Society: Series B (Statistical
  Methodology)\/}, (in press).

\bibitem[\protect\citeauthoryear{Duan, Johndrow, and Dunson}{Duan
  et~al.}{2018}]{duan2018scaling}
Duan, L.~L., J.~E. Johndrow, and D.~B. Dunson (2018).
\newblock {Scaling Up Data Augmentation MCMC via Calibration}.
\newblock {\em The Journal of Machine Learning Research\/}~{\em 19\/}(1),
  2575--2608.

\bibitem[\protect\citeauthoryear{Dunson, Pillai, and Park}{Dunson
  et~al.}{2007}]{dunson2007Bayesian}
Dunson, D.~B., N.~Pillai, and J.-H. Park (2007).
\newblock {Bayesian Density Regression}.
\newblock {\em Journal of the Royal Statistical Society: Series B (Statistical
  Methodology)\/}~{\em 69\/}(2), 163--183.

\bibitem[\protect\citeauthoryear{El~Moselhy and Marzouk}{El~Moselhy and
  Marzouk}{2012}]{el2012Bayesian}
El~Moselhy, T.~A. and Y.~M. Marzouk (2012).
\newblock {Bayesian Inference with Optimal Maps}.
\newblock {\em Journal of Computational Physics\/}~{\em 231\/}(23), 7815--7850.

\bibitem[\protect\citeauthoryear{Fearnhead, Bierkens, Pollock, and
  Roberts}{Fearnhead et~al.}{2018}]{fearnhead2018piecewise}
Fearnhead, P., J.~Bierkens, M.~Pollock, and G.~O. Roberts (2018).
\newblock {Piecewise Deterministic Markov Processes for Continuous-time Monte
  Carlo}.
\newblock {\em Statistical Science\/}~{\em 33\/}(3), 386--412.

\bibitem[\protect\citeauthoryear{Giordano, Broderick, and Jordan}{Giordano
  et~al.}{2018}]{giordano2018covariances}
Giordano, R., T.~Broderick, and M.~I. Jordan (2018).
\newblock {Covariances, Robustness and Variational Bayes}.
\newblock {\em Journal of Machine Learning Research\/}~{\em 19\/}(1),
  1981--2029.

\bibitem[\protect\citeauthoryear{Girolami and Calderhead}{Girolami and
  Calderhead}{2011}]{girolami2011riemann}
Girolami, M. and B.~Calderhead (2011).
\newblock {Riemann Manifold Langevin and Hamiltonian Monte Carlo Methods}.
\newblock {\em Journal of the Royal Statistical Society: Series B (Statistical
  Methodology)\/}~{\em 73\/}(2), 123--214.

\bibitem[\protect\citeauthoryear{Ishwaran and Zarepour}{Ishwaran and
  Zarepour}{2002}]{ishwaran2002exact}
Ishwaran, H. and M.~Zarepour (2002).
\newblock {Exact and Approximate Sum Representations for the Dirichlet
  Process}.
\newblock {\em Canadian Journal of Statistics\/}~{\em 30\/}(2), 269--283.

\bibitem[\protect\citeauthoryear{Johndrow, Smith, Pillai, and Dunson}{Johndrow
  et~al.}{2019}]{johndrow2019mcmc}
Johndrow, J.~E., A.~Smith, N.~Pillai, and D.~B. Dunson (2019).
\newblock {MCMC for Imbalanced Categorical Data}.
\newblock {\em Journal of the American Statistical Association\/}~{\em
  114\/}(527), 1394--1403.

\bibitem[\protect\citeauthoryear{Kantorovich}{Kantorovich}{1942}]{kantorovich1942translocation}
Kantorovich, L.~V. (1942).
\newblock {On the Translocation of Masses}.
\newblock In {\em Dokl. Akad. Nauk. USSR (NS)}, Volume~37, pp.\  199--201.

\bibitem[\protect\citeauthoryear{Kingma and Ba}{Kingma and
  Ba}{2014}]{kingma2014adam}
Kingma, D.~P. and J.~Ba (2014).
\newblock {ADAM: a Method for Stochastic Optimization}.
\newblock In {\em International Conference on Learning Representations}.

\bibitem[\protect\citeauthoryear{Kingma and Dhariwal}{Kingma and
  Dhariwal}{2018}]{kingma2018glow}
Kingma, D.~P. and P.~Dhariwal (2018).
\newblock {Glow: Generative Flow with Invertible 1x1 Convolutions}.
\newblock In {\em Advances in Neural Information Processing Systems}, pp.\
  10215--10224.

\bibitem[\protect\citeauthoryear{Kingma, Salimans, Jozefowicz, Chen, Sutskever,
  and Welling}{Kingma et~al.}{2016}]{kingma2016improved}
Kingma, D.~P., T.~Salimans, R.~Jozefowicz, X.~Chen, I.~Sutskever, and
  M.~Welling (2016).
\newblock {Improved Variational Inference with Inverse Autoregressive Flow}.
\newblock In {\em Advances in Neural Information Processing Systems}, pp.\
  4743--4751.

\bibitem[\protect\citeauthoryear{Kingma and Welling}{Kingma and
  Welling}{2014}]{kingma2013auto}
Kingma, D.~P. and M.~Welling (2014).
\newblock {Auto-Encoding Variational Bayes}.
\newblock In {\em International Conference on Learning Representations}.

\bibitem[\protect\citeauthoryear{Kolouri, Nadjahi, Simsekli, Badeau, and
  Rohde}{Kolouri et~al.}{2019}]{NEURIPS2019_f0935e4c}
Kolouri, S., K.~Nadjahi, U.~Simsekli, R.~Badeau, and G.~Rohde (2019).
\newblock {Generalized Sliced Wasserstein Distances}.
\newblock In H.~Wallach, H.~Larochelle, A.~Beygelzimer, F.~d\textquotesingle
  Alch\'{e}-Buc, E.~Fox, and R.~Garnett (Eds.), {\em Advances in Neural
  Information Processing Systems}, Volume~32, pp.\  261--272. Curran
  Associates, Inc.

\bibitem[\protect\citeauthoryear{Kong and Chaudhuri}{Kong and
  Chaudhuri}{2020}]{pmlr-v108-kong20a}
Kong, Z. and K.~Chaudhuri (2020).
\newblock {The Expressive Power of a Class of Normalizing Flow Models}.
\newblock In {\em International Conference on Artificial Intelligence and
  Statistics}, Volume 108, pp.\  3599--3609.

\bibitem[\protect\citeauthoryear{Liang}{Liang}{2005}]{liang2005generalized}
Liang, F. (2005).
\newblock {A Generalized Wang--Landau Algorithm for Monte Carlo Computation}.
\newblock {\em Journal of the American Statistical Association\/}~{\em
  100\/}(472), 1311--1327.

\bibitem[\protect\citeauthoryear{Mengersen and Tweedie}{Mengersen and
  Tweedie}{1996}]{mengersen1996rates}
Mengersen, K.~L. and R.~L. Tweedie (1996).
\newblock {Rates of Convergence of the Hastings and Metropolis Algorithms}.
\newblock {\em Annals of Statistics\/}~{\em 24\/}(1), 101--121.

\bibitem[\protect\citeauthoryear{Miller, Foti, and Adams}{Miller
  et~al.}{2017}]{miller2017variational}
Miller, A.~C., N.~J. Foti, and R.~P. Adams (2017).
\newblock {Variational Boosting: Iteratively Refining Posterior
  Approximations}.
\newblock In {\em International Conference on Machine Learning}, pp.\
  2420--2429. PMLR.

\bibitem[\protect\citeauthoryear{Monge}{Monge}{1781}]{monge1781memoire}
Monge, G. (1781).
\newblock {M{\'e}moire sur la th{\'e}orie des d{\'e}blais et des remblais}.
\newblock {\em Histoire de l'Acad{\'e}mie Royale des Sciences de Paris\/}.

\bibitem[\protect\citeauthoryear{Neal}{Neal}{2011}]{neal2011mcmc}
Neal, R.~M. (2011).
\newblock {MCMC using Hamiltonian Dynamics}.
\newblock {\em Handbook of Markov Chain Monte Carlo\/}~{\em 2\/}(11), 2.

\bibitem[\protect\citeauthoryear{Nishimura, Dunson, and Lu}{Nishimura
  et~al.}{2020}]{nishimura2020discontinuous}
Nishimura, A., D.~B. Dunson, and J.~Lu (2020).
\newblock {Discontinuous Hamiltonian Monte Carlo for Discrete Parameters and
  Discontinuous Likelihoods}.
\newblock {\em Biometrika\/}~{\em 107\/}(2), 365--380.

\bibitem[\protect\citeauthoryear{Pakman and Paninski}{Pakman and
  Paninski}{2013}]{NIPS2013_a7d8ae45}
Pakman, A. and L.~Paninski (2013).
\newblock {Auxiliary-variable Exact Hamiltonian Monte Carlo Samplers for Binary
  Distributions}.
\newblock In C.~J.~C. Burges, L.~Bottou, M.~Welling, Z.~Ghahramani, and K.~Q.
  Weinberger (Eds.), {\em Advances in Neural Information Processing Systems},
  Volume~26, pp.\  2490--2498. Curran Associates, Inc.

\bibitem[\protect\citeauthoryear{Papamakarios, Pavlakou, and
  Murray}{Papamakarios et~al.}{2017}]{papamakarios2017masked}
Papamakarios, G., T.~Pavlakou, and I.~Murray (2017).
\newblock {Masked Autoregressive Flow for Density Estimation}.
\newblock In {\em Advances in Neural Information Processing Systems}, pp.\
  2338--2347.

\bibitem[\protect\citeauthoryear{Parno and Marzouk}{Parno and
  Marzouk}{2018}]{parno2018transport}
Parno, M.~D. and Y.~M. Marzouk (2018).
\newblock {Transport Map Accelerated Markov Chain Monte Carlo}.
\newblock {\em SIAM/ASA Journal on Uncertainty Quantification\/}~{\em 6\/}(2),
  645--682.

\bibitem[\protect\citeauthoryear{Piironen and Vehtari}{Piironen and
  Vehtari}{2017}]{piironen2017sparsity}
Piironen, J. and A.~Vehtari (2017).
\newblock {Sparsity Information and Regularization in the Horseshoe and Other
  Shrinkage Priors}.
\newblock {\em Electronic Journal of Statistics\/}~{\em 11\/}(2), 5018--5051.

\bibitem[\protect\citeauthoryear{Rajaratnam and Sparks}{Rajaratnam and
  Sparks}{2015}]{rajaratnam2015mcmc}
Rajaratnam, B. and D.~Sparks (2015).
\newblock {MCMC-based Inference in the Era of Big Data: A Fundamental Analysis
  of the Convergence Complexity of High-Dimensional Chains}.
\newblock {\em arXiv preprint arXiv:1508.00947\/}.

\bibitem[\protect\citeauthoryear{Rezende and Mohamed}{Rezende and
  Mohamed}{2015}]{pmlr-v37-rezende15}
Rezende, D. and S.~Mohamed (2015, 07--09 Jul).
\newblock {Variational Inference with Normalizing Flows}.
\newblock In {\em Proceedings of the 32nd International Conference on Machine
  Learning}, Volume~37, pp.\  1530--1538.

\bibitem[\protect\citeauthoryear{Robert and Casella}{Robert and
  Casella}{2013}]{casella1999monte}
Robert, C. and G.~Casella (2013).
\newblock {\em {Monte Carlo Statistical Methods}}.
\newblock Springer Science \& Business Media.

\bibitem[\protect\citeauthoryear{Robert, Elvira, Tawn, and Wu}{Robert
  et~al.}{2018}]{robert2018accelerating}
Robert, C.~P., V.~Elvira, N.~Tawn, and C.~Wu (2018).
\newblock {Accelerating MCMC Algorithms}.
\newblock {\em Wiley Interdisciplinary Reviews: Computational
  Statistics\/}~{\em 10\/}(5), e1435.

\bibitem[\protect\citeauthoryear{Roberts and Tweedie}{Roberts and
  Tweedie}{1996}]{roberts1996exponential}
Roberts, G.~O. and R.~L. Tweedie (1996).
\newblock {Exponential Convergence of Langevin Distributions and Their Discrete
  Approximations}.
\newblock {\em Bernoulli\/}~{\em 2\/}(4), 341--363.

\bibitem[\protect\citeauthoryear{Schilling}{Schilling}{2017}]{schilling2017measures}
Schilling, R.~L. (2017).
\newblock {\em {Measures, Integrals and Martingales}}.
\newblock Cambridge University Press.

\bibitem[\protect\citeauthoryear{Sojoudi}{Sojoudi}{2016}]{sojoudi2016equivalence}
Sojoudi, S. (2016).
\newblock {Equivalence of Graphical Lasso and Thresholding for Sparse Graphs}.
\newblock {\em Journal of Machine Learning Research\/}~{\em 17\/}(1),
  3943--3963.

\bibitem[\protect\citeauthoryear{Solomon, De~Goes, Peyr{\'e}, Cuturi, Butscher,
  Nguyen, Du, and Guibas}{Solomon et~al.}{2015}]{solomon2015convolutional}
Solomon, J., F.~De~Goes, G.~Peyr{\'e}, M.~Cuturi, A.~Butscher, A.~Nguyen,
  T.~Du, and L.~Guibas (2015).
\newblock {Convolutional Wasserstein Distances: Efficient Optimal
  Transportation on Geometric Domains}.
\newblock {\em ACM Transactions on Graphics\/}~{\em 34\/}(4), 1--11.

\bibitem[\protect\citeauthoryear{Spantini, Bigoni, and Marzouk}{Spantini
  et~al.}{2018}]{spantini2018inference}
Spantini, A., D.~Bigoni, and Y.~Marzouk (2018).
\newblock {Inference via Low-Dimensional Couplings}.
\newblock {\em The Journal of Machine Learning Research\/}~{\em 19\/}(1),
  2639--2709.

\bibitem[\protect\citeauthoryear{Tierney}{Tierney}{1994}]{tierney1994markov}
Tierney, L. (1994).
\newblock {Markov Chains for Exploring Posterior Distributions}.
\newblock {\em The Annals of Statistics\/}, 1701--1728.

\end{thebibliography}

\end{document}